\documentclass[%
 reprint,
 amsmath,amssymb,
 aps,
pra,
]{revtex4-2}
\DeclareMathAlphabet{\msfsl}{OT1}{cmss}{m}{sl}
\usepackage{graphicx}
\usepackage{dcolumn}
\usepackage{bm}
\usepackage[normalem]{ulem}
\usepackage{amsmath}    
\usepackage{amsthm}
\usepackage{natbib}
\newtheorem{theorem}{Theorem}

\usepackage{hyperref}
\usepackage{comment}
\usepackage{xcolor}
\usepackage{mathtools}

\newcommand{\pa}[1]{\left \langle #1 \right \rangle}

\newcommand{\Nv} {{N_\mathrm{v}}}
\newcommand{\Nh} {{N_\mathrm{h}}}

\makeatletter
\def\@makefntext#1{\noindent\@makefnmark~#1} 
\makeatother

\definecolor{myblue}{RGB}{12, 12, 158}
\definecolor{myred}{RGB}{158, 19, 22}
\definecolor{myorange}{RGB}{245, 150, 12}
\definecolor{mygreen}{RGB}{26, 148, 49}
\definecolor{Prune}{RGB}{99,0,60}
\definecolor{Purple}{RGB}{75, 0, 130}
\definecolor{Pink}{RGB}{255, 105, 180}
\definecolor{deepskyblue}{RGB}{0, 191,255}
\definecolor{limegreen}{RGB}{50, 205, 50}
\definecolor{crimson}{rgb}{0.86, 0.08, 0.24}
\definecolor{coral}{rgb}{1.0, 0.5, 0.31}
\definecolor{blue(ncs)}{rgb}{0.0, 0.53, 0.74}

\newcommand{\figpanel}[1]{(\textbf{\lowercase{#1}})}

\begin{document}

\preprint{APS/123-QED}

\title{Inferring Higher-Order Couplings with Neural Networks}

\author{Aurélien Decelle}
 \affiliation{Escuela Técnica Superior de Ingenieros Industriales, Universidad Politécnica de Madrid, Calle de José Gutiérrez Abascal 2, Madrid 28006,
Spain}
 \affiliation{Departamento de Física Teórica, Universidad Complutense de Madrid, 28040 Madrid, Spain}

\author{Alfonso de Jesús Navas Gómez}
\email{alfonn01@ucm.es}
\affiliation{Departamento de Física Teórica, Universidad Complutense de Madrid, 28040 Madrid, Spain}
\affiliation{Université Paris-Saclay, CNRS, INRIA Tau team, LISN, 91190 Gif-sur-Yvette, France}

\author{Beatriz Seoane}
\affiliation{Departamento de Física Teórica, Universidad Complutense de Madrid, 28040 Madrid, Spain}

\begin{abstract}
Maximum entropy methods, rooted in the inverse Ising or Potts problem of statistical physics, have been fundamental in inferring pairwise interaction models of complex systems. These models have found impactful applications in fields such as bioinformatics, where they are used to infer protein structural contacts, and in neuroscience, for describing patterns of neural activity. Despite their considerable success, these approaches fail to capture higher-order interactions that could be crucial for a full description of collective behavior. In contrast, modern machine learning methods can model such interactions, but their interpretability usually comes at a prohibitive computational cost. Restricted Boltzmann Machines (RBMs) provide a computationally efficient alternative by encoding statistical correlations through hidden units in a bipartite architecture. In this work, we introduce a method that maps RBMs onto generalized Potts models, enabling the systematic extraction of interactions up to arbitrary order. Leveraging large-$N$ approximations, made tractable by the RBM’s structure, we extract effective many-body couplings with minimal computational effort.
We further propose a robust framework for recovering higher-order interactions in more complex generative models, and introduce a simple gauge-fixing scheme for the effective Potts representation. Validation on synthetic data demonstrates accurate recovery of two- and three-body interactions. Applied to protein sequence data, our method reconstructs contact maps with high fidelity and outperforms state-of-the-art inverse Potts models. These results establish RBMs as a powerful and efficient tool for modeling higher-order structure for high-dimensional categorical data.
\end{abstract}

\maketitle

\textit{Introduction.---}
The Maximum Entropy (ME) principle in data-driven modeling consists of fitting empirical low-order moments of the data, such as means and covariances, while making minimal assumptions about unobserved variables. In the case of network inference with categorical variables, ME data modeling is formulated as an inverse Potts problem of statistical mechanics, i.e., we seek the fields and pairwise couplings of a Potts Hamiltonian that best describe the observed data. This approach is interpretable, easy to use, and provides meaningful insights even in situations with limited data. It has provided a framework for understanding the underlying dynamics of many complex systems~\cite{nguyen2017inverse}, including neural circuits~\cite{schneidman2006weak,roudi2009statistical,meshulam2017collective,maoz2020learning}, gene networks~\cite{de2018introduction,lezon2006using,tkavcik2008information}, and protein structures~\cite{ weigt2009identification, ekeberg2013improved, ekeberg2014fast, cocco2018inverse}. However, the expressiveness of pairwise models is inherently limited because they reduce all group interactions to pairwise terms, which precludes capturing the higher-order interactions crucial in real-world systems~\cite{battiston2021physics,roudi2009pairwise}.
On the other hand, although modern generative machine learning models have recently achieved significant breakthroughs, such as predicting protein structures from sequences~\cite{jumper2021highly,abramson2024accurate}, they often struggle in data-scarce scenarios and lack interpretability. Furthermore, extracting meaningful and understandable insights from their parameters remains challenging, with significant ongoing efforts~\cite{rao2020transformer, feinauer2022mean, feinauer2022interpretable, caredda2024direct, merger2023learning}.

An alternative that effectively mediates between simple pairwise modeling and modern deep learning frameworks is the Restricted Boltzmann Machine (RBM)~\cite{Smolensky, ackley1985learning}. In contrast to traditional pairwise models, an RBM operates on a bipartite graph, where only one layer represents the observable data. Introducing latent or hidden variables enables RBMs to act as universal approximators~\cite{le2008representational}, significantly increasing their expressive power while keeping the number of parameters to be learned under control. Recent studies have shown that binary RBMs can be reinterpreted as a generalized Ising or lattice gas model with interactions that go beyond pairwise to include higher-order terms~\cite{cossu2019machine, beentjes2020higher, bulso2021restricted, feng2023statistical, decelle2024inferring}. Furthermore, this mapping has proven useful for inference applications, as it enables the extraction of coupling parameters from simulated data with multi-body interactions using equal or fewer parameters than analogous inverse Ising models~\cite{decelle2024inferring}. This highlights the superior ability of RBMs to model complex systems, offering enhanced performance over standard maximum entropy techniques without additional computational costs and providing practical techniques to assess the explainability of such models~\cite{di2025unbearable}. 

Most applications of ME focus on categorical variables rather than binary ones. A prominent example is Direct Coupling Analysis (DCA)~\cite{morcos2011direct,weigt2009identification}, which uses inverse Potts models to infer epistatic couplings from the Multiple Sequence Alignment (MSA) of protein~\cite{cocco2018inverse} or RNA families~\cite{cocco2023statistical}. MSAs are alignments of sequences from homologous families, groups of proteins or RNAs with a common evolutionary ancestor, to identify conserved regions and reveal structural, functional, and evolutionary relationships.
Recent studies introduced methods to identify the closest pairwise model distribution to a trained probabilistic model, enabling the extraction of effective pairwise couplings from modern neural networks~\cite{feinauer2022interpretable,tubiana2019learning}, including RBMs~\cite{tubiana2019learning}. However, these approaches are computationally intensive, requiring either averaging over numerous single-site data permutations or performing sampling, which limits their applicability to pairwise interactions. In this work, we propose a generic theoretical expression to infer \(n\)-th order couplings from generic categorical probabilistic models and introduce a fast, reliable framework to approximate high-order couplings in Potts-Bernoulli RBMs~\cite{tubiana2019learning, decelle2023unsupervised}, leveraging their straightforward marginal energy function structure. We demonstrate that the derived expressions accurately infer pairwise and three-wise couplings in controlled experiments, enabling protein contact predictions from trained RBMs with accuracy comparable to that of state-of-the-art DCA methods.

\begin{figure*}[t!]
\centering
\raisebox{1.0cm}{\includegraphics[height=2.4cm,trim=20 0 20 0,clip]{figs/rbm_architecture.pdf}}
\includegraphics[height=4.cm,trim=10 0 0 0,clip]{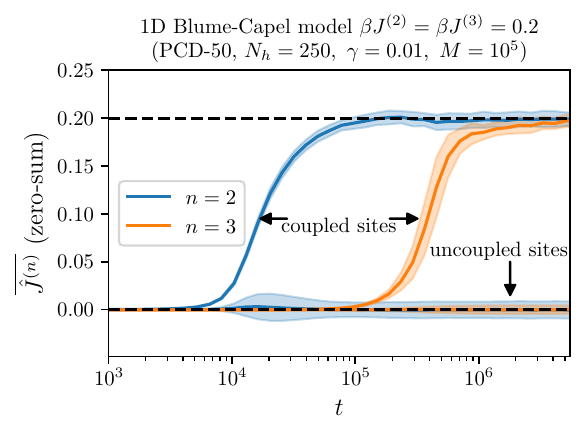}
\raisebox{0.25cm}{\includegraphics[height=3.4cm,trim=0 0 0 0,clip]{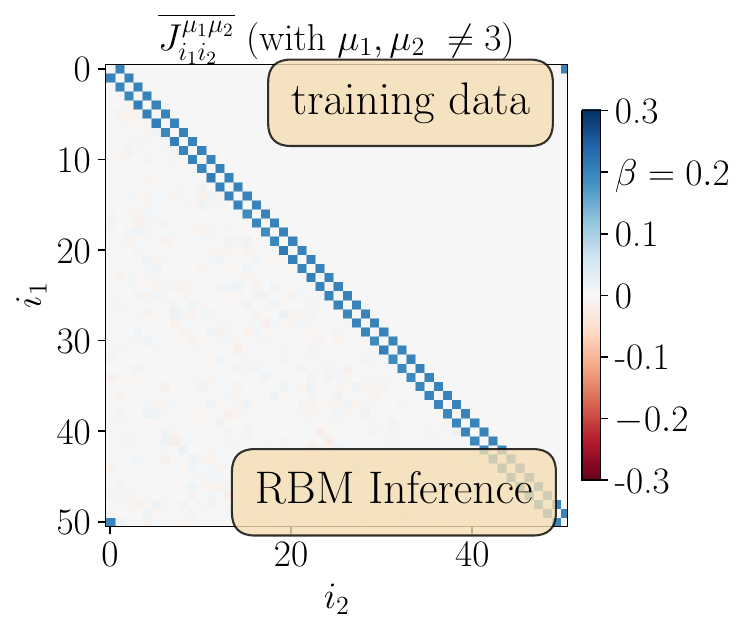}}
\raisebox{0.15cm}{
\includegraphics[height=3.6cm,trim=0 0 0 0,clip]{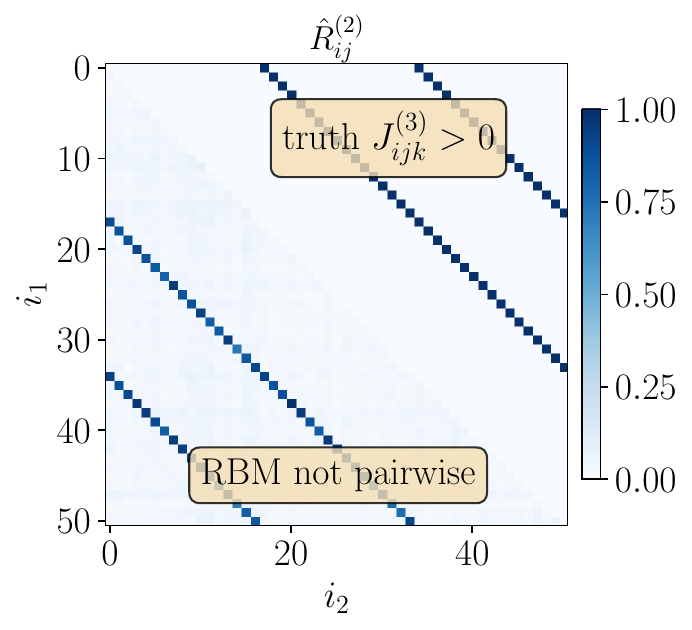}}
\put(-500, 110){\figpanel{a}}
\put(-380, 110){\figpanel{b}}
\put(-220, 110){\figpanel{c}}
\put(-100, 110){\figpanel{d}}
\caption{{\figpanel{a} Potts-Bernoulli RBM architecture with binary hidden units $h_a$ and $q$-state visible units $v_i$.
\figpanel{b} Time evolution of effective couplings inferred by an RBM trained on equilibrium configurations from the spin-1 Blume-Capel model (Eq.~\eqref{blume-capel_hamiltonian}, $\beta J^{(2)} = \beta J^{(3)} = 0.2$). Solid lines show mean inferred values in the zero-sum gauge for $n=2$ and $n=3$ interactions, averaged separately over coupled and uncoupled terms; shaded areas indicate standard deviations among pairs/triplets. Dashed lines mark ground-truth couplings. Training hyperparameters are noted in the title: PCD-$k$ (persistent contrastive divergence with $k$ steps), $\Nh$ (number of hidden units), and $\gamma$ (learning rate).
\figpanel{c} Inferred pairwise coupling matrix (below diagonal) vs. ground truth (above diagonal).
\figpanel{d} Normalized signal $\hat{R}^{(2)}_{i_1,i_2}$ indicating deviations from a purely pairwise model, compared to the true three-body interactions, displayed as nonzero entries across the corresponding pairs in a $2 \times 2$ grid.}}
\label{fig:blumecapel}
\end{figure*}

\textit{The Potts-Bernoulli RBM.---} Consider an undirected stochastic neural network defined on a bipartite lattice, see Fig. \ref{fig:blumecapel}--\figpanel{a},
a visible layer $\boldsymbol{v} \!=\! \{ v_i \}_{i=1}^{\Nv}$, which represents the data, and a hidden layer $\boldsymbol{h} \!=\! \{ h_a \}_{a=1}^{\Nh}$, which encodes the interactions among the visible variables. As in other energy-based generative models, the joint probability of any given configuration $\{ \boldsymbol{v}, \boldsymbol{h} \}$ is given by the Boltzmann distribution
\begin{gather}
    p (\boldsymbol{v}, \boldsymbol{h})\!=\!\frac{1}{\mathcal{Z}} e^{-\mathcal{H}(\boldsymbol{v}, \boldsymbol{h})} \ \mathrm{where,} \ \mathcal{Z} \!=\! \!\sum_{\{\boldsymbol{v}, \boldsymbol{h}\}} e^{-\mathcal{H}(\boldsymbol{v}, \boldsymbol{h})}
    \label{Boltzmann_distribution_RBM}
\end{gather}
with \(\mathcal{H}(\boldsymbol{v}, \boldsymbol{h})\) being the \textit{energy function} or the \textit{Hamiltonian} of the model. We define the hidden nodes as Bernoulli variables, i.e., \( h_a \!\in\! \{ 0, 1 \} \), and the visible nodes as categorical variables, or Potts ``spins", that can take on $q$ states or colors, i.e., \( v_i \!\in\! \{ 1, \dots, q \} \). Hence, the Hamiltonian of such a system is given by
\begin{equation}
    \mathcal{H}(\boldsymbol{v}, \boldsymbol{h}) 
    \!=\! - \! \sum_{i, a, \mu } \delta_\mu^{v_i} W_{i a}^\mu h_a \!-\! \sum_{i, \mu} \delta_\mu^{v_i}  b_i^\mu \!-\! \sum_{a} c_a h_a
    \label{RBM_potts}
\end{equation}
where, \(\delta_i^j \) denotes the Kronecker delta ($\delta_{i}^{j} \!=\!1$ if $i\!=\!j$, and  $\delta_i^j\!=\!0$ otherwise), and $\bm \Theta \!\coloneqq \! \{\bm W, \bm b,\bm c\}$ are the  model parameters: the \textit{weight tensor} $\boldsymbol{W} \!=\! \{ W_{ia}^\mu \}, $ which is a rank-3 tensor that models the interactions between the visible and hidden layers; and the \textit{visible} and \textit{hidden} \textit{biases}, denoted by $\boldsymbol{b} \!= \! \{ b_i^\mu \}$ and $\boldsymbol{c} \!=\! \{ c_a \}$, respectively. For clarity, we have used Latin indexes $i \!\in\! \{ 1, \dots, \Nv \}$ and $a \!\in\! \{ 1, \dots, \Nh \}$ to denote visible and hidden sites, respectively, and Greek ones $\mu \!\in\! \{1, \dots, q \}$ to indicate the possible colors of the visible variables. 

The parameters of an RBM are typically trained through likelihood maximization to ensure that the marginal distribution over the visible units, \( \msfsl{p}(\boldsymbol{v}) \!=\! \sum_{\bm h}  p(\boldsymbol{v}, \boldsymbol{h}) \), closely matches the empirical distribution of a dataset \( \mathcal{D}=\{\bm v^{(m)}\}_{m=1}^M \) containing $M$ entries, i.e., \( p_\mathcal{D}(\boldsymbol{v}) = \frac{1}{M}\sum_{m=1}^M \delta(\bm v - \bm v^{(m)}) \), where \(\delta\) represents the \(\Nv\)-dimensional Dirac delta distribution. For details about the training process, see Appendix~\ref{sec:training}.

It is easy to verify that the following transformation:  
\[
\textstyle W_{ia}^\mu \rightarrow W_{ia}^\mu + A_{ia}, \;\; b_i^\mu \rightarrow b_i^\mu + B_i, \;\; c_a \rightarrow c_a - \sum_i A_{ia},
\]  
leaves the Boltzmann distribution in \eqref{Boltzmann_distribution_RBM} unchanged. {Such invariance is a natural consequence of the over-parametrization of models involving generalized Potts spins \cite{cocco2018inverse, tubiana2019learning}}. Fixing the model gauge is helpful in uniquely defining the inference problem and ensuring efficient convergence to optimal parameters during training. The most common gauges are the \textit{lattice gas} gauge, which assumes \(\dot{b}_i^q \!=\! \dot{W}_{ia}^q\! =\! 0, \ \forall i, a,\) and the \textit{zero-sum} gauge, defined by \(\sum_\mu \hat{b}_i^\mu = \sum_\mu \hat{W}_{i a}^\mu=0, \  \forall i, a\). In the above, we used dot and hat symbols to denote couplings in the lattice-gas and zero-sum gauges, respectively. From here on, we will maintain this convention and use plain symbols for expressions that hold independently of the gauge. During RBM training, we did not observe any clear advantage in using one gauge over the other. For consistency, we adopted the zero-sum gauge in all experiments presented.

\textit{Generalized Multi-body Potts Model.---} The binary RBM is a universal approximator~\cite{le2008representational}, capable of modeling any multivariate random variable over a binary vector space given sufficient hidden units. Thus, when the dataset comprises equilibrium configurations of a multi-body Ising Hamiltonian, RBM learning can capture interactions of any order among visible units, provided that enough data is available. This theoretical property has been demonstrated in practice: recent studies~\cite{cossu2019machine, beentjes2020higher, feng2023statistical, decelle2024inferring} have shown that the energy function of a binary RBM can be precisely mapped to an effective Ising model with interactions of all orders, represented by coupling tensors dependent on the RBM parameters $\bm \Theta$. Additionally, Ref.~\cite{decelle2024inferring} demonstrated that the mapping between RBMs and interacting physical models can accurately reconstruct the underlying interactions, including three-body terms, used to generate synthetic data in controlled inverse experiments. Building on this foundation, we propose a theoretical and practical approach to map any categorical RBM onto a Potts Hamiltonian.
We begin with the marginalized Hamiltonian of the Potts-Bernoulli RBM, obtained by summing over the hidden units, i.e.,  
\(
\msfsl{p}(\boldsymbol{v})=\sum_{\bm h}p(\bm v,\bm h) \propto e^{ -\msfsl{H}(\boldsymbol{v})},
\)
which is given by  
\begin{equation}
    \msfsl{H}(\boldsymbol{v}) 
    \!=\! - \sum_{i,\mu} b_i^\mu \delta_\mu^{v_i} \!-\! \sum_a \ln \left(1 + e^{ c_a + \sum_{i, \mu} W_{ia}^{\mu} \delta_\mu^{v_i}} \right),
    \label{RBM_Potts_marginalized}    
\end{equation}
and map it into an effective Potts Hamiltonian with multi-body interactions,  
 \begin{equation}
    \mathcal{H} (\boldsymbol{v}) 
    \!=\! -\! \sum_{i, \mu} H_i^\mu \delta_\mu^{v_i} - \! \! \! \! \! \! \!\sum_{1 \le i_1 < i_2 \le \Nv} \sum_{\mu_1, \mu_2} \! \! J_{ i_1 i_2}^{\mu_1 \mu_2} \delta_{\mu_1}^{v_{i_1}} \delta_{\mu_2}^{v_{i_2}} + \dots
    \label{general_Potts}
\end{equation}
The goal is to derive explicit expressions for the fields \(H_i^\mu\) and \(n\)-th order couplings \(J_{i_1 \dots i_n}^{\mu_1 \dots \mu_n}\) in terms of the RBM parameters \(\bm\Theta\), providing a physical interpretation of the learned model. We note that model \eqref{general_Potts} needs gauge fixing, just as we did with the Potts-Bernoulli Hamiltonian. In this case, the lattice-gas gauge condition is 
$
   \dot{H}_i^q = \dot{J}_{i_1 i_2}^{\mu_1 q} \!= \dots = \! \dot{J}_{i_1 \dots i_{\Nv}}^{\mu_1 \dots q}\! =\! 0,$
while that of the zero-sum gauge is
\begin{equation}
    \textstyle\sum_{\mu'} \hat H_i^{\mu'}\! =\! \sum_{\mu'} \hat J_{i_1 i_2}^{\mu_1 \mu'} \! = \dots =\! \sum_{\mu'} \hat J_{i_1 \dots i_{\Nv} }^{\mu_1 \dots {\mu'}} \!=\! 0,
    \label{zero_sum_condition_gm}
\end{equation}
for all $ i, i_1, \dots, i_\Nv$ and $\mu_1, \dots, \mu_{\Nv}$. 

{Gauge transformations of the couplings leave the Boltzmann distribution unchanged, but quantities that depend explicitly on the couplings may be gauge-dependent}. In DCA, the zero-sum gauge is often preferred because it minimizes the Frobenius norm of the couplings \footnote{ {The Frobenius norm of the pairwise couplings, given by $ F_{ij}^{(2)} \coloneqq \!\!\! \sqrt{\sum_{\mu, \nu} \left( J_{ij}^{\mu \nu} \right)^2}$, quantifies the interaction strength between sites $i$ and $j$.}}, thereby emphasizing field contributions and reducing the role of pairwise terms \cite{cocco2018inverse}. As shown in Appendix \ref{gauge_appendix}, this property extends to higher-order Potts models. Empirically, we observed improved inference performance across all tested scenarios, consistent with previous findings for binary variables~\cite{decelle2024inferring} and protein contact prediction~\cite{feinauer2014improving}. Accordingly, we adopt the zero-sum gauge in the main text; results for the lattice-gas formulation are reported in Appendix \ref{latticegas_appendix}.

\textit{Effective Couplings of an RBM.---} 
The marginal Hamiltonian in Eq.~\eqref{RBM_Potts_marginalized} admits a natural expansion into explicit $n$-body interaction terms, thereby defining an effective generalized Potts model in the lattice-gas gauge. The full derivation, including explicit expressions for the corresponding effective fields and couplings as functions of $\bm{\Theta}$, is presented in Appendix~\ref{derivation_lattice_gauge_appendix}. These expressions can be algebraically transformed to the zero-sum gauge, which we adopt for inference purposes; the details of this transformation are given in Appendix~\ref{derivation_2_appendix}. The resulting $n$-th order effective coupling in the zero-sum gauge reads:
\begin{gather}
   \textstyle \hat{J}_{i_1 \dots i_n}^{\mu_1 \dots \mu_n} \!
    = \! \sum_{K \subseteq [n] }  (-1)^{n - |K|} \Bigg[
    {q^{-\Nv}} \! \sum_{\mu'_1\dots\mu'_{\Nv}} \!\!\!  \nonumber \\ \textstyle
     \sum_a \ln \! \left( 1\! + \!e^{c_a +  \sum_{ k \in K } \hat{W}_{i_k a}^{\mu_k} + \sum_{ l \in [\Nv] \setminus K} \hat{W}_{i_l a}^{\mu'_l} } \right) \Bigg].
     \label{n-body_formula_zs}
\end{gather}

Here, we introduced the set $[n] = {1, 2, \dots, n}$ and use $|A|$ to denote the cardinality (i.e., number of elements) of a set $A$. Throughout, we use sub-indices $k, l \in [\Nv]$ to indicate arbitrary sites in the system. Finally, we note that in the case $n\! =\! 1$, the effective fields in all gauges are given by $H_{i_1}^{\mu_1} \!=\! b_{i_1}^{\mu_1}\! +\! J_{i_1}^{\mu_1}$
where both $b_{i_1}^{\mu_1}$ and $J_{i_1}^{\mu_1}$ are defined in the same gauge as $H_{i_1}^{\mu_1}$.

Eq.~\eqref{n-body_formula_zs} introduces an average over all possible $q^{\Nv}$ configurations, rendering the direct computation of effective interactions in the zero-sum gauge intractable even for moderately sized systems. Nevertheless, we will show that this expression can be efficiently approximated using the Central Limit Theorem (CLT). In practice, inferring the effective model in the lattice-gas gauge is computationally much faster (See equation \eqref{n-body_formula}) in Appendix~\ref{derivation_lattice_gauge_appendix}, which can be evaluated directly without making approximations. Yet, the zero-sum gauge consistently yields better inference performance as we show in Appendix~\ref{latticegas_appendix}.

\textit{Effective Couplings of Generic Probabilistic Models.---} 
The formulas derived for the effective $n$-body couplings of the marginalized RBM can be generalized to any probabilistic model $\pi \colon \boldsymbol{\Omega} \to (0,1) \subset \mathbb{R}$ defined over a high-dimensional categorical space $\boldsymbol{\Omega} \coloneqq \{1, 2, \dots, q\}^{N_v}$. Specifically, such a model can be mapped to a multibody generalized Potts model \eqref{general_Potts}, where the $n$-th order couplings are given by
\begin{gather}
J_{i_1\dots i_n}^{\mu_1 \dots \mu_n} \!\!= \!\!\!\!\! \sum_{K \subseteq [n]}\! \!\! (-\!1)^{n - |K|}\! \!\ \mathbb{E}_{\boldsymbol{u} \sim G } \ln \pi \left( \boldsymbol{u} \big| {u_{i_{k}} \!\!=\! \mu_{k} \!:\! k \!\in\! K } \right) .\!\!
\label{n-body_formula_general}
\end{gather}
In the above we denoted by $\pi \left( \boldsymbol{u} \big| {u_{i_{k}} \!\!=\! \mu_{k} \!:\! k \!\in\! K } \right)$ the value of the probability mass function $\pi$ evaluated in a mutated configuration from $\boldsymbol{u}$ in which $u_{i_k} \!=\! \mu_k$, with $k\!\in\!K$. Besides, we used $\mathbb{E}_{\boldsymbol{u} \sim G}[ \cdots ]$ to denote the average of the variable $\boldsymbol{u}$ distributed over an arbitrary probability measure $G$ of $\boldsymbol{\Omega}$. This arbitrariness reflects the gauge invariance of generalized Potts models mentioned earlier. Thus, we have the lattice gas gauge when $G$ is the degenerate measure that assigns probability 1 to the configuration $(q, q, \dots, q)$ and 0 to any other $\boldsymbol{v} \!\in\! \boldsymbol{\Omega}$; and the zero-sum gauge if $G$ is uniformly distributed over $\boldsymbol{\Omega}$. It is easy to check that we can recover expressions \eqref{n-body_formula} and \eqref{n-body_formula_zs} by replacing $\pi(\boldsymbol{v}) \!\propto \! \exp \left[ \sum_{i,\mu} b_i^\mu \delta_\mu^{v_i} + \! \sum_a \ln \left(1 + e^{ c_a + \sum_{i, \mu} W_{ia}^{\mu} \delta_\mu^{v_i}} \right) \right]$ in \eqref{n-body_formula_general} considering $G$ either as the degenerate or the uniform measure, respectively. A general proof of the Potts equivalence and gauge fixing using \eqref{n-body_formula_general} is provided in Appendix~\ref{gauge_appendix}. It is worth adding that for energy-based models, i.e., when $\pi (\boldsymbol{v}) \!\propto\! e^{- \msfsl{H}(\boldsymbol{v})}$, we can replace $\ln \pi (\boldsymbol{v})$ in \eqref{n-body_formula_general} by the negative energy function $-\msfsl{H}(\boldsymbol{v})$ because the partition function cancels out in the sum. 

\textit{Higher-order interactions---} Computing all interactions beyond second order is typically computationally prohibitive. To probe the relevance of higher-order interactions, we introduce an $\mathcal{O}(\Nv^2)$ projection. Given an arbitrary categorical probability mass function $\pi \colon \Omega \to (0,1)$, we define the likelihood differential ratio as follows:
\begin{equation*}
\hat{r}_{{i}{j}}^{\mu \nu}(\boldsymbol{v}) \! \coloneqq \! \frac{1}{q^2}\! \sum_{\mu'\!,\! \nu'}\! \ln\! \left[\! \frac{\pi \big(\boldsymbol{v}|v_i \!=\!\mu,\! v_j \!=\! \nu \big) \pi \big(\boldsymbol{v} |  v _i \!=\! \mu'\!,\! v_j \!=\! \nu' \big)}{\pi(\boldsymbol{v}|v_i \!=\!\mu,\! v_j = \nu') \pi(\boldsymbol{v}| v_i\!=\!\mu'\!,\! v_j \!=\!\nu)}\! \right]\!.
\end{equation*}
For a model $\pi(\boldsymbol{v})$ with strictly pairwise interactions between sites $i_1$ and $i_2$, it is straightforward to verify that $\hat{r}_{i_1 i_2}^{\mu \nu}(\boldsymbol{v}) = \hat{J}_{i_1 i_2}^{\mu \nu}$. The log-likelihood ratio $r_{ij}^{\mu \nu}(\boldsymbol{v})$ has been used to infer effective pairwise couplings in models where these are not explicitly defined, including auto-regressive models \cite{trinquier2021efficient, feinauer2022interpretable}, variational autoencoders \cite{feinauer2022interpretable}, and RBMs \cite{tubiana2019learning}. Refs. \cite{tubiana2019learning} and \cite{feinauer2022interpretable} computed averages of $r_{ij}^{\mu \nu}(\boldsymbol{v})$, with the former using dataset samples and the latter employing uniform samples to impose a zero-sum gauge, while also developing sampling-based estimators for other gauge choices. In fact, a direct comparison with Eq.~\eqref{n-body_formula_zs} at $n\!=\!2$ (or Eq.~\eqref{2-body_couplings_zs}) shows that $
    \hat{J}_{{i}{j}}^{\mu \nu} = \mathbb{E}_{\boldsymbol{u} \sim U}\big[ \hat{r}_{{i}{j}}^{\mu \nu} (\boldsymbol{u}) \big]$, which agrees with Ref. \cite{feinauer2022interpretable}.  Additionally, to assess whether sites $i$ and $j$ interact through higher-than-pairwise effective terms, one can compute $
\mathrm{Var}_{ \boldsymbol{u} \sim {U}} \big[ \hat{r}_{{i}{j}}^{\mu \nu} (\boldsymbol{u}) \big] \!=\! \mathbb{E}_{\boldsymbol{u} \sim U} \left[ \hat{J}_{{i}{j}}^{\mu \nu} - \hat{r}_{{i}{j}}^{\mu \nu} (\boldsymbol{u}) \right]^2$, 
and define the quantity
\begin{equation}
    \hat{{R}}_{{i}{j}}^{(2)} = \sqrt{\frac{1}{q^2}\sum_{\mu, \nu}  \mathrm{Var}_{ \boldsymbol{u} \sim {U}} \big[ \hat{r}_{{i}{j}}^{\mu \nu} (\boldsymbol{u}) \big] },
    \label{Frobenius_high_order}
\end{equation}
where $ \hat{{R}}_{{i_1}{i_2}}^{(2)} \!\approx \! 0$ indicates that sites $i_1$ and $i_2$ interact predominantly pair-wisely in the effective model. Analogous quantities can be defined to quantify deviations from an independent-site model, such as $\hat{{R}}_{i}^{(1)}$, or from higher-order interactions, such as $\hat{{R}}_{i_1 \dots i_n}^{(n)}$. For completeness, we include the complete formulas for $\hat{r}_{i}^{\mu} (\boldsymbol{v})$ and $\hat{r}_{i_1 \dots i_n}^{\mu_1 \dots \mu_n }(\boldsymbol{v})$:
\begin{align*}
    \hat{r}_{i}^{\mu} (\boldsymbol{v}) 
    &\coloneq {\textstyle \frac{1}{q} \sum_{\mu'} \ln \frac{\pi(\boldsymbol{v}|v_i = \mu)}{\pi(\boldsymbol{v}|v_i = \mu')}} \\
    \hat{r}_{i_1 \dots i_n}^{\mu_1 \dots \mu_n} (\boldsymbol{v})  &\coloneq {\textstyle \sum_{K \subseteq [n]} (-1)^{n-|K|} \bigg[ \frac{1}{q^n} \sum_{\mu'_1, \dots \mu'_n}} \\
    & \qquad \ln \pi \big(\boldsymbol{v}|v_{i_k} \!\!\!=\! \mu_{i_k} \!, v_{i_l} \!\!\!=\! \mu_{i_l}' \colon k \!\in\! K, l \!\in\! [n]\!\setminus\!K \big) \bigg].
\end{align*}

\textit{Large-$N$ Approximations.---} Computing effective couplings in the zero-sum gauge requires averaging over $q^{\Nv}$ configurations, which is intractable for realistic system sizes. Previous works have approximated this average using samples to extract pairwise interactions~\cite{feinauer2022interpretable, tubiana2019learning}, but the method can be too computationally demanding to extend to higher-order terms.
Ref.~\cite{decelle2024inferring} introduced an approximation scheme that leverages the structure of the RBM energy function and the CLT to approximate the {average inside square brackets} in \eqref{n-body_formula_zs}. We outline the main steps of the approximation below; full details are provided in Appendix~\ref{CLT_appendix}. By employing this approach to compute the \(n\)-th order couplings, we initially reduce the sum over all possible configurations, which contains \(q^{\Nv}\) elements, to a sum over \(q^n\) elements {using the following identity}
\begin{align*}
   & \frac{1}{q^{\Nv}} \!\!\sum_{\mu'_1\dots\mu'_{\Nv}} \!\! \ln \left(  1 + e^{c_a +  \sum_{k=1}^{\Nv} \hat{W}_{i_k a}^{\mu'_k} }\right) \nonumber \\ \!\!
    &\ \!=\! \frac{1}{q^{n}} \!\!\! \sum_{\mu'_1\dots\mu'_n} \!\!\! \mathbb{E}_{x \sim {X_a^{\setminus \{ i_1, \dots, i_n \}}}
    } \!\! \left[ \ln \left(  1 + e^{c_a +  \sum_{k=1}^{n} \hat{W}_{i_k a}^{\mu'_k} + x} \right) \right]\!,\!\!
\end{align*}
where, $X_a^{ \setminus \left\{ i_1, \dots, i_n \right\}} \! \coloneqq \!  \sum_{l = n+1}^{\Nv}\! \hat{W}_{i_l a}^\ast$  is a random variable with each $\hat{W}_{i_l a}^\ast$ drawn uniformly from $\{ \hat{W}_{i_l a}^\mu \colon \mu \in [q] \}$. {By the CLT, when $n \ll \Nv$, the variables ${X_a^{\setminus \{ i_1, \dots, i_n \}}}$ are approximately Gaussian, allowing the sum to be approximated by a Gaussian integral, which we compute numerically. In practice, discretizing the integral using $\sim$20 points provides accurate estimates. However, the CLT may not apply in certain instances, particularly when the weight matrix $\hat{W}_{i a}^\mu$ is sparse, allowing the sum to be computed exactly, or when a few entries dominate, in which case the distribution is better captured by a Gaussian mixture, as discussed in Appendix~\ref{CLT_appendix}. In all cases considered in this work, corrections to the effective parameters due to deviations from Gaussianity were either negligible or absent. An efficient parallelization scheme to compute the couplings, including corrections to the Gaussian approximation, is available on ~\href{https://github.com/DsysDML/couplings_inference}{GitHub}~\cite{github:couplings_inference}. Table~\ref{tab:computational_times} shows detailed extraction times for fields, 2-body, and 3-body couplings.

\begin{figure*}[t]
    \centering
    \includegraphics[height=4cm]{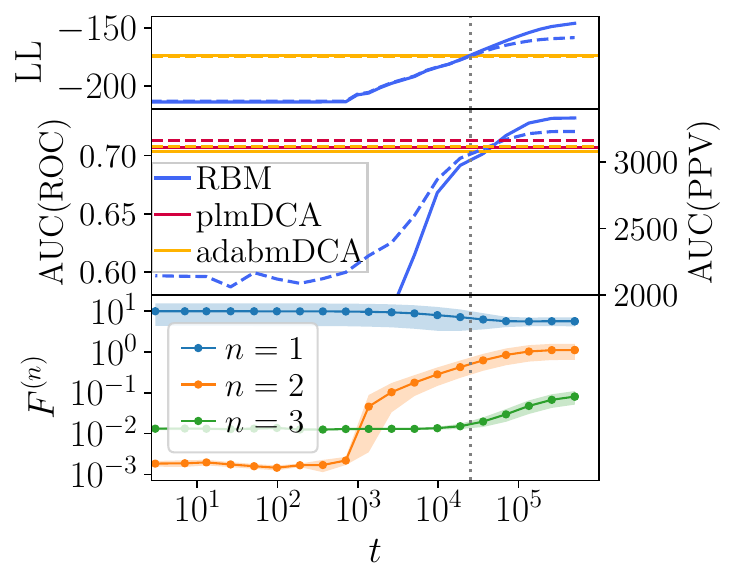}
    \includegraphics[height=4cm]{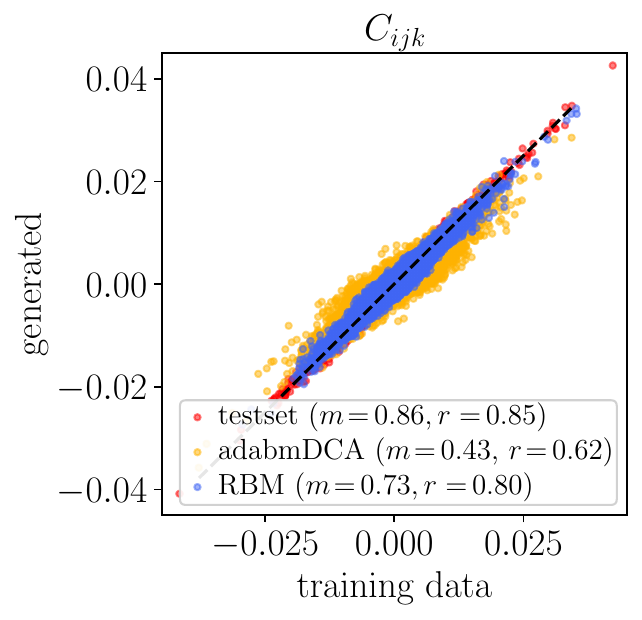}
    \raisebox{0.15cm}{\includegraphics[height=3.6cm]{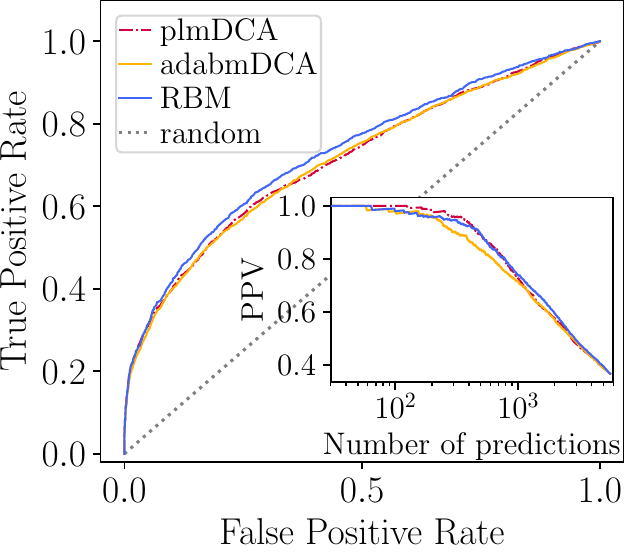}}
    \includegraphics[height=4cm]{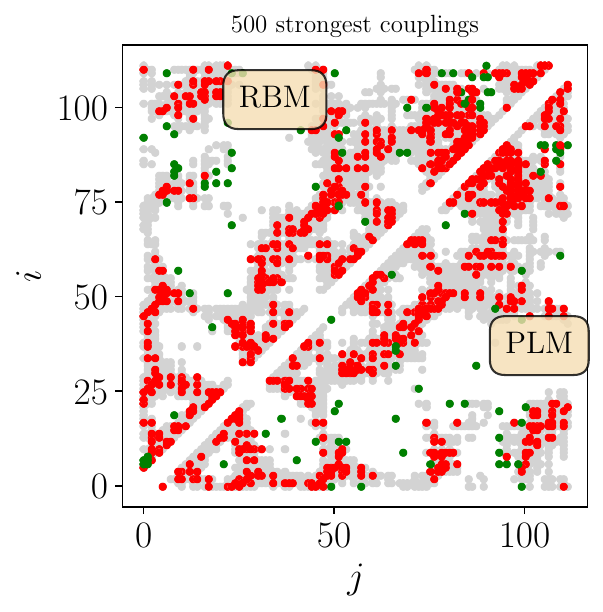}
    \put(-470, 103){\figpanel{a1}}
    \put(-410, 60){\figpanel{a2}}
    \put(-410, 25){\figpanel{a3}}
    \put(-345, 110){\figpanel{b}}
    \put(-235, 110){\figpanel{c1}}
    \put(-165, 45){\figpanel{c2}}
    \put(-110, 110){\figpanel{d}}
    \caption{{\textbf{RBM-based contact prediction on the Response Regulator Receiver Domain (Pfam: PF00072).}
We evaluate the RBM on the PF00072 MSA for residue-residue contact prediction, comparing its performance to DCA-based methods (plmDCA~\cite{ekeberg2013improved, ekeberg2014fast} and adabmDCA~\cite{rosset2025adabmdca}) using public implementations.
\figpanel{a1}: RBM log-likelihood (LL) (solid: train, dashed: test) and final adamDCA LL (horizontal line) during training.
\figpanel{a2}: AUC for ROC (solid) and PPV (dashed) during training; horizontal lines indicate plmDCA and adabmDCA.
\figpanel{a3}: Frobenius norms of effective $n$-body interactions ($n\!=\!1,2,3$) vs. training time; RBM surpasses baselines as 3-body terms emerge.
\figpanel{b}: Comparison of 3-point statistics from RBM- and adabmDCA-generated samples vs. training data; red: test vs. train. Legend: Pearson correlation ($r$), slope of linear fit ($m$).
\figpanel{c1}: ROC curves; inset \figpanel{c2}: PPV vs. number of top-ranked predictions (being 1 for the top 50 for all methods).
\figpanel{d}: Contact maps, experimental (light gray), RBM (upper triangle), plmDCA (lower); red: true positives, green: false positives.
}}
\label{fig:contact_prediction}
\end{figure*}

\textit{Inverse Numerical Experiments.---} We evaluate the reliability of our approach through an inverse experiment using data generated from a predefined spin-1 Blume-Capel model with 2-body and 3-body interactions. The ground-truth Hamiltonian is:
\begin{equation}
   \textstyle \mathcal{H}_\mathcal{D} (\boldsymbol{s})= - J^{(3)} \sum_{\langle i, j, k \rangle}  s_{i} s_{j} s_{k} - J^{(2)} \sum_{\langle i, j \rangle} s_{i} s_{j},
    \label{blume-capel_hamiltonian}
\end{equation}
where \(s_i \in \{-1, 0, 1\}\). The dataset is generated by sampling \(M=10^5\) equilibrium configurations from the Boltzmann distribution of model \eqref{blume-capel_hamiltonian} at an inverse temperature of \(\beta\!=\!0.2\). In particular, we considered a 1D chain of \(N\! =\! 51\) spins with periodic boundary conditions (\(s_{N+1} \equiv s_1\)), setting pairwise nearest-neighbor interactions with \(J^{(2)} \! =\! 1\) and adding sparse three-spin interactions at distant sites with \(J^{(3)} \! =\!1\). Specifically, triads involve different \(i, j, k\) sites satisfying \(i \equiv j \equiv k \ (\mathrm{mod}\ 17)\), {excluding self coupling interactions}.
This 3-state Potts model is particularly appealing, as it simultaneously satisfies both the zero-sum and lattice-gas gauge conditions.

We train a binary-Potts RBM and extract two- and three-body effective couplings in the zero-sum gauge using Eq.~\eqref{n-body_formula_zs}.
We then compare the inferred values with the true ones, separately analyzing coupling and non-coupling links ($J=0$) in the original model. Fig.~\ref{fig:blumecapel}~\figpanel{b} shows that RBMs sequentially learn interactions from lower to higher orders, independent of the gauge used for inference (the curves in the lattice-gas gauge are given in Fig.~\ref{fig:latticegas} in the Appendix~\ref{latticegas_appendix}). This aligns with findings in binary RBMs~\cite{decelle2024inferring} and transformers on language data~\cite{rende2024distributional}.  Fig.\ref{fig:blumecapel} \figpanel{c} illustrates the recovery of the pairwise structure and \figpanel{d} shows how the $\hat{R}^{(2)}_{ij}$ terms directly identify the presence of specific three-body interactions, without the need to compute them explicitly.

\textit{rbmDCA.---} {A key application of the inverse Potts problem is residue–residue contact prediction in protein structures from MSAs \cite{cocco2018inverse}, which involves modeling 21 discrete states (20 amino acids plus a gap). Hence, we trained a Potts-Bernoulli RBM on sequences from a single homologous family, the response regulator domain (Pfam: \href{https://www.ebi.ac.uk/interpro/entry/pfam/PF00072/}{PF00072}), using the MSA provided in Ref.~\cite{trinquier2021efficient}. The dataset consists of 49,527 aligned sequences, each of length 112 amino acids.} Due to limited sampling in protein data, a small $\ell_2$ penalty was essential to ensure convergence \footnote{Without regularization, strong effective couplings emerged early in training, degrading MCMC-based gradient estimates and leading to non-monotonic inference performance.}. To monitor overfitting, we split the data into 60\% training and 40\% test sets, and used the same partition across all methods for fair comparison. Figure~\ref{fig:contact_prediction}--\figpanel{a1} shows the evolution of train and test log-likelihoods during RBM training \cite{bereux2024fast}; a vertical line marks when the RBM surpasses the adabmDCA test log-likelihood~\cite{muntoni2021adabmdca, rosset2025adabmdca}. RBM-generated samples accurately capture third-order statistics (Fig.\ref{fig:contact_prediction}--\figpanel{b}), unlike {adabmDCA}. Lower-order statistics are reported in Appendix~\ref{generation_appendix}.

Once trained, we apply the zero-sum mapping~\eqref{n-body_formula_zs} to extract the effective multibody Potts parameters. Figure~\ref{fig:contact_prediction}--\figpanel{a3} shows the Frobenius norms of the effective fields, pairwise couplings, and three-body interactions over training time. The RBM begins to outperform adabmDCA precisely when it starts encoding three-body interactions, which pairwise models cannot capture. Using the inferred two-body terms, we compute rbmDCA contact predictions and compare them to known structural contacts (Fig.\ref{fig:contact_prediction}--\figpanel{d}); a schematic of the full pipeline is shown in Appendix Fig.~\ref{fig:rbmDCA}. We benchmark against plmDCA~\cite{ekeberg2013improved, ekeberg2014fast} and adabmDCA, comparing receiver operating characteristic (ROC) and positive predictive value (PPV) curves in Figs.\ref{fig:contact_prediction}--\figpanel{c1,c2}. The evolution of AUC for both metrics during training is shown in Fig.\ref{fig:contact_prediction}--\figpanel{a2}. The RBM outperforms both baselines, but only once third-order interactions emerge. This aligns with Ref.~\cite{schmidt2017three}, where three-body terms improved couplings derived from mean-field models, although, as in our case, contact prediction still relied solely on two-body couplings. Leveraging higher-order terms to enhance contact inference remains an open direction. Throughout, we evaluate contacts using only non-trivial residue pairs ($|i - j| > 5$ and distance $r < 7.5$\AA). A detailed description of DCA methods is provided in the Appendix~\ref{appendix_DCA}.

\textit{Conclusions.---} We introduced a general framework for interpreting generative models over high-dimensional categorical data by mapping them onto effective Potts models with many-body interactions. Within this framework, we developed a protocol that leverages the energy landscape of RBMs to extract multi-body couplings from data efficiently. {We demonstrated that this approach accurately recovers high-order interactions in controlled inverse experiments and significantly improves contact prediction in protein families, outperforming state-of-the-art methods. These results position RBMs as powerful tools for interpretable modeling of categorical data, combining high-quality generative performance, illustrated in Appendix~\ref{generation_appendix} for protein sequences, with enhanced interpretability that extends beyond the pairwise scope of traditional maximum entropy models. 

Moreover, we observed that RBMs learn interactions hierarchically: an $(n+1)-$body interaction emerges only after the $n$-body terms are captured. This sequential learning structure, recently reported in deep transformer architectures~\cite{rende2024distributional}, suggests that the effective mappings developed here provide a principled approach to dissect the learning dynamics of generative models and their relationship to the data structure.
}

\textit{Acknowledgments.--- }The authors gratefully acknowledge the invaluable support of Nicolas Béreux and Lorenzo Rosset for their assistance with the RBM and adabmDCA training and analysis code. {Additionally, we thank Carlos A. Gandarilla-Pérez, Martin Weigt, and Giovanni Catania for fruitful discussions during the development of this project. We also thank the anonymous referees for their constructive comments, which significantly improved the quality of the manuscript.}
The authors acknowledge financial support by the Comunidad de
Madrid and the Complutense University of Madrid through the Atracción de Talento program (Refs. 2019-T1/TIC-13298 and Refs. 2023-5A/TIC-28934), and to projects PID2021-125506NA-I00 and PID2024-158623NB-C21 financed by 
the ``Ministerio de Econom\'{\i}a y Competitividad, Agencia Estatal de Investigaci\'on" (MICIU/AEI/10.13039/501100011033), and the Fondo Europeo de Desarrollo Regional (FEDER, UE).

\textit{Code and Data Availability.--- } The code for training the RBMs is available at~\cite{github:rbms}, while the code used to extract the couplings can be found at~\cite{github:couplings_inference}. Both the Blume-Capel simulations and the MSA data for the PF00072 family employed to train the models are provided in~\cite{github:couplings_inference}. The original PF00072 MSA was obtained from~\cite{trinquier2021efficient}.

\bibliographystyle{apsrev4-2}
\bibliography{apssamp}

\begin{thebibliography}{55}%
\makeatletter
\providecommand \@ifxundefined [1]{%
 \@ifx{#1\undefined}
}%
\providecommand \@ifnum [1]{%
 \ifnum #1\expandafter \@firstoftwo
 \else \expandafter \@secondoftwo
 \fi
}%
\providecommand \@ifx [1]{%
 \ifx #1\expandafter \@firstoftwo
 \else \expandafter \@secondoftwo
 \fi
}%
\providecommand \natexlab [1]{#1}%
\providecommand \enquote  [1]{``#1''}%
\providecommand \bibnamefont  [1]{#1}%
\providecommand \bibfnamefont [1]{#1}%
\providecommand \citenamefont [1]{#1}%
\providecommand \href@noop [0]{\@secondoftwo}%
\providecommand \href [0]{\begingroup \@sanitize@url \@href}%
\providecommand \@href[1]{\@@startlink{#1}\@@href}%
\providecommand \@@href[1]{\endgroup#1\@@endlink}%
\providecommand \@sanitize@url [0]{\catcode `\\12\catcode `\$12\catcode `\&12\catcode `\#12\catcode `\^12\catcode `\_12\catcode `\%12\relax}%
\providecommand \@@startlink[1]{}%
\providecommand \@@endlink[0]{}%
\providecommand \url  [0]{\begingroup\@sanitize@url \@url }%
\providecommand \@url [1]{\endgroup\@href {#1}{\urlprefix }}%
\providecommand \urlprefix  [0]{URL }%
\providecommand \Eprint [0]{\href }%
\providecommand \doibase [0]{https://doi.org/}%
\providecommand \selectlanguage [0]{\@gobble}%
\providecommand \bibinfo  [0]{\@secondoftwo}%
\providecommand \bibfield  [0]{\@secondoftwo}%
\providecommand \translation [1]{[#1]}%
\providecommand \BibitemOpen [0]{}%
\providecommand \bibitemStop [0]{}%
\providecommand \bibitemNoStop [0]{.\EOS\space}%
\providecommand \EOS [0]{\spacefactor3000\relax}%
\providecommand \BibitemShut  [1]{\csname bibitem#1\endcsname}%
\let\auto@bib@innerbib\@empty
\bibitem [{\citenamefont {Nguyen}\ \emph {et~al.}(2017)\citenamefont {Nguyen}, \citenamefont {Zecchina},\ and\ \citenamefont {Berg}}]{nguyen2017inverse}%
  \BibitemOpen
  \bibfield  {author} {\bibinfo {author} {\bibfnamefont {H.~C.}\ \bibnamefont {Nguyen}}, \bibinfo {author} {\bibfnamefont {R.}~\bibnamefont {Zecchina}},\ and\ \bibinfo {author} {\bibfnamefont {J.}~\bibnamefont {Berg}},\ }\href {https://doi.org/10.1080/00018732.2017.1341604} {\bibfield  {journal} {\bibinfo  {journal} {Advances in Physics}\ }\textbf {\bibinfo {volume} {66}},\ \bibinfo {pages} {197} (\bibinfo {year} {2017})},\ \Eprint {https://arxiv.org/abs/https://doi.org/10.1080/00018732.2017.1341604} {https://doi.org/10.1080/00018732.2017.1341604} \BibitemShut {NoStop}%
\bibitem [{\citenamefont {Schneidman}\ \emph {et~al.}(2006)\citenamefont {Schneidman}, \citenamefont {Berry}, \citenamefont {Segev},\ and\ \citenamefont {Bialek}}]{schneidman2006weak}%
  \BibitemOpen
  \bibfield  {author} {\bibinfo {author} {\bibfnamefont {E.}~\bibnamefont {Schneidman}}, \bibinfo {author} {\bibfnamefont {M.~J.}\ \bibnamefont {Berry}}, \bibinfo {author} {\bibfnamefont {R.}~\bibnamefont {Segev}},\ and\ \bibinfo {author} {\bibfnamefont {W.}~\bibnamefont {Bialek}},\ }\href {https://doi.org/10.1038/nature04701} {\bibfield  {journal} {\bibinfo  {journal} {Nature}\ }\textbf {\bibinfo {volume} {440}},\ \bibinfo {pages} {1007} (\bibinfo {year} {2006})}\BibitemShut {NoStop}%
\bibitem [{\citenamefont {Roudi}\ \emph {et~al.}(2009{\natexlab{a}})\citenamefont {Roudi}, \citenamefont {Aurell},\ and\ \citenamefont {Hertz}}]{roudi2009statistical}%
  \BibitemOpen
  \bibfield  {author} {\bibinfo {author} {\bibfnamefont {Y.}~\bibnamefont {Roudi}}, \bibinfo {author} {\bibfnamefont {E.}~\bibnamefont {Aurell}},\ and\ \bibinfo {author} {\bibfnamefont {J.~A.}\ \bibnamefont {Hertz}},\ }\href {https://doi.org/10.3389/neuro.10.022.2009} {\bibfield  {journal} {\bibinfo  {journal} {Frontiers in Computational Neuroscience}\ }\textbf {\bibinfo {volume} {3}},\ \bibinfo {pages} {652} (\bibinfo {year} {2009}{\natexlab{a}})}\BibitemShut {NoStop}%
\bibitem [{\citenamefont {Meshulam}\ \emph {et~al.}(2017)\citenamefont {Meshulam}, \citenamefont {Gauthier}, \citenamefont {Brody}, \citenamefont {Tank},\ and\ \citenamefont {Bialek}}]{meshulam2017collective}%
  \BibitemOpen
  \bibfield  {author} {\bibinfo {author} {\bibfnamefont {L.}~\bibnamefont {Meshulam}}, \bibinfo {author} {\bibfnamefont {J.~L.}\ \bibnamefont {Gauthier}}, \bibinfo {author} {\bibfnamefont {C.~D.}\ \bibnamefont {Brody}}, \bibinfo {author} {\bibfnamefont {D.~W.}\ \bibnamefont {Tank}},\ and\ \bibinfo {author} {\bibfnamefont {W.}~\bibnamefont {Bialek}},\ }\href {https://doi.org/https://doi.org/10.1016/j.neuron.2017.10.027} {\bibfield  {journal} {\bibinfo  {journal} {Neuron}\ }\textbf {\bibinfo {volume} {96}},\ \bibinfo {pages} {1178} (\bibinfo {year} {2017})}\BibitemShut {NoStop}%
\bibitem [{\citenamefont {Maoz}\ \emph {et~al.}(2020)\citenamefont {Maoz}, \citenamefont {Tkačik}, \citenamefont {Esteki}, \citenamefont {Kiani},\ and\ \citenamefont {Schneidman}}]{maoz2020learning}%
  \BibitemOpen
  \bibfield  {author} {\bibinfo {author} {\bibfnamefont {O.}~\bibnamefont {Maoz}}, \bibinfo {author} {\bibfnamefont {G.}~\bibnamefont {Tkačik}}, \bibinfo {author} {\bibfnamefont {M.~S.}\ \bibnamefont {Esteki}}, \bibinfo {author} {\bibfnamefont {R.}~\bibnamefont {Kiani}},\ and\ \bibinfo {author} {\bibfnamefont {E.}~\bibnamefont {Schneidman}},\ }\href {https://doi.org/10.1073/pnas.1912804117} {\bibfield  {journal} {\bibinfo  {journal} {Proceedings of the National Academy of Sciences}\ }\textbf {\bibinfo {volume} {117}},\ \bibinfo {pages} {25066} (\bibinfo {year} {2020})}\BibitemShut {NoStop}%
\bibitem [{\citenamefont {{De Martino}}\ and\ \citenamefont {{De Martino}}(2018)}]{de2018introduction}%
  \BibitemOpen
  \bibfield  {author} {\bibinfo {author} {\bibfnamefont {A.}~\bibnamefont {{De Martino}}}\ and\ \bibinfo {author} {\bibfnamefont {D.}~\bibnamefont {{De Martino}}},\ }\href {https://doi.org/https://doi.org/10.1016/j.heliyon.2018.e00596} {\bibfield  {journal} {\bibinfo  {journal} {Heliyon}\ }\textbf {\bibinfo {volume} {4}},\ \bibinfo {pages} {e00596} (\bibinfo {year} {2018})}\BibitemShut {NoStop}%
\bibitem [{\citenamefont {Lezon}\ \emph {et~al.}(2006)\citenamefont {Lezon}, \citenamefont {Banavar}, \citenamefont {Cieplak}, \citenamefont {Maritan},\ and\ \citenamefont {Fedoroff}}]{lezon2006using}%
  \BibitemOpen
  \bibfield  {author} {\bibinfo {author} {\bibfnamefont {T.~R.}\ \bibnamefont {Lezon}}, \bibinfo {author} {\bibfnamefont {J.~R.}\ \bibnamefont {Banavar}}, \bibinfo {author} {\bibfnamefont {M.}~\bibnamefont {Cieplak}}, \bibinfo {author} {\bibfnamefont {A.}~\bibnamefont {Maritan}},\ and\ \bibinfo {author} {\bibfnamefont {N.~V.}\ \bibnamefont {Fedoroff}},\ }\href {https://doi.org/10.1073/pnas.0609152103} {\bibfield  {journal} {\bibinfo  {journal} {Proceedings of the National Academy of Sciences}\ }\textbf {\bibinfo {volume} {103}},\ \bibinfo {pages} {19033} (\bibinfo {year} {2006})}\BibitemShut {NoStop}%
\bibitem [{\citenamefont {Tkačik}\ \emph {et~al.}(2008)\citenamefont {Tkačik}, \citenamefont {Callan},\ and\ \citenamefont {Bialek}}]{tkavcik2008information}%
  \BibitemOpen
  \bibfield  {author} {\bibinfo {author} {\bibfnamefont {G.}~\bibnamefont {Tkačik}}, \bibinfo {author} {\bibfnamefont {C.~G.}\ \bibnamefont {Callan}},\ and\ \bibinfo {author} {\bibfnamefont {W.}~\bibnamefont {Bialek}},\ }\href {https://doi.org/10.1073/pnas.0806077105} {\bibfield  {journal} {\bibinfo  {journal} {Proceedings of the National Academy of Sciences}\ }\textbf {\bibinfo {volume} {105}},\ \bibinfo {pages} {12265} (\bibinfo {year} {2008})}\BibitemShut {NoStop}%
\bibitem [{\citenamefont {Weigt}\ \emph {et~al.}(2009)\citenamefont {Weigt}, \citenamefont {White}, \citenamefont {Szurmant}, \citenamefont {Hoch},\ and\ \citenamefont {Hwa}}]{weigt2009identification}%
  \BibitemOpen
  \bibfield  {author} {\bibinfo {author} {\bibfnamefont {M.}~\bibnamefont {Weigt}}, \bibinfo {author} {\bibfnamefont {R.~A.}\ \bibnamefont {White}}, \bibinfo {author} {\bibfnamefont {H.}~\bibnamefont {Szurmant}}, \bibinfo {author} {\bibfnamefont {J.~A.}\ \bibnamefont {Hoch}},\ and\ \bibinfo {author} {\bibfnamefont {T.}~\bibnamefont {Hwa}},\ }\href {https://doi.org/10.1073/pnas.0805923106} {\bibfield  {journal} {\bibinfo  {journal} {Proceedings of the National Academy of Sciences}\ }\textbf {\bibinfo {volume} {106}},\ \bibinfo {pages} {67} (\bibinfo {year} {2009})}\BibitemShut {NoStop}%
\bibitem [{\citenamefont {Ekeberg}\ \emph {et~al.}(2013)\citenamefont {Ekeberg}, \citenamefont {L\"ovkvist}, \citenamefont {Lan}, \citenamefont {Weigt},\ and\ \citenamefont {Aurell}}]{ekeberg2013improved}%
  \BibitemOpen
  \bibfield  {author} {\bibinfo {author} {\bibfnamefont {M.}~\bibnamefont {Ekeberg}}, \bibinfo {author} {\bibfnamefont {C.}~\bibnamefont {L\"ovkvist}}, \bibinfo {author} {\bibfnamefont {Y.}~\bibnamefont {Lan}}, \bibinfo {author} {\bibfnamefont {M.}~\bibnamefont {Weigt}},\ and\ \bibinfo {author} {\bibfnamefont {E.}~\bibnamefont {Aurell}},\ }\href {https://doi.org/10.1103/PhysRevE.87.012707} {\bibfield  {journal} {\bibinfo  {journal} {Phys. Rev. E}\ }\textbf {\bibinfo {volume} {87}},\ \bibinfo {pages} {012707} (\bibinfo {year} {2013})}\BibitemShut {NoStop}%
\bibitem [{\citenamefont {Ekeberg}\ \emph {et~al.}(2014)\citenamefont {Ekeberg}, \citenamefont {Hartonen},\ and\ \citenamefont {Aurell}}]{ekeberg2014fast}%
  \BibitemOpen
  \bibfield  {author} {\bibinfo {author} {\bibfnamefont {M.}~\bibnamefont {Ekeberg}}, \bibinfo {author} {\bibfnamefont {T.}~\bibnamefont {Hartonen}},\ and\ \bibinfo {author} {\bibfnamefont {E.}~\bibnamefont {Aurell}},\ }\href {https://doi.org/https://doi.org/10.1016/j.jcp.2014.07.024} {\bibfield  {journal} {\bibinfo  {journal} {Journal of Computational Physics}\ }\textbf {\bibinfo {volume} {276}},\ \bibinfo {pages} {341} (\bibinfo {year} {2014})}\BibitemShut {NoStop}%
\bibitem [{\citenamefont {Cocco}\ \emph {et~al.}(2018)\citenamefont {Cocco}, \citenamefont {Feinauer}, \citenamefont {Figliuzzi}, \citenamefont {Monasson},\ and\ \citenamefont {Weigt}}]{cocco2018inverse}%
  \BibitemOpen
  \bibfield  {author} {\bibinfo {author} {\bibfnamefont {S.}~\bibnamefont {Cocco}}, \bibinfo {author} {\bibfnamefont {C.}~\bibnamefont {Feinauer}}, \bibinfo {author} {\bibfnamefont {M.}~\bibnamefont {Figliuzzi}}, \bibinfo {author} {\bibfnamefont {R.}~\bibnamefont {Monasson}},\ and\ \bibinfo {author} {\bibfnamefont {M.}~\bibnamefont {Weigt}},\ }\href {https://doi.org/10.1088/1361-6633/aa9965} {\bibfield  {journal} {\bibinfo  {journal} {Reports on Progress in Physics}\ }\textbf {\bibinfo {volume} {81}},\ \bibinfo {pages} {032601} (\bibinfo {year} {2018})}\BibitemShut {NoStop}%
\bibitem [{\citenamefont {Battiston}\ \emph {et~al.}(2021)\citenamefont {Battiston}, \citenamefont {Amico}, \citenamefont {Barrat}, \citenamefont {Bianconi}, \citenamefont {Ferraz~de Arruda}, \citenamefont {Franceschiello}, \citenamefont {Iacopini}, \citenamefont {K{\'e}fi}, \citenamefont {Latora}, \citenamefont {Moreno} \emph {et~al.}}]{battiston2021physics}%
  \BibitemOpen
  \bibfield  {author} {\bibinfo {author} {\bibfnamefont {F.}~\bibnamefont {Battiston}}, \bibinfo {author} {\bibfnamefont {E.}~\bibnamefont {Amico}}, \bibinfo {author} {\bibfnamefont {A.}~\bibnamefont {Barrat}}, \bibinfo {author} {\bibfnamefont {G.}~\bibnamefont {Bianconi}}, \bibinfo {author} {\bibfnamefont {G.}~\bibnamefont {Ferraz~de Arruda}}, \bibinfo {author} {\bibfnamefont {B.}~\bibnamefont {Franceschiello}}, \bibinfo {author} {\bibfnamefont {I.}~\bibnamefont {Iacopini}}, \bibinfo {author} {\bibfnamefont {S.}~\bibnamefont {K{\'e}fi}}, \bibinfo {author} {\bibfnamefont {V.}~\bibnamefont {Latora}}, \bibinfo {author} {\bibfnamefont {Y.}~\bibnamefont {Moreno}}, \emph {et~al.},\ }\href {https://doi.org/10.1038/s41567-021-01371-4} {\bibfield  {journal} {\bibinfo  {journal} {Nat. Phys.}\ }\textbf {\bibinfo {volume} {17}},\ \bibinfo {pages} {1093} (\bibinfo {year} {2021})}\BibitemShut {NoStop}%
\bibitem [{\citenamefont {Roudi}\ \emph {et~al.}(2009{\natexlab{b}})\citenamefont {Roudi}, \citenamefont {Nirenberg},\ and\ \citenamefont {Latham}}]{roudi2009pairwise}%
  \BibitemOpen
  \bibfield  {author} {\bibinfo {author} {\bibfnamefont {Y.}~\bibnamefont {Roudi}}, \bibinfo {author} {\bibfnamefont {S.}~\bibnamefont {Nirenberg}},\ and\ \bibinfo {author} {\bibfnamefont {P.~E.}\ \bibnamefont {Latham}},\ }\href {https://doi.org/10.1371/journal.pcbi.1000380} {\bibfield  {journal} {\bibinfo  {journal} {PLOS Computational Biology}\ }\textbf {\bibinfo {volume} {5}},\ \bibinfo {pages} {1} (\bibinfo {year} {2009}{\natexlab{b}})}\BibitemShut {NoStop}%
\bibitem [{\citenamefont {Jumper}\ \emph {et~al.}(2021)\citenamefont {Jumper}, \citenamefont {Evans}, \citenamefont {Pritzel}, \citenamefont {Green}, \citenamefont {Figurnov}, \citenamefont {Ronneberger}, \citenamefont {Tunyasuvunakool}, \citenamefont {Bates}, \citenamefont {{\v{Z}}{\'\i}dek}, \citenamefont {Potapenko} \emph {et~al.}}]{jumper2021highly}%
  \BibitemOpen
  \bibfield  {author} {\bibinfo {author} {\bibfnamefont {J.}~\bibnamefont {Jumper}}, \bibinfo {author} {\bibfnamefont {R.}~\bibnamefont {Evans}}, \bibinfo {author} {\bibfnamefont {A.}~\bibnamefont {Pritzel}}, \bibinfo {author} {\bibfnamefont {T.}~\bibnamefont {Green}}, \bibinfo {author} {\bibfnamefont {M.}~\bibnamefont {Figurnov}}, \bibinfo {author} {\bibfnamefont {O.}~\bibnamefont {Ronneberger}}, \bibinfo {author} {\bibfnamefont {K.}~\bibnamefont {Tunyasuvunakool}}, \bibinfo {author} {\bibfnamefont {R.}~\bibnamefont {Bates}}, \bibinfo {author} {\bibfnamefont {A.}~\bibnamefont {{\v{Z}}{\'\i}dek}}, \bibinfo {author} {\bibfnamefont {A.}~\bibnamefont {Potapenko}}, \emph {et~al.},\ }\href {https://doi.org/10.1038/s41586-021-03819-2} {\bibfield  {journal} {\bibinfo  {journal} {Nature}\ }\textbf {\bibinfo {volume} {596}},\ \bibinfo {pages} {583} (\bibinfo {year} {2021})}\BibitemShut {NoStop}%
\bibitem [{\citenamefont {Abramson}\ \emph {et~al.}(2024)\citenamefont {Abramson}, \citenamefont {Adler}, \citenamefont {Dunger}, \citenamefont {Evans}, \citenamefont {Green}, \citenamefont {Pritzel}, \citenamefont {Ronneberger}, \citenamefont {Willmore}, \citenamefont {Ballard}, \citenamefont {Bambrick} \emph {et~al.}}]{abramson2024accurate}%
  \BibitemOpen
  \bibfield  {author} {\bibinfo {author} {\bibfnamefont {J.}~\bibnamefont {Abramson}}, \bibinfo {author} {\bibfnamefont {J.}~\bibnamefont {Adler}}, \bibinfo {author} {\bibfnamefont {J.}~\bibnamefont {Dunger}}, \bibinfo {author} {\bibfnamefont {R.}~\bibnamefont {Evans}}, \bibinfo {author} {\bibfnamefont {T.}~\bibnamefont {Green}}, \bibinfo {author} {\bibfnamefont {A.}~\bibnamefont {Pritzel}}, \bibinfo {author} {\bibfnamefont {O.}~\bibnamefont {Ronneberger}}, \bibinfo {author} {\bibfnamefont {L.}~\bibnamefont {Willmore}}, \bibinfo {author} {\bibfnamefont {A.~J.}\ \bibnamefont {Ballard}}, \bibinfo {author} {\bibfnamefont {J.}~\bibnamefont {Bambrick}}, \emph {et~al.},\ }\href {https://doi.org/10.1038/s41586-024-07487-w} {\bibfield  {journal} {\bibinfo  {journal} {Nature}\ ,\ \bibinfo {pages} {1}} (\bibinfo {year} {2024})}\BibitemShut {NoStop}%
\bibitem [{\citenamefont {Rao}\ \emph {et~al.}(2021)\citenamefont {Rao}, \citenamefont {Meier}, \citenamefont {Sercu}, \citenamefont {Ovchinnikov},\ and\ \citenamefont {Rives}}]{rao2020transformer}%
  \BibitemOpen
  \bibfield  {author} {\bibinfo {author} {\bibfnamefont {R.}~\bibnamefont {Rao}}, \bibinfo {author} {\bibfnamefont {J.}~\bibnamefont {Meier}}, \bibinfo {author} {\bibfnamefont {T.}~\bibnamefont {Sercu}}, \bibinfo {author} {\bibfnamefont {S.}~\bibnamefont {Ovchinnikov}},\ and\ \bibinfo {author} {\bibfnamefont {A.}~\bibnamefont {Rives}},\ }in\ \href {https://doi.org/10.1101/2020.12.15.422761} {\emph {\bibinfo {booktitle} {Proceedings of the 9th International Conference on Learning Representations}}}\ (\bibinfo {year} {2021})\BibitemShut {NoStop}%
\bibitem [{\citenamefont {Feinauer}\ and\ \citenamefont {Borgonovo}(2022)}]{feinauer2022mean}%
  \BibitemOpen
  \bibfield  {author} {\bibinfo {author} {\bibfnamefont {C.}~\bibnamefont {Feinauer}}\ and\ \bibinfo {author} {\bibfnamefont {E.}~\bibnamefont {Borgonovo}},\ }\bibfield  {journal} {\bibinfo  {journal} {bioRxiv}\ }\href {https://doi.org/10.1101/2022.12.12.520028} {10.1101/2022.12.12.520028} (\bibinfo {year} {2022})\BibitemShut {NoStop}%
\bibitem [{\citenamefont {Feinauer}\ \emph {et~al.}(2022)\citenamefont {Feinauer}, \citenamefont {Meynard-Piganeau},\ and\ \citenamefont {Lucibello}}]{feinauer2022interpretable}%
  \BibitemOpen
  \bibfield  {author} {\bibinfo {author} {\bibfnamefont {C.}~\bibnamefont {Feinauer}}, \bibinfo {author} {\bibfnamefont {B.}~\bibnamefont {Meynard-Piganeau}},\ and\ \bibinfo {author} {\bibfnamefont {C.}~\bibnamefont {Lucibello}},\ }\href {https://doi.org/10.1371/journal.pcbi.1010219} {\bibfield  {journal} {\bibinfo  {journal} {PLOS Computational Biology}\ }\textbf {\bibinfo {volume} {18}},\ \bibinfo {pages} {1} (\bibinfo {year} {2022})}\BibitemShut {NoStop}%
\bibitem [{\citenamefont {Caredda}\ and\ \citenamefont {Pagnani}(2025)}]{caredda2024direct}%
  \BibitemOpen
  \bibfield  {author} {\bibinfo {author} {\bibfnamefont {F.}~\bibnamefont {Caredda}}\ and\ \bibinfo {author} {\bibfnamefont {A.}~\bibnamefont {Pagnani}},\ }\href {https://doi.org/10.1186/s12859-025-06062-y} {\bibfield  {journal} {\bibinfo  {journal} {BMC Bioinformatics}\ ,\ \bibinfo {pages} {2024}} (\bibinfo {year} {2025})}\BibitemShut {NoStop}%
\bibitem [{\citenamefont {Merger}\ \emph {et~al.}(2023)\citenamefont {Merger}, \citenamefont {Ren\'e}, \citenamefont {Fischer}, \citenamefont {Bouss}, \citenamefont {Nestler}, \citenamefont {Dahmen}, \citenamefont {Honerkamp},\ and\ \citenamefont {Helias}}]{merger2023learning}%
  \BibitemOpen
  \bibfield  {author} {\bibinfo {author} {\bibfnamefont {C.}~\bibnamefont {Merger}}, \bibinfo {author} {\bibfnamefont {A.}~\bibnamefont {Ren\'e}}, \bibinfo {author} {\bibfnamefont {K.}~\bibnamefont {Fischer}}, \bibinfo {author} {\bibfnamefont {P.}~\bibnamefont {Bouss}}, \bibinfo {author} {\bibfnamefont {S.}~\bibnamefont {Nestler}}, \bibinfo {author} {\bibfnamefont {D.}~\bibnamefont {Dahmen}}, \bibinfo {author} {\bibfnamefont {C.}~\bibnamefont {Honerkamp}},\ and\ \bibinfo {author} {\bibfnamefont {M.}~\bibnamefont {Helias}},\ }\href {https://doi.org/10.1103/PhysRevX.13.041033} {\bibfield  {journal} {\bibinfo  {journal} {Phys. Rev. X}\ }\textbf {\bibinfo {volume} {13}},\ \bibinfo {pages} {041033} (\bibinfo {year} {2023})}\BibitemShut {NoStop}%
\bibitem [{\citenamefont {Smolensky}(1986)}]{Smolensky}%
  \BibitemOpen
  \bibfield  {author} {\bibinfo {author} {\bibfnamefont {P.}~\bibnamefont {Smolensky}},\ }in\ \href {https://doi.org/10.7551/mitpress/5236.003.0009} {\emph {\bibinfo {booktitle} {Parallel Distributed Processing, Volume 1: Explorations in the Microstructure of Cognition: Foundations}}}\ (\bibinfo  {publisher} {The MIT Press},\ \bibinfo {year} {1986})\BibitemShut {NoStop}%
\bibitem [{\citenamefont {Ackley}\ \emph {et~al.}(1985)\citenamefont {Ackley}, \citenamefont {Hinton},\ and\ \citenamefont {Sejnowski}}]{ackley1985learning}%
  \BibitemOpen
  \bibfield  {author} {\bibinfo {author} {\bibfnamefont {D.~H.}\ \bibnamefont {Ackley}}, \bibinfo {author} {\bibfnamefont {G.~E.}\ \bibnamefont {Hinton}},\ and\ \bibinfo {author} {\bibfnamefont {T.~J.}\ \bibnamefont {Sejnowski}},\ }\href {https://doi.org/https://doi.org/10.1016/S0364-0213(85)80012-4} {\bibfield  {journal} {\bibinfo  {journal} {Cognitive Science}\ }\textbf {\bibinfo {volume} {9}},\ \bibinfo {pages} {147} (\bibinfo {year} {1985})}\BibitemShut {NoStop}%
\bibitem [{\citenamefont {Le~Roux}\ and\ \citenamefont {Bengio}(2008)}]{le2008representational}%
  \BibitemOpen
  \bibfield  {author} {\bibinfo {author} {\bibfnamefont {N.}~\bibnamefont {Le~Roux}}\ and\ \bibinfo {author} {\bibfnamefont {Y.}~\bibnamefont {Bengio}},\ }\href {https://doi.org/10.1162/neco.2008.04-07-510} {\bibfield  {journal} {\bibinfo  {journal} {Neural Computation}\ }\textbf {\bibinfo {volume} {20}},\ \bibinfo {pages} {1631} (\bibinfo {year} {2008})}\BibitemShut {NoStop}%
\bibitem [{\citenamefont {Cossu}\ \emph {et~al.}(2019)\citenamefont {Cossu}, \citenamefont {Del~Debbio}, \citenamefont {Giani}, \citenamefont {Khamseh},\ and\ \citenamefont {Wilson}}]{cossu2019machine}%
  \BibitemOpen
  \bibfield  {author} {\bibinfo {author} {\bibfnamefont {G.}~\bibnamefont {Cossu}}, \bibinfo {author} {\bibfnamefont {L.}~\bibnamefont {Del~Debbio}}, \bibinfo {author} {\bibfnamefont {T.}~\bibnamefont {Giani}}, \bibinfo {author} {\bibfnamefont {A.}~\bibnamefont {Khamseh}},\ and\ \bibinfo {author} {\bibfnamefont {M.}~\bibnamefont {Wilson}},\ }\href {https://doi.org/10.1103/PhysRevB.100.064304} {\bibfield  {journal} {\bibinfo  {journal} {Phys. Rev. B}\ }\textbf {\bibinfo {volume} {100}},\ \bibinfo {pages} {064304} (\bibinfo {year} {2019})}\BibitemShut {NoStop}%
\bibitem [{\citenamefont {Beentjes}\ and\ \citenamefont {Khamseh}(2020)}]{beentjes2020higher}%
  \BibitemOpen
  \bibfield  {author} {\bibinfo {author} {\bibfnamefont {S.~V.}\ \bibnamefont {Beentjes}}\ and\ \bibinfo {author} {\bibfnamefont {A.}~\bibnamefont {Khamseh}},\ }\href {https://doi.org/10.1103/PhysRevE.102.053314} {\bibfield  {journal} {\bibinfo  {journal} {Phys. Rev. E}\ }\textbf {\bibinfo {volume} {102}},\ \bibinfo {pages} {053314} (\bibinfo {year} {2020})}\BibitemShut {NoStop}%
\bibitem [{\citenamefont {Bulso}\ and\ \citenamefont {Roudi}(2021)}]{bulso2021restricted}%
  \BibitemOpen
  \bibfield  {author} {\bibinfo {author} {\bibfnamefont {N.}~\bibnamefont {Bulso}}\ and\ \bibinfo {author} {\bibfnamefont {Y.}~\bibnamefont {Roudi}},\ }\href {https://doi.org/10.1162/neco_a_01420} {\bibfield  {journal} {\bibinfo  {journal} {Neural Computation}\ }\textbf {\bibinfo {volume} {33}},\ \bibinfo {pages} {2646} (\bibinfo {year} {2021})}\BibitemShut {NoStop}%
\bibitem [{\citenamefont {Feng}\ \emph {et~al.}(2023)\citenamefont {Feng}, \citenamefont {Kong},\ and\ \citenamefont {Trivedi}}]{feng2023statistical}%
  \BibitemOpen
  \bibfield  {author} {\bibinfo {author} {\bibfnamefont {S.}~\bibnamefont {Feng}}, \bibinfo {author} {\bibfnamefont {D.}~\bibnamefont {Kong}},\ and\ \bibinfo {author} {\bibfnamefont {N.}~\bibnamefont {Trivedi}},\ }\href {https://arxiv.org/abs/2302.03212} {\bibinfo {title} {A statistical approach to topological entanglement: Boltzmann machine representation of high-order irreducible correlation}} (\bibinfo {year} {2023}),\ \Eprint {https://arxiv.org/abs/2302.03212} {arXiv:2302.03212 [quant-ph]} \BibitemShut {NoStop}%
\bibitem [{\citenamefont {Decelle}\ \emph {et~al.}(2024)\citenamefont {Decelle}, \citenamefont {Furtlehner}, \citenamefont {Navas~G{\'o}mez},\ and\ \citenamefont {Seoane}}]{decelle2024inferring}%
  \BibitemOpen
  \bibfield  {author} {\bibinfo {author} {\bibfnamefont {A.}~\bibnamefont {Decelle}}, \bibinfo {author} {\bibfnamefont {C.}~\bibnamefont {Furtlehner}}, \bibinfo {author} {\bibfnamefont {A.~d.~J.}\ \bibnamefont {Navas~G{\'o}mez}},\ and\ \bibinfo {author} {\bibfnamefont {B.}~\bibnamefont {Seoane}},\ }\href {https://doi.org/10.21468/SciPostPhys.16.4.095} {\bibfield  {journal} {\bibinfo  {journal} {SciPost Phys.}\ }\textbf {\bibinfo {volume} {16}},\ \bibinfo {pages} {095} (\bibinfo {year} {2024})}\BibitemShut {NoStop}%
\bibitem [{\citenamefont {di~Sarra}\ \emph {et~al.}(2025)\citenamefont {di~Sarra}, \citenamefont {Bravi},\ and\ \citenamefont {Roudi}}]{di2025unbearable}%
  \BibitemOpen
  \bibfield  {author} {\bibinfo {author} {\bibfnamefont {G.}~\bibnamefont {di~Sarra}}, \bibinfo {author} {\bibfnamefont {B.}~\bibnamefont {Bravi}},\ and\ \bibinfo {author} {\bibfnamefont {Y.}~\bibnamefont {Roudi}},\ }\href {https://doi.org/10.1209/0295-5075/ada636} {\bibfield  {journal} {\bibinfo  {journal} {Europhysics Letters}\ }\textbf {\bibinfo {volume} {149}},\ \bibinfo {pages} {21002} (\bibinfo {year} {2025})}\BibitemShut {NoStop}%
\bibitem [{\citenamefont {Morcos}\ \emph {et~al.}(2011)\citenamefont {Morcos}, \citenamefont {Pagnani}, \citenamefont {Lunt}, \citenamefont {Bertolino}, \citenamefont {Marks}, \citenamefont {Sander}, \citenamefont {Zecchina}, \citenamefont {Onuchic}, \citenamefont {Hwa},\ and\ \citenamefont {Weigt}}]{morcos2011direct}%
  \BibitemOpen
  \bibfield  {author} {\bibinfo {author} {\bibfnamefont {F.}~\bibnamefont {Morcos}}, \bibinfo {author} {\bibfnamefont {A.}~\bibnamefont {Pagnani}}, \bibinfo {author} {\bibfnamefont {B.}~\bibnamefont {Lunt}}, \bibinfo {author} {\bibfnamefont {A.}~\bibnamefont {Bertolino}}, \bibinfo {author} {\bibfnamefont {D.~S.}\ \bibnamefont {Marks}}, \bibinfo {author} {\bibfnamefont {C.}~\bibnamefont {Sander}}, \bibinfo {author} {\bibfnamefont {R.}~\bibnamefont {Zecchina}}, \bibinfo {author} {\bibfnamefont {J.~N.}\ \bibnamefont {Onuchic}}, \bibinfo {author} {\bibfnamefont {T.}~\bibnamefont {Hwa}},\ and\ \bibinfo {author} {\bibfnamefont {M.}~\bibnamefont {Weigt}},\ }\href {https://doi.org/10.1073/pnas.1111471108} {\bibfield  {journal} {\bibinfo  {journal} {Proceedings of the National Academy of Sciences}\ }\textbf {\bibinfo {volume} {108}},\ \bibinfo {pages} {E1293} (\bibinfo {year} {2011})}\BibitemShut {NoStop}%
\bibitem [{\citenamefont {Cocco}\ \emph {et~al.}(2023)\citenamefont {Cocco}, \citenamefont {Martino}, \citenamefont {Pagnani}, \citenamefont {Weigt},\ and\ \citenamefont {Ritort}}]{cocco2023statistical}%
  \BibitemOpen
  \bibfield  {author} {\bibinfo {author} {\bibfnamefont {S.}~\bibnamefont {Cocco}}, \bibinfo {author} {\bibfnamefont {A.~D.}\ \bibnamefont {Martino}}, \bibinfo {author} {\bibfnamefont {A.}~\bibnamefont {Pagnani}}, \bibinfo {author} {\bibfnamefont {M.}~\bibnamefont {Weigt}},\ and\ \bibinfo {author} {\bibfnamefont {F.}~\bibnamefont {Ritort}},\ }\bibinfo {title} {Statistical physics of biological molecules},\ in\ \href {https://doi.org/10.1142/9789811273926_0026} {\emph {\bibinfo {booktitle} {Spin Glass Theory and Far Beyond}}}\ (\bibinfo  {publisher} {World Scientific},\ \bibinfo {year} {2023})\ Chap.\ \bibinfo {chapter} {Chapter 26}, pp.\ \bibinfo {pages} {523--559}\BibitemShut {NoStop}%
\bibitem [{\citenamefont {Tubiana}\ \emph {et~al.}(2019)\citenamefont {Tubiana}, \citenamefont {Cocco},\ and\ \citenamefont {Monasson}}]{tubiana2019learning}%
  \BibitemOpen
  \bibfield  {author} {\bibinfo {author} {\bibfnamefont {J.}~\bibnamefont {Tubiana}}, \bibinfo {author} {\bibfnamefont {S.}~\bibnamefont {Cocco}},\ and\ \bibinfo {author} {\bibfnamefont {R.}~\bibnamefont {Monasson}},\ }\href {https://doi.org/10.7554/eLife.39397} {\bibfield  {journal} {\bibinfo  {journal} {eLife}\ }\textbf {\bibinfo {volume} {8}},\ \bibinfo {pages} {e39397} (\bibinfo {year} {2019})}\BibitemShut {NoStop}%
\bibitem [{\citenamefont {Decelle}\ \emph {et~al.}(2023)\citenamefont {Decelle}, \citenamefont {Seoane},\ and\ \citenamefont {Rosset}}]{decelle2023unsupervised}%
  \BibitemOpen
  \bibfield  {author} {\bibinfo {author} {\bibfnamefont {A.}~\bibnamefont {Decelle}}, \bibinfo {author} {\bibfnamefont {B.}~\bibnamefont {Seoane}},\ and\ \bibinfo {author} {\bibfnamefont {L.}~\bibnamefont {Rosset}},\ }\href {https://doi.org/10.1103/PhysRevE.108.014110} {\bibfield  {journal} {\bibinfo  {journal} {Phys. Rev. E}\ }\textbf {\bibinfo {volume} {108}},\ \bibinfo {pages} {014110} (\bibinfo {year} {2023})}\BibitemShut {NoStop}%
\bibitem [{Note1()}]{Note1}%
  \BibitemOpen
  \bibinfo {note} {{The Frobenius norm of the pairwise couplings, given by $ F_{ij}^{(2)} \mathrel {\mathop :}\mathrel {\mkern -1.2mu}=\protect \!\protect \!\protect \! \protect \sqrt {\DOTSB \sum@ \slimits@ _{\mu , \nu } \left ( J_{ij}^{\mu \nu } \right )^2}$, quantifies the interaction strength between sites $i$ and $j$.}}\BibitemShut {Stop}%
\bibitem [{\citenamefont {Feinauer}\ \emph {et~al.}(2014)\citenamefont {Feinauer}, \citenamefont {Skwark}, \citenamefont {Pagnani},\ and\ \citenamefont {Aurell}}]{feinauer2014improving}%
  \BibitemOpen
  \bibfield  {author} {\bibinfo {author} {\bibfnamefont {C.}~\bibnamefont {Feinauer}}, \bibinfo {author} {\bibfnamefont {M.~J.}\ \bibnamefont {Skwark}}, \bibinfo {author} {\bibfnamefont {A.}~\bibnamefont {Pagnani}},\ and\ \bibinfo {author} {\bibfnamefont {E.}~\bibnamefont {Aurell}},\ }\href {https://doi.org/10.1371/journal.pcbi.1003847} {\bibfield  {journal} {\bibinfo  {journal} {PLOS Computational Biology}\ }\textbf {\bibinfo {volume} {10}},\ \bibinfo {pages} {1} (\bibinfo {year} {2014})}\BibitemShut {NoStop}%
\bibitem [{\citenamefont {Trinquier}\ \emph {et~al.}(2021)\citenamefont {Trinquier}, \citenamefont {Uguzzoni}, \citenamefont {Pagnani}, \citenamefont {Zamponi},\ and\ \citenamefont {Weigt}}]{trinquier2021efficient}%
  \BibitemOpen
  \bibfield  {author} {\bibinfo {author} {\bibfnamefont {J.}~\bibnamefont {Trinquier}}, \bibinfo {author} {\bibfnamefont {G.}~\bibnamefont {Uguzzoni}}, \bibinfo {author} {\bibfnamefont {A.}~\bibnamefont {Pagnani}}, \bibinfo {author} {\bibfnamefont {F.}~\bibnamefont {Zamponi}},\ and\ \bibinfo {author} {\bibfnamefont {M.}~\bibnamefont {Weigt}},\ }\href {https://doi.org/10.1038/s41467-021-25756-4} {\bibfield  {journal} {\bibinfo  {journal} {Nat commun}\ }\textbf {\bibinfo {volume} {12}},\ \bibinfo {pages} {5800} (\bibinfo {year} {2021})}\BibitemShut {NoStop}%
\bibitem [{\citenamefont {Navas~Gómez}(2025)}]{github:couplings_inference}%
  \BibitemOpen
  \bibfield  {author} {\bibinfo {author} {\bibfnamefont {A.~d.~J.}\ \bibnamefont {Navas~Gómez}},\ }\href {https://github.com/DsysDML/couplings_inference} {\bibinfo {title} {{DsysDML}/{couplings{\_}inference}}},\ \bibinfo {howpublished} {GitHub repository} (\bibinfo {year} {2025})\BibitemShut {NoStop}%
\bibitem [{\citenamefont {Rosset}\ \emph {et~al.}(2012)\citenamefont {Rosset}, \citenamefont {Netti}, \citenamefont {Muntoni}, \citenamefont {Weigt},\ and\ \citenamefont {Zamponi}}]{rosset2025adabmdca}%
  \BibitemOpen
  \bibfield  {author} {\bibinfo {author} {\bibfnamefont {L.}~\bibnamefont {Rosset}}, \bibinfo {author} {\bibfnamefont {R.}~\bibnamefont {Netti}}, \bibinfo {author} {\bibfnamefont {A.~P.}\ \bibnamefont {Muntoni}}, \bibinfo {author} {\bibfnamefont {M.}~\bibnamefont {Weigt}},\ and\ \bibinfo {author} {\bibfnamefont {F.}~\bibnamefont {Zamponi}},\ }\bibinfo {title} {adabmdca 2.0—a flexible but easy-to-use package for direct coupling analysis},\ in\ \href {https://doi.org/10.1007/978-1-0716-4828-5_6} {\emph {\bibinfo {booktitle} {Protein Evolution}}}\ (\bibinfo  {publisher} {Springer US},\ \bibinfo {year} {2012})\ p.\ \bibinfo {pages} {83–104}\BibitemShut {NoStop}%
\bibitem [{\citenamefont {Rende}\ \emph {et~al.}(2024)\citenamefont {Rende}, \citenamefont {Gerace}, \citenamefont {Laio},\ and\ \citenamefont {Goldt}}]{rende2024distributional}%
  \BibitemOpen
  \bibfield  {author} {\bibinfo {author} {\bibfnamefont {R.}~\bibnamefont {Rende}}, \bibinfo {author} {\bibfnamefont {F.}~\bibnamefont {Gerace}}, \bibinfo {author} {\bibfnamefont {A.}~\bibnamefont {Laio}},\ and\ \bibinfo {author} {\bibfnamefont {S.}~\bibnamefont {Goldt}},\ }in\ \href {https://doi.org/10.48550/arXiv.2410.19637} {\emph {\bibinfo {booktitle} {Proceedings of the 38th Annual Conference on Neural Information Processing Systems}}}\ (\bibinfo {year} {2024})\BibitemShut {NoStop}%
\bibitem [{Note2()}]{Note2}%
  \BibitemOpen
  \bibinfo {note} {Without regularization, strong effective couplings emerged early in training, degrading MCMC-based gradient estimates and leading to non-monotonic inference performance.}\BibitemShut {Stop}%
\bibitem [{\citenamefont {B{\'e}reux}\ \emph {et~al.}(2025)\citenamefont {B{\'e}reux}, \citenamefont {Decelle}, \citenamefont {Furtlehner}, \citenamefont {Rosset},\ and\ \citenamefont {Seoane}}]{bereux2024fast}%
  \BibitemOpen
  \bibfield  {author} {\bibinfo {author} {\bibfnamefont {N.}~\bibnamefont {B{\'e}reux}}, \bibinfo {author} {\bibfnamefont {A.}~\bibnamefont {Decelle}}, \bibinfo {author} {\bibfnamefont {C.}~\bibnamefont {Furtlehner}}, \bibinfo {author} {\bibfnamefont {L.}~\bibnamefont {Rosset}},\ and\ \bibinfo {author} {\bibfnamefont {B.}~\bibnamefont {Seoane}},\ }in\ \href {https://doi.org/10.48550/arXiv.2405.15376} {\emph {\bibinfo {booktitle} {The 13th International Conference on Learning Representations}}}\ (\bibinfo {year} {2025})\BibitemShut {NoStop}%
\bibitem [{\citenamefont {Muntoni}\ \emph {et~al.}(2021)\citenamefont {Muntoni}, \citenamefont {Pagnani}, \citenamefont {Weigt},\ and\ \citenamefont {Zamponi}}]{muntoni2021adabmdca}%
  \BibitemOpen
  \bibfield  {author} {\bibinfo {author} {\bibfnamefont {A.~P.}\ \bibnamefont {Muntoni}}, \bibinfo {author} {\bibfnamefont {A.}~\bibnamefont {Pagnani}}, \bibinfo {author} {\bibfnamefont {M.}~\bibnamefont {Weigt}},\ and\ \bibinfo {author} {\bibfnamefont {F.}~\bibnamefont {Zamponi}},\ }\href {https://doi.org/10.1186/s12859-021-04441-9} {\bibfield  {journal} {\bibinfo  {journal} {BMC bioinformatics}\ }\textbf {\bibinfo {volume} {22}},\ \bibinfo {pages} {1} (\bibinfo {year} {2021})}\BibitemShut {NoStop}%
\bibitem [{\citenamefont {Schmidt}\ and\ \citenamefont {Hamacher}(2017)}]{schmidt2017three}%
  \BibitemOpen
  \bibfield  {author} {\bibinfo {author} {\bibfnamefont {M.}~\bibnamefont {Schmidt}}\ and\ \bibinfo {author} {\bibfnamefont {K.}~\bibnamefont {Hamacher}},\ }\href {https://doi.org/10.1103/PhysRevE.96.052405} {\bibfield  {journal} {\bibinfo  {journal} {Phys. Rev. E}\ }\textbf {\bibinfo {volume} {96}},\ \bibinfo {pages} {052405} (\bibinfo {year} {2017})}\BibitemShut {NoStop}%
\bibitem [{\citenamefont {B{\'e}reux}\ and\ \citenamefont {Seoane}(2025)}]{github:rbms}%
  \BibitemOpen
  \bibfield  {author} {\bibinfo {author} {\bibfnamefont {N.}~\bibnamefont {B{\'e}reux}}\ and\ \bibinfo {author} {\bibfnamefont {B.}~\bibnamefont {Seoane}},\ }\href {https://github.com/DsysDML/rbms} {\bibinfo {title} {{DsysDML}/{rbms}}},\ \bibinfo {howpublished} {GitHub repository} (\bibinfo {year} {2025})\BibitemShut {NoStop}%
\bibitem [{\citenamefont {Tieleman}(2008)}]{tieleman2008training}%
  \BibitemOpen
  \bibfield  {author} {\bibinfo {author} {\bibfnamefont {T.}~\bibnamefont {Tieleman}},\ }in\ \href {https://doi.org/10.1145/1390156.1390290} {\emph {\bibinfo {booktitle} {Proceedings of the 25th International Conference on Machine Learning}}},\ \bibinfo {series and number} {ICML '08}\ (\bibinfo  {publisher} {Association for Computing Machinery},\ \bibinfo {address} {New York, NY, USA},\ \bibinfo {year} {2008})\ p.\ \bibinfo {pages} {1064–1071}\BibitemShut {NoStop}%
\bibitem [{\citenamefont {Decelle}\ \emph {et~al.}(2021)\citenamefont {Decelle}, \citenamefont {Furtlehner},\ and\ \citenamefont {Seoane}}]{decelle2021equilibrium}%
  \BibitemOpen
  \bibfield  {author} {\bibinfo {author} {\bibfnamefont {A.}~\bibnamefont {Decelle}}, \bibinfo {author} {\bibfnamefont {C.}~\bibnamefont {Furtlehner}},\ and\ \bibinfo {author} {\bibfnamefont {B.}~\bibnamefont {Seoane}},\ }in\ \href {https://proceedings.neurips.cc/paper_files/paper/2021/file/2aedcba61ca55ceb62d785c6b7f10a83-Paper.pdf} {\emph {\bibinfo {booktitle} {Advances in Neural Information Processing Systems}}},\ Vol.~\bibinfo {volume} {34},\ \bibinfo {editor} {edited by\ \bibinfo {editor} {\bibfnamefont {M.}~\bibnamefont {Ranzato}}, \bibinfo {editor} {\bibfnamefont {A.}~\bibnamefont {Beygelzimer}}, \bibinfo {editor} {\bibfnamefont {Y.}~\bibnamefont {Dauphin}}, \bibinfo {editor} {\bibfnamefont {P.}~\bibnamefont {Liang}},\ and\ \bibinfo {editor} {\bibfnamefont {J.~W.}\ \bibnamefont {Vaughan}}}\ (\bibinfo  {publisher} {Curran Associates, Inc.},\ \bibinfo {year} {2021})\ pp.\ \bibinfo {pages} {5345--5359}\BibitemShut {NoStop}%
\bibitem [{\citenamefont {Agoritsas}\ \emph {et~al.}(2023)\citenamefont {Agoritsas}, \citenamefont {Catania}, \citenamefont {Decelle},\ and\ \citenamefont {Seoane}}]{agoritsas2023explaining}%
  \BibitemOpen
  \bibfield  {author} {\bibinfo {author} {\bibfnamefont {E.}~\bibnamefont {Agoritsas}}, \bibinfo {author} {\bibfnamefont {G.}~\bibnamefont {Catania}}, \bibinfo {author} {\bibfnamefont {A.}~\bibnamefont {Decelle}},\ and\ \bibinfo {author} {\bibfnamefont {B.}~\bibnamefont {Seoane}},\ }in\ \href {https://doi.org/10.48550/arXiv.2301.09428} {\emph {\bibinfo {booktitle} {Proceedings of the 40th International Conference on Machine Learning}}},\ \bibinfo {series} {Proceedings of Machine Learning Research}, Vol.\ \bibinfo {volume} {202}\ (\bibinfo  {publisher} {PMLR},\ \bibinfo {year} {2023})\ pp.\ \bibinfo {pages} {322--336}\BibitemShut {NoStop}%
\bibitem [{\citenamefont {Coolen}\ \emph {et~al.}(2005)\citenamefont {Coolen}, \citenamefont {Kühn},\ and\ \citenamefont {Sollich}}]{coolen2005theory}%
  \BibitemOpen
  \bibfield  {author} {\bibinfo {author} {\bibfnamefont {A.~C.~C.}\ \bibnamefont {Coolen}}, \bibinfo {author} {\bibfnamefont {R.}~\bibnamefont {Kühn}},\ and\ \bibinfo {author} {\bibfnamefont {P.}~\bibnamefont {Sollich}},\ }\href {https://doi.org/10.1093/oso/9780198530237.001.0001} {\emph {\bibinfo {title} {Theory of Neural Information Processing Systems}}}\ (\bibinfo  {publisher} {Oxford University Press},\ \bibinfo {year} {2005})\BibitemShut {NoStop}%
\bibitem [{\citenamefont {Blume}(1966)}]{blume1966theory}%
  \BibitemOpen
  \bibfield  {author} {\bibinfo {author} {\bibfnamefont {M.}~\bibnamefont {Blume}},\ }\href {https://doi.org/10.1103/PhysRev.141.517} {\bibfield  {journal} {\bibinfo  {journal} {Phys. Rev.}\ }\textbf {\bibinfo {volume} {141}},\ \bibinfo {pages} {517} (\bibinfo {year} {1966})}\BibitemShut {NoStop}%
\bibitem [{\citenamefont {Capel}(1966)}]{capel1966possibility}%
  \BibitemOpen
  \bibfield  {author} {\bibinfo {author} {\bibfnamefont {H.}~\bibnamefont {Capel}},\ }\href {https://doi.org/10.1016/0031-8914(66)90027-9} {\bibfield  {journal} {\bibinfo  {journal} {Physica}\ }\textbf {\bibinfo {volume} {32}},\ \bibinfo {pages} {966} (\bibinfo {year} {1966})}\BibitemShut {NoStop}%
\bibitem [{\citenamefont {Durbin}\ \emph {et~al.}(1998)\citenamefont {Durbin}, \citenamefont {Eddy}, \citenamefont {Krogh},\ and\ \citenamefont {Mitchison}}]{durbin1998biological}%
  \BibitemOpen
  \bibfield  {author} {\bibinfo {author} {\bibfnamefont {R.}~\bibnamefont {Durbin}}, \bibinfo {author} {\bibfnamefont {S.~R.}\ \bibnamefont {Eddy}}, \bibinfo {author} {\bibfnamefont {A.}~\bibnamefont {Krogh}},\ and\ \bibinfo {author} {\bibfnamefont {G.}~\bibnamefont {Mitchison}},\ }\href {https://doi.org/10.1017/CBO9780511790492} {\emph {\bibinfo {title} {Biological Sequence Analysis: Probabilistic Models of Proteins and Nucleic Acids}}}\ (\bibinfo  {publisher} {Cambridge University Press},\ \bibinfo {year} {1998})\BibitemShut {NoStop}%
\bibitem [{\citenamefont {Dunn}\ \emph {et~al.}(2007)\citenamefont {Dunn}, \citenamefont {Wahl},\ and\ \citenamefont {Gloor}}]{dunn2008mutual}%
  \BibitemOpen
  \bibfield  {author} {\bibinfo {author} {\bibfnamefont {S.}~\bibnamefont {Dunn}}, \bibinfo {author} {\bibfnamefont {L.}~\bibnamefont {Wahl}},\ and\ \bibinfo {author} {\bibfnamefont {G.}~\bibnamefont {Gloor}},\ }\href {https://doi.org/10.1093/bioinformatics/btm604} {\bibfield  {journal} {\bibinfo  {journal} {Bioinformatics}\ }\textbf {\bibinfo {volume} {24}},\ \bibinfo {pages} {333} (\bibinfo {year} {2007})}\BibitemShut {NoStop}%
\bibitem [{\citenamefont {Burger}\ and\ \citenamefont {van Nimwegen}(2010)}]{burger2010disentangling}%
  \BibitemOpen
  \bibfield  {author} {\bibinfo {author} {\bibfnamefont {L.}~\bibnamefont {Burger}}\ and\ \bibinfo {author} {\bibfnamefont {E.}~\bibnamefont {van Nimwegen}},\ }\href {https://doi.org/10.1371/journal.pcbi.1000633} {\bibfield  {journal} {\bibinfo  {journal} {PLOS Computational Biology}\ }\textbf {\bibinfo {volume} {6}},\ \bibinfo {pages} {1} (\bibinfo {year} {2010})}\BibitemShut {NoStop}%
\bibitem [{\citenamefont {Feinauer}(2016)}]{feinauer2016statistical}%
  \BibitemOpen
  \bibfield  {author} {\bibinfo {author} {\bibfnamefont {C.}~\bibnamefont {Feinauer}},\ }\emph {\bibinfo {title} {The Statistical Mechanics Approach to Protein Sequence Data: Beyond Contact Prediction}},\ \href {https://hdl.handle.net/11583/2640930} {Ph.D. thesis},\ \bibinfo  {school} {Politecnico di Torino}, \bibinfo {address} {Turin, Italy} (\bibinfo {year} {2016})\BibitemShut {NoStop}%
\end{thebibliography}%

\newpage
\appendix
\begin{widetext}
\section{Training of the Potts-Bernoulli RBM}\label{sec:training}

Our RBMs are trained by maximizing the log-likelihood function \(\mathcal{L}(\boldsymbol{\Theta} | \mathcal{D})\), which quantifies the probability of observing a given dataset \(\mathcal{D} = \{\boldsymbol{v}^{(1)}, \dots, \boldsymbol{v}^{(M)}\}\) under a probabilistic model parameterized by \(\boldsymbol{\Theta}\). The optimization objective is then formally expressed as:
\begin{align}\label{eq:LL}
    \mathcal{L}( \mathcal{D}|\boldsymbol{\Theta} ) &=  \frac{1}{M} \sum_{m=1}^{M} \log p_{\boldsymbol{\Theta}}\left( \boldsymbol{v}^{(m)} \right) = \frac{1}{M} \sum_{m=1}^{M} \log \sum_{\boldsymbol{h}} e^{- \mathcal{H}_{\boldsymbol{\Theta}} (\boldsymbol{v}^{(m)}, \boldsymbol{h})} - \log \mathcal{Z}_{\boldsymbol{\Theta}}.
\end{align}
We optimize \(\mathcal{L}(\boldsymbol{\Theta} | \mathcal{D})\) using the (stochastic) gradient descent algorithm. For our RBM, the gradient is given by:
\begin{align}
    \frac{\partial \mathcal{L}}{\partial b_i^\mu} &= \pa{\delta_\mu^{v_i}}_{\mathcal{D}} - \pa{\delta_\mu^{v_i}}_{\mathcal{H}}, \quad \frac{\partial \mathcal{L}}{\partial c_a} = \pa{h_a}_{\mathcal{D}} - \pa{h_a}_{\mathcal{H}}, \nonumber\\ &\quad \text{and} \quad  \frac{\partial \mathcal{L}}{\partial W_{i a}^\mu} = \pa{h_a \delta_\mu^{v_i}}_{\mathcal{D}} - \pa{h_a \delta_\mu^{v_i}}_{\mathcal{H}},
\end{align}
where
\begin{equation}
    \langle f(\bm v,\bm h) \rangle_{\mathcal{D}} = M^{-1} \sum_m \sum_{\{\bm h\}} f(\bm v^{(m)},\bm h) p(\bm h \| \bm v^{(m)})
    \label{marginal gradient}
\end{equation}
represents the expectation of \(f(\bm v,\bm h) \) over the dataset, and \(\langle f(\bm v,\bm h) \rangle_{\mathcal{H}}\) denotes the expectation with respect to the Boltzmann distribution of the model, whose Hamiltonian is defined in Eq.~\eqref{RBM_potts} in the main text. Note also that we use $p(\bm h \| \bm v)$ to denote the probability of a given state of the hidden layer $\boldsymbol{h}$ given a visible activation $\boldsymbol{v}$. Since computing the partition function is typically intractable, the second average is approximated using Markov Chain Monte Carlo (MCMC) and the Alternating Gibbs Sampling algorithm. This algorithm exploits the bipartite structure of the RBM by iteratively sampling the visible and hidden variables conditioned on the other layer. This approach enables efficient parallelization, significantly accelerating the sampling process. Then, with samples of visible layer, we can use Eq.~\eqref{marginal gradient} to obtain an accurate estimation of the negative part of the gradient.

In this work, we trained RBMs using the persistent contrastive divergence (PCD-\(k\)) algorithm~\cite{tieleman2008training}, where the Markov chains for gradient estimation are initialized from the final state of the previous update and evolved for \(k\) Gibbs steps. This approach has been shown to yield quasi-equilibrium models~\cite{decelle2021equilibrium}. We further verified that the learned models operated in the equilibrium regime~\cite{agoritsas2023explaining, decelle2021equilibrium}, exhibiting no memory of the training scheme when sampled for generation.

In the experiments conducted on the protein family, we empirically observed that $\ell_2$ regularization significantly improved the results. This is done by adding a term $- \lambda_{\ell} \sum_{i,\mu,a} \left(W_{ia}^\mu \right)^2- \lambda_{\ell} \sum_{i,\mu} \left(b_{i}^\mu \right)^2- \lambda_{\ell} \sum_{a} \left(c_{a} \right)^2$ in the likelihood, which effective changes the gradient on the weights as 
\begin{equation}
    \quad  \frac{\partial \mathcal{L}}{\partial W_{i a}^\mu} = \pa{h_a \delta_\mu^{v_i}}_{\mathcal{D}} - \pa{h_a \delta_\mu^{v_i}}_{\mathcal{H}} - 2\lambda_{\ell} W_{ia}^\mu,
\end{equation}
and same for the rest of the parameters.

\paragraph{Training parameters for the Blume-Capel model:} the experiments illustrated on the Blume-Capel model were done using the following parameters for the training of the RBM: PCD-$50$ (persistent contrastive divergence with $50$ steps), $\Nh=250$ (number of hidden units), batch size 1000 and $\gamma=0.01$ (learning rate) and $M=10^5$ training samples.

\paragraph{Training parameters for the PF00072 protein:} the experiments illustrated the protein PF00072 were done using the following parameters for the training of the RBM: PCD-100, $\Nh \!=\! 10^3$ hidden units, batch size 5000, $\ell_2$ regularization $\lambda_\ell \!=\! 0.001$, learning rate $\gamma\! =\! 0.01$.

\section{Gauge Invariance in Generalized Multibody Potts Models}\label{gauge_invariance_appendix}

\subsection{Gauge Transformation of Models with Higher-Order Interaction Terms}
To familiarize the reader with the gauge transformations of the generalized Potts model defined in Eq.~\eqref{general_Potts}, we now compute explicitly how the parameters of the Hamiltonian transform under a generic gauge transformation in the case where interactions up to three bodies are included:
\begin{gather}
    \mathcal{H} (\boldsymbol{v}) 
    = - \sum_{i, \mu} H_i^\mu \delta_\mu^{v_i} - \sum_{1 \le i < j \le N_v} \sum_{\mu, \nu} J_{i j}^{\mu \nu} \delta_{\mu}^{v_{i}} \delta_{\nu}^{v_{j}} - \sum_{1 \le i < j < k \le N_v} \sum_{\mu, \nu, \eta} J_{i j k}^{\mu \nu \eta} \delta_{\mu}^{v_{i}} \delta_{\nu}^{v_{j}} \delta_{\eta}^{v_{k}}.
    \label{general_Potts_3}
\end{gather}
Any gauge transformation for the coupling tensors can be written as:
\begin{subequations}
\begin{gather}
    J_{ijk}^{\mu \nu \eta} \rightarrow J_{ijk}^{\mu \nu \eta} + K_{ij k}^{\mu \nu} + K_{jk i}^{\nu \eta} + K_{ki j}^{\eta \mu}, \\
    J_{ij}^{\mu \nu} \rightarrow J_{ij}^{\mu \nu} + K_{ij}^{\mu} + K_{ji}^{\nu} - \sum_{k \ne i,j} K_{ij k}^{\mu \nu}, \\
    H_i^\mu \rightarrow H_i^\mu + K_i - \sum_{j \ne i} K_{ij}^{\mu}.
\end{gather}
\label{3-body_gauge_transformation}
\end{subequations}
These transformations leave the Hamiltonian~\eqref{general_Potts_3} invariant up to a constant. To see this, let us denote the transformed Hamiltonian by \(\tilde{\mathcal{H}}(\boldsymbol{v})\) and compute the difference \(\tilde{\mathcal{H}}(\boldsymbol{v}) - \mathcal{H}(\boldsymbol{v})\). Substituting the transformed couplings from~\eqref{3-body_gauge_transformation} into~\eqref{general_Potts_3} yields:
\begin{align}
    \tilde{\mathcal{H}}(\boldsymbol{v}) - \mathcal{H}(\boldsymbol{v}) 
    &= - \sum_{i, \mu} \left( H_i^\mu + K_i - \sum_{j \ne i} K_{ij}^{\mu} - H_i^\mu \right) \delta_\mu^{v_i} \nonumber \\
    &\quad - \sum_{i < j} \sum_{\mu,\nu} \left( J_{ij}^{\mu \nu} + K_{ij}^{\mu} + K_{ji}^{\nu} - \sum_{k \ne i,j} K_{ij k}^{\mu \nu} - J_{ij}^{\mu \nu} \right) \delta_\mu^{v_i} \delta_\nu^{v_j} \nonumber \\
    &\quad - \sum_{i < j < k} \sum_{\mu,\nu,\eta} \left( J_{ijk}^{\mu \nu \eta} + K_{ij k}^{\mu \nu} + K_{jk i}^{\nu \eta} + K_{ki j}^{\eta \mu} - J_{ijk}^{\mu \nu \eta} \right) \delta_\mu^{v_i} \delta_\nu^{v_j} \delta_\eta^{v_k} \nonumber \\
    &= - \sum_i K_i \delta_\mu^{v_i} + \sum_{i \ne j} K_{ij}^{\mu} \delta_\mu^{v_i} 
    - \sum_{i<j} \sum_{\mu,\nu} \left( K_{ij}^{\mu} + K_{ji}^{\nu} \right) \delta_\mu^{v_i} \delta_\nu^{v_j} \nonumber \\
    & \quad  + \sum_{i<j} \sum_{\mu,\nu} \sum_{k \ne i,j} K_{ij k}^{\mu \nu} \delta_\mu^{v_i} \delta_\nu^{v_j}  - \sum_{i<j<k} \sum_{\mu,\nu,\eta} \left( K_{ij k}^{\mu \nu} + K_{jk i}^{\nu \eta} + K_{ki j}^{\eta \mu} \right) \delta_\mu^{v_i} \delta_\nu^{v_j} \delta_\eta^{v_k}. 
    \label{gauge_difference}
\end{align}

We now show that all configuration-dependent terms in the right-hand side of~\eqref{gauge_difference} cancel out. For the terms involving the two-body parameters, we observe:
\begin{align*}
    \sum_{i \ne j} \sum_{\mu} K_{ij}^{\mu} \delta_\mu^{v_i} 
    &= \sum_{i < j} \sum_{\mu} K_{ij}^{\mu} \delta_\mu^{v_i} + \sum_{j < i} \sum_{\mu} K_{ij}^{\mu} \delta_\mu^{v_i} \\
    &= \sum_{i < j} \left[ \sum_{\mu} K_{ij}^{\mu} \delta_\mu^{v_i} + \sum_{\nu} K_{ji}^{\nu} \delta_\nu^{v_j} \right],
\end{align*}
which exactly cancels the second and third term in the first line of~\eqref{gauge_difference}. Similarly, for the three-body terms, we reorganize:
\begin{align*}
    \sum_{\substack{i < j, \\ k \ne i,j}} \sum_{\mu,\nu} K_{ij k}^{\mu \nu} \delta_\mu^{v_i} \delta_\nu^{v_j}
    &= \sum_{i<j<k} \sum_{\mu,\nu} K_{ij k}^{\mu \nu} \delta_\mu^{v_i} \delta_\nu^{v_j}
     + \sum_{j<k<i} \sum_{\mu,\nu} K_{ij k}^{\mu \nu} \delta_\mu^{v_i} \delta_\nu^{v_j}
     + \sum_{k<i<j} \sum_{\mu,\nu} K_{ij k}^{\mu \nu} \delta_\mu^{v_i} \delta_\nu^{v_j} \\
    &= \sum_{i<j<k} \sum_{\mu,\nu} \left[
        K_{ij k}^{\mu \nu} \delta_\mu^{v_i} \delta_\nu^{v_j}
        + K_{ki j}^{\eta \mu} \delta_\eta^{v_k} \delta_\mu^{v_i}
        + K_{jk i}^{\nu \eta} \delta_\nu^{v_j} \delta_\eta^{v_k}
    \right],
\end{align*}
which exactly cancels the second line of~\eqref{gauge_difference}. As a result, the difference between the transformed and the original Hamiltonian reduces to:
\begin{equation}
    \hat{\mathcal{H}}(\boldsymbol{v}) - \mathcal{H}(\boldsymbol{v}) = - \sum_{i} K_i .
\end{equation}
Since this final expression is independent of the configuration \(\boldsymbol{v} = (v_1, v_2, \dots, v_{N_v})\), as it only involves a constant shift summed over all sites, we can safely say that this the transformation preserves the form of the Hamiltonian up to an additive constant which implies and invariance in its associated Boltzmann distribution.

Finally, we emphasize that the structure of the transformation given in~\eqref{3-body_gauge_transformation} is the most general one that preserves the form of the Hamiltonian with up to three-body interactions. If one were to introduce a more general transformation, such as:
\(
    J_{ijk}^{\mu \nu \eta} \rightarrow J_{ijk}^{\mu \nu \eta} + K_{ijk}^{\mu \nu \eta},
\)
with an arbitrary tensor \( K_{ijk}^{\mu \nu \eta} \), the new terms would produce three-body contributions in the transformed Hamiltonian that cannot be canceled by induced two-body or one-body terms. Therefore, the transformations in~\eqref{3-body_gauge_transformation} exhaust the class of gauge transformations that preserve the interaction structure of the generalized Potts Hamiltonian in~\eqref{general_Potts_3}. Thus, in general, one can write the transformation rules for a Hamiltonian with up to $n$-order interaction as:
\begin{subequations}
\begin{gather}
    J_{i_1 i_2 \dots i_{n-1} i_n}^{\mu_1 \mu_2 \dots \mu_{n-1} \mu_n} \rightarrow J_{i_1 i_2 \dots i_{n-1} i_n}^{\mu_1 \mu_2 \dots \mu_{n-1} \mu_n} + K_{i_1 i_2 \dots i_{n-1} i_n}^{\mu_1 \mu_2 \dots \mu_{n-1} } + K_{i_2 i_3 \dots i_n i_1}^{\mu_2 \mu_3 \dots \mu_n} + \dots + K_{i_n i_1 \dots i_{n-2} i_{n-1}}^{\mu_n \mu_1 \dots \mu_{n-2}}, \\
    \vdots \nonumber \\
    J_{i_1 i_2 \dots i_k}^{\mu_1 \mu_2 \dots \mu_k} \rightarrow J_{i_1 i_2 \dots i_{k-1} i_k}^{\mu_1 \mu_2 \dots \mu_{k-1} \mu_k}  + K_{i_1 i_2 \dots i_{k-1} i_k}^{\mu_1 \mu_2 \dots \mu_{k-1}} + K_{i_2 i_3 \dots i_k i_1}^{\mu_2 \mu_3 \dots \mu_{k}} + \dots + K_{i_n i_1 \dots i_{n-2} i_{n-1}}^{\mu_n \mu_1 \dots \mu_{n-2} } + \sum_{i_{k+1} \neq i_1,\dots, i_k } K_{i_1 i_2 \dots + i_k i_{k+1}}^{\mu_1 \mu_2 \dots \mu_k}, \\
    \vdots \nonumber \\ 
    H_i^\mu \rightarrow H_i^\mu + K_i - \sum_{j \ne i} K_{ij}^{\mu}.
\end{gather} 
\end{subequations}

\subsection{Gauge Choice for Coupling Inference}
\label{minimization_frobenius}

{The zero-sum gauge is considered optimal for the inference of pairwise Potts models, as it relies on local fields as much as possible while minimizing the use of pairwise couplings~\cite{cocco2018inverse}. We show here that this rationale extends naturally to the generalized Potts model with higher-order interaction terms. In practice, it is computationally unfeasible to determine all the interaction terms in the effective Hamiltonian Eq.~\eqref{general_Potts} when the interaction order is \( n = N_v \). Therefore, the model must be truncated at a small order, typically \( n \leq 3 \). In this context, it is desirable to adopt a gauge that captures as many of the interactions as possible using lower-order coupling tensors, introducing non-vanishing higher-order terms only when strictly necessary. For \( n = 2 \), a natural and widely adopted choice is the zero-sum gauge, which minimizes the Frobenius norm of the pairwise coupling tensors:
\begin{equation}
    F_{ij}^{(2)} = \sqrt{\sum_{\mu, \nu} \left( J_{ij}^{\mu \nu} \right)^2 }.
    \label{frobenius_order_2}
\end{equation}

In the following, we will show that this argument generalizes to higher-order couplings. Let us consider a generic \(n\)-th order tensor \( J_{i_1 \dots i_n}^{\mu_1 \dots \mu_n} \), and introduce an infinitesimal gauge transformation:
\begin{align*}
    J_{i_1 \dots i_n}^{\mu_1 \dots \mu_n}
    \rightarrow J_{i_1 \dots i_n}^{\mu_1 \dots \mu_n} 
    &+ \Delta_{i_1 \dots i_{n-1} i_n}^{\mu_1 \dots \mu_{n-1}} 
    + \Delta_{i_2 \dots i_n i_1}^{\mu_2 \dots \mu_n} 
    + \Delta_{i_3 \dots i_n i_1 i_2}^{\mu_3 \dots \mu_n \mu_1} 
    + \dots 
    + \Delta_{i_n i_1 \dots i_{n-2} i_{n-1}}^{\mu_n \mu_1 \dots \mu_{n-2}}.
\end{align*}
We now compute the variation of the squared Frobenius norm induced by this transformation:
\begin{align*}
    \Delta \sum_{\mu_1, \dots, \mu_n} \left( J_{i_1 \dots i_n}^{\mu_1 \dots \mu_n} \right)^2
    &= \sum_{\mu_1, \dots, \mu_n} \left( J_{i_1 \dots i_n}^{\mu_1 \dots \mu_n} + \Delta_{i_1 \dots i_{n-1} i_n}^{\mu_1 \dots \mu_{n-1}} + \cdots + \Delta_{i_n i_1 \dots i_{n-2} i_{n-1}}^{\mu_n \mu_1 \dots \mu_{n-2}} \right)^2 
    - \sum_{\mu_1, \dots, \mu_n} \left( J_{i_1 \dots i_n}^{\mu_1 \dots \mu_n} \right)^2 \\
    &= 2 \sum_{\mu_1, \dots, \mu_n} J_{i_1 \dots i_n}^{\mu_1 \dots \mu_n} 
    \left( \Delta_{i_1 \dots i_{n-1} i_n}^{\mu_1 \dots \mu_{n-1}} 
    + \Delta_{i_2 \dots i_n i_1}^{\mu_2 \dots \mu_n} 
    + \dots 
    + \Delta_{i_n i_1 \dots i_{n-2} i_{n-1}}^{\mu_n \mu_1 \dots \mu_{n-2}} \right) 
    + \mathcal{O}(\Delta^2).
\end{align*}
Neglecting second-order terms in \(\Delta\), and regrouping by unique index blocks:
\begin{align*}
    \Delta \sum_{\mu_1, \dots, \mu_n} \left( J_{i_1 \dots i_n}^{\mu_1 \dots \mu_n} \right)^2
    &= 2 \sum_{\mu_1, \dots, \mu_{n-1}} \left( \sum_{\mu_n} J_{i_1 \dots i_n}^{\mu_1 \dots \mu_n} \right) \Delta_{i_1 \dots i_{n-1} i_n}^{\mu_1 \dots \mu_{n-1}} \\
    &\quad + 2 \sum_{\mu_2, \dots, \mu_n} \left( \sum_{\mu_1} J_{i_1 \dots i_n}^{\mu_1 \dots \mu_n} \right) \Delta_{i_2 \dots i_n i_1}^{\mu_2 \dots \mu_n} \\
    &\quad + 2 \sum_{\mu_3, \dots, \mu_n, \mu_1} \left( \sum_{\mu_2} J_{i_1 \dots i_n}^{\mu_1 \dots \mu_n} \right) \Delta_{i_3 \dots i_n i_1 i_2}^{\mu_3 \dots \mu_n \mu_1} \\
    &\quad + \dots \\
    &\quad + 2 \sum_{\mu_n, \mu_1, \dots, \mu_{n-2}} \left( \sum_{\mu_{n-1}} J_{i_1 \dots i_n}^{\mu_1 \dots \mu_n} \right) \Delta_{i_n i_1 \dots i_{n-2} i_{n-1}}^{\mu_n \mu_1 \dots \mu_{n-2}}.
\end{align*}
If the zero-sum gauge condition holds, i.e.,
\[
    \sum_{\mu_k} J_{i_1 \dots i_k \dots i_n}^{\mu_1 \dots \mu_k \dots \mu_n} = 0, \quad \forall k \in \{1, \dots, n\},
\]
then each term above vanishes, and the total variation is zero:
\[
    \Delta \sum_{\mu_1, \dots, \mu_n} \left( J_{i_1 \dots i_n}^{\mu_1 \dots \mu_n} \right)^2 = 0.
\]
This implies that the zero-sum gauge minimizes the Frobenius norm at \(n\)-th order, and therefore suppresses the strength of the highest-order interactions when the Hamiltonian in Eq.~\eqref{general_Potts} is truncated at \( ( n-1) \)-th order. Repeating this argument recursively for each lower order, we conclude that the zero-sum gauge provides the most compact and efficient parametrization of the model for inference purposes. In particular, this gauge choice is advantageous for applications such as protein contact prediction, where the inference of lower-order effective couplings must compensate for the truncation of higher-order terms.}

\section{Deriving the Effective Lattice-Gas Model } 
\label{derivation_lattice_gauge_appendix}
In Ref. ~\cite{cossu2019machine}, a Bernoulli-Bernoulli RBM (i.e., \( v_i, h_a \in \{0, 1\} \)) was mapped onto a multi-body lattice-gas model of binary variables. The following section employs this approach to derive a mapping for an arbitrary number of colors, \(q\). Marginalizing the Boltzmann distribution in Eq.~\eqref{Boltzmann_distribution_RBM} yields:
\begin{align}
    p(\boldsymbol{v})= \frac{1}{\mathcal{Z}} \sum_{\boldsymbol{h}} e^{-\mathcal{H}(\boldsymbol{v}, \boldsymbol{h})} = \frac{1}{\mathcal{Z}} e^{-\msfsl{H}(\boldsymbol{v})}.
    \label{marginalized_distribution}
\end{align}
By substituting the RBM Hamiltonian from Eq.~\eqref{RBM_potts} into this expression, and summing over the hidden nodes, we obtain:
\begin{align}
    \msfsl{H}(\boldsymbol{v}) 
    &= - \sum_{i, \mu} b_i^\mu \delta_\mu^{v_i}  - \sum_a \ln \sum_{h_a} e^{c_a h_a + h_a \sum_{i,\mu} W_{ia}^\mu \delta_\mu^{v_i} }.
    \label{marginalized_hamiltonian}
\end{align}
Now we define
\begin{equation}
    q(h_a) \coloneqq e^{c_a h_a} \ \mathrm{and} \ t \coloneqq \sum_{i,\mu} W_{ia}^\mu \delta_\mu^{v_i},
    \label{aux_var}
\end{equation}
to introduce the following {\em cumulant}-generating function
\begin{subequations}
\begin{align}
    K_a (t) 
    &\coloneqq \ln \sum_{h_a} q(h_a) \ e^{h_a t} = \sum_{k=0}^{\infty} \frac{\kappa_a^{(k)} t^k}{k!} 
     \label{cumulant_gen_function} \\
    &= \ln \left( 1 + e^{c_a + t} \right),
    \label{cumulant_gen_function2}
\end{align}
\end{subequations}
where the $k$-th cumulant is 
\begin{equation}
    \kappa_a^{(k)} = \left. \frac{\partial^k K_a (t)}{\partial t^k} \right|_{t=0}.
    \label{cumulant_function}
\end{equation}
Replacing Eq.~\eqref{cumulant_gen_function} into Eq.~\eqref{marginalized_hamiltonian} yields
\begin{align*}
    \msfsl{H}(\boldsymbol{v}) 
    &= - \sum_{i,\mu} b_i^\mu \delta_\mu^{v_i} - \sum_{a,k} \frac{\kappa_a^{(k)} t^k}{k!}.
\end{align*}
 Considering the definition \eqref{aux_var}, we can rewrite the above expression as
\begin{equation}
     \msfsl{H}(\boldsymbol{v}) = - \sum_a \kappa_a^{(0)} - \sum_{i,a} \left( b_i^{\mu} + \sum_a \kappa_i^{(1)} W_{ia}^\mu \right) \delta_\mu^{v_i} - \sum_{k > 1} \frac{1}{k!} \sum_{i_1\dotsi_k} \sum_{\mu_1\dots\mu_k} \left( \sum_a \kappa_a^{(k)} W_{i_1 a}^{\mu_1} \cdots W_{i_k a}^{\mu_k}  \right) \delta_{\mu_1}^{v_{i_1}} \cdots \delta_{\mu_k}^{v_{i_k}}.
     \label{marginalized_hamiltonian_2}
 \end{equation}
 Here, we note that high-order terms in $k$ also contribute to the $n$-body coupling constants, where $n$ is the number of distinct sites. Here we note that high-order terms cancel out if they include factors of visible variables with different colors on the same site, i.e.,
 $$
 \delta_{\mu_1}^{v_i} \delta_{\mu_2}^{v_i} \cdots \delta_{\mu_k}^{v_i} =
 \begin{cases}
     \delta_{\mu}^{v_i} & \text{if } \mu = \mu_1 =\mu_2 = \cdots =\mu_k, \\
     0 & \text{otherwise.}
 \end{cases}
 $$ 
Then, for each $k$ in \eqref{marginalized_hamiltonian_2}, the sum over the visible sites $i_1, \dots, i_k$ can be grouped by the number $n$ of distinct sites being considered. Thus, the term of \eqref{marginalized_hamiltonian_2} that includes solely sites $i_1$ and $i_2$, considering the contributions of all orders in $k$, is given by
 \begin{gather*}
\sum_{k>1} \frac{1}{k!} \sum_{l=1}^{k-1} \sum_{1\le i_1 < i_2 \le \Nv} \sum_{\mu_1, \mu_2} \left( \sum_a \kappa_a^{(k)} \binom{k}{l} \big(W_{i_1 a}^{\mu_1} \big)^l \big( W_{i_2 a}^{\mu_2} \big)^{k-l} \right) \delta_{\mu_1}^{v_{i_1}} \delta_{\mu_2}^{v_{i_2}} \nonumber \\
= \sum_{k>1} \frac{1}{k!} \sum_{1 \le i_1 < i_2 \le \Nv} \sum_{\mu_1, \mu_2}\left( \sum_a \kappa_a^{(k)} \left[ \big(W_{i_1 a}^{\mu_1} + W_{i_2 a}^{\mu_2} \big)^k - \big( W_{i_1 a}^{\mu_1} \big)^k -\big( W_{i_2 a}^{\mu_2} \big)^k \right] \right)  \delta_{\mu_1}^{v_{i_1}} \delta_{\mu_2}^{v_{i_2}}.
\end{gather*}
The direct comparison of the above with Eq.~\eqref{general_Potts} implies that
\begin{align}
    J_{i_1 i_2}^{\mu_1 \mu_2} 
    &= \sum_{k>1} \frac{1}{k!} \sum_a \kappa_a^{(k)} \left[ \left(W_{i_1 a}^{\mu_1} + W_{i_2 a}^{\mu_2} \right)^k - \left( W_{i_1 a}^{\mu_1} \right)^k -\left( W_{i_2 a}^{\mu_2} \right)^k \right].
    \label{2-body_couplings_intermediate}
\end{align}
Replacing \eqref{cumulant_function} in \eqref{2-body_couplings_intermediate} gives
\begin{align}
    J_{i_1 i_2}^{\mu_1 \mu_2} 
    &= \sum_{k>1} \frac{1}{k!} \sum_a \left[ ( W_{i_1 a}^{\mu_1} + W_{i_2 a}^{\mu_2})^k - (W_{i_1 a}^{\mu_1})^k - (W_{i_2 a}^{\mu_2})^k \right] \left. \frac{\partial^k K_a(t)}{{\partial t}^k} \right|_{t=0} \nonumber \\
    &= \sum_a \sum_k \frac{1}{k!} \left[ ( W_{i_1 a}^{\mu_1} + W_{i_2 a}^{\mu_2} )^k - (W_{i_1 a}^{\mu_1})^k - (W_{i_2 a}^{\mu_2})^k \right] \left. \frac{\partial^k K_a(t)}{{\partial t}^k} \right|_{t=0}  +  \sum_a K_a (0) \nonumber \\
    & =  \sum_i \left. \left[ e^{(W_{i_1 a}^{\mu_1} + W_{i_2 a}^{\mu_2}) \partial_t} - e^{W_{i_1 a}^{\mu_1} \partial_t} - e^{W_{i_2 a}^{\mu_2} \partial_t} + 1 \right]  K_i(t) \right|_{t=0}.
   \nonumber
\end{align}
In the above expression, we identify the shift (or translation) operator
\begin{equation}
    e^{a\partial_x } f(x) \coloneqq f(a+x),
    \label{shift_opertor}
\end{equation}
which gives us
\begin{equation} 
    J_{i_1 i_2}^{\mu_1 \mu_2} = \sum_a \left[ K_a \left(W_{i_1 a}^{\mu_1} + W_{i_2 a}^{\mu_2} \right) - K_a \left( W_{i_1 a}^{\mu_1} \right) - K_a\left(W_{i_2 a}^{\mu_2}\right) + K_a(0)  \right].
    \label{2-body-couplings_intermediate_2}
\end{equation}
Using definition \eqref{cumulant_gen_function2} in \eqref{2-body-couplings_intermediate_2} leads automatically to
\begin{equation} \boxed{
    J_{ i_1 i_2}^{\mu_1 \mu_2}
   \! = \sum_a \ln \frac{\left( 1\! +\! e^{c_a \!+\! W_{i_1 a}^{\mu_1} \!+\! W_{i_2 a}^{\mu_2}} \right)\left(1\! +\! e^{c_a} \right)}{\left(1 + e^{c_a + W_{i_1 a}^{\mu_1}} \right) \left( 1 \!+ \!e^{c_a \!+ \!W_{i_2 a}^{\mu_2}} \right)}.
    \label{2-body_couplings_formula} }
\end{equation}
Similarly, the field contributions in \eqref{marginalized_hamiltonian_2} can be written as
\begin{equation}
    -\sum_{i, \mu} \left( b_i^\mu + \sum_a \kappa_a^{(1)} W_{i_1 a}^\mu \right) \delta_\mu^{v_i} - \sum_{k>1} \frac{1}{k!} \sum_{i, \mu} \left( \sum_a \kappa_a^{(k)} \left( W_{i_1 a}^\mu \right)^k \right) \delta_{\mu}^{v_i}.
    \label{fields_intermediate_1}
\end{equation}
Developing the left-side term in the above:
\begin{align}
    \sum_{k>1} \frac{1}{k!} \sum_a \kappa_a^{(k)} \left( W_{i_1 a}^\mu \right)^k 
    &= \sum_{k \ge 0} \frac{1}{k!} \sum_a \left( W_{i a}^\mu \right)^k \left. \frac{\partial^k K_a(t)}{{\partial t}^k} \right|_{t=0} - \sum_a K_a (0) - \sum_a \kappa_a^{(1)} W_{ia}^\mu \nonumber \\
    &= \sum_a \left. e^{W_{ia}^\mu \partial_t } K_a(t) \right|_{t=0} - \sum_a K_a (0) - \sum_a \kappa_a^{(1)} W_{ia}^\mu \nonumber \\
    &= \sum_a \left[ K_i \left( W_{ia}^\mu \right) - K_a (0) - \kappa_i^{(1)} W_{ia}^\mu \right] \nonumber \\
    &= \sum_a \left[ \ln \left( \frac{1+e^{c_i + W_{ia}^\mu }}{1 + e^{c_i}} \right) - \kappa_i^{(1)} W_{ia}^\mu \right].
    \label{fields_intermediate_2}
\end{align}
Finally, substituting \eqref{fields_intermediate_2} in \eqref{fields_intermediate_1} and comparing with Eq.~\eqref{general_Potts}, we obtain that the effective fields are given by
\begin{equation}\boxed{
    H_i^\mu = b_i^\mu + \sum_a \ln \left( \frac{1 + e^{c_a + W_{ia}^\mu }}{1+e^{c_a}} \right) \coloneqq b_i^\mu + J_i^\mu,}
    \label{fields_formula_lg}
\end{equation}
where we denoted by $J_i^\mu$ the contribution to the effective fields due to the interaction with the hidden variables. An analogous procedure can be applied by considering the contributions from all \(k\)-order terms to any \(n\)-th order interaction in Eq.~\eqref{marginalized_hamiltonian_2}, allowing us to derive the corresponding coupling constant. The general expression is
\begin{equation}\label{n-body_formula}
     \boxed{ J_{ i_1 \dots i_n}^{\mu_1 \dots \mu_n}
     \! = \!\!\! \sum_{K \subseteq [n] } \!\! (-1)^{n - |K|} \! \! \left[ \sum_a \ln \! \left( 1\! +\! e^{c_a\! +\!  \sum_{ k \in K } \!W_{i_k a}^{\mu_k}\! }\right)\! \right] \!.}
\end{equation}
where \( [n] \!=\! \{1, 2, \dots, n \} \) labels coupled sites, $K$ is a subset of \( n \) and $|{K}|$ denotes the number of elements of \( K \). Also, the coupled sites are labeled with sub-indexes \(k \!\in\! [n]\) when they appear in the given subset of $K$. Expression~\eqref{n-body_formula} maps the weights of an RBM to the physical couplings of a generalized Potts model in the lattice gas gauge. For the binary case (\( q=2 \)), it reduces to previously reported formulas in Refs.~\cite{cossu2019machine, beentjes2020higher, decelle2024inferring}. Finally, to verify that the couplings written in this formulation satisfy the lattice gas condition, we set the RBM in its corresponding lattice gas gauge (i.e., $\dot{b}_j^q= \dot{W}_{ij}^q = 0$ for all $i,j$) then it is easy to see that
\begin{gather*}
    \dot{H}_i^{q} = \dot{b_i^q} + \sum_a \ln \left( \frac{1+e^{c_a + {\dot{W}_{ia}^q}}}{1+e^{c_a}} \right) = 0,  \ \mathrm{and}  \\
    \dot{J}_{i_1 i_2}^{\mu_1 q} 
    = \sum_a \ln \frac{\left( 1 + e^{c_a + \dot{W}_{i_1 a}^{\mu_1} + \dot{W}_{i_2 a}^q} \right)\left(1 + e^{c_a} \right)}{\left(1 + e^{c_a + \dot{W}_{i_1 a}^{\mu_1}} \right) \left( 1 + e^{c_a + \dot{W}_{i_2 a}^q} \right)} = 0. 
\end{gather*}
Thus, in general, we have
\begin{equation*}
    \dot{J}_{i_1 \dots i_n}^{\mu_1 \dots q} = \sum_{K \subseteq [n-1] } \sum_a \left[ (-1)^{n - |K|} \ln \left( 1 + e^{c_a +  \sum_{ k \in K } \dot{W}_{i_k a}^{\mu_k} }\right)  - (-1)^{n - |K|} \ln \left( 1 + e^{c_a +  \sum_{ n \in K } \dot{W}_{i_k a}^{\mu_k} + \dot{W}_{i_n a}^q }\right) \right] = 0.
\end{equation*}

\section{Deriving the Effective Zero-Sum Model} 
\label{derivation_2_appendix}
\begin{figure}
    \centering
    \includegraphics[width=0.5\linewidth]{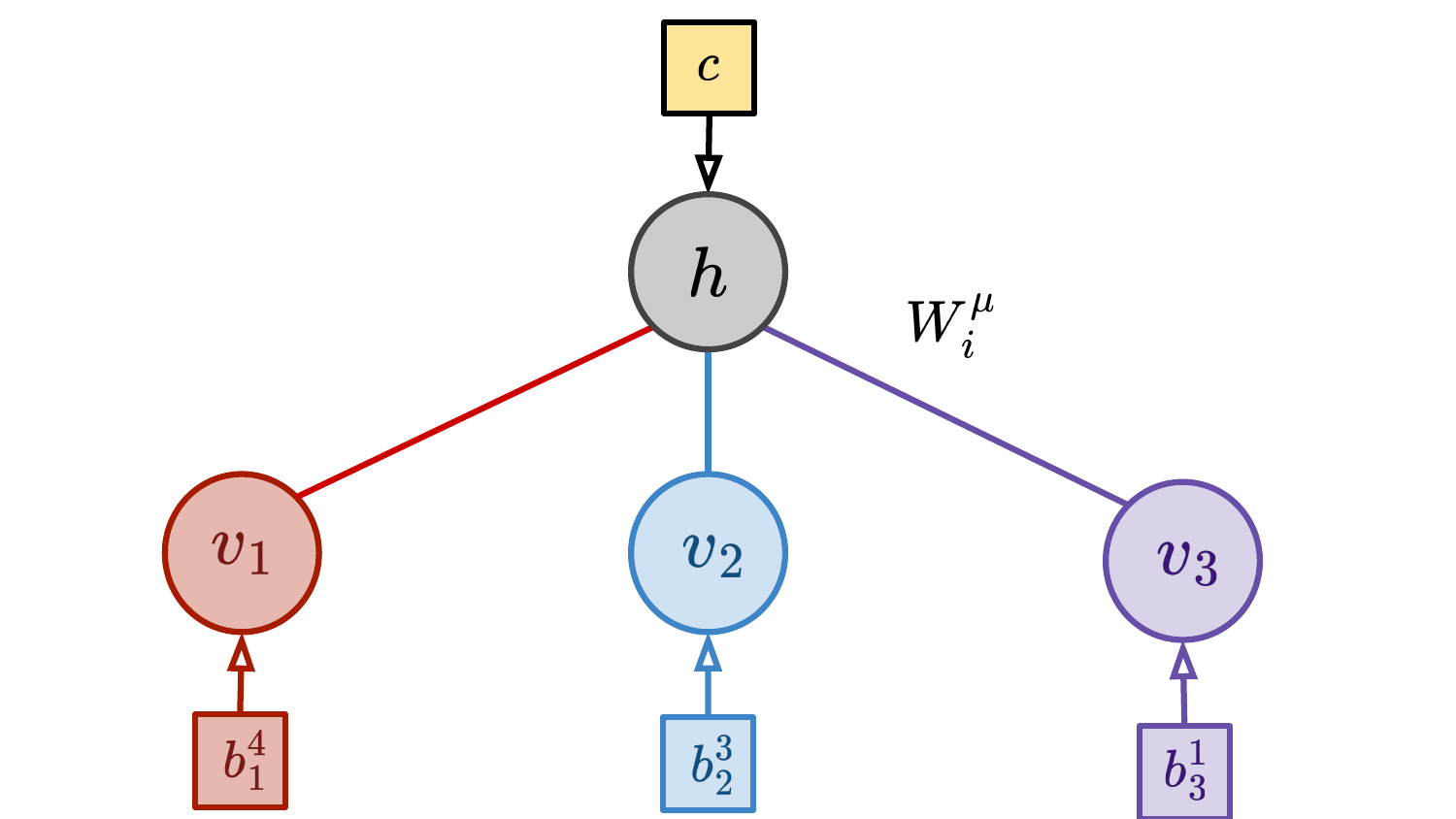}
    \caption{Diagram of the 3-visible hidden nodes RBM we use to derive the exact mapping in the zero-sum Gauge.}
    \label{fig:3-visible_RBM}
\end{figure}
Since directly addressing the gauge transformation of model (4) for a general RBM can be challenging, because the gauge transformation introduces modifications to interaction terms at all possible orders (see Appendix~\ref{gauge_invariance_appendix}), we begin with a a fully tractable system: a hidden node connected to only three visible variables (see Fig.~\ref{fig:3-visible_RBM}). In this case, the effective Potts Hamiltonian includes up to three-body interactions:
\begin{equation}
    \mathcal{H}(\boldsymbol{v}) = - \sum_{i, \mu} H_i^\mu  \delta_{\mu}^{v_{i_1}} - \sum_{i_1 < i_2 } \sum_{\mu_1, \mu_2} J_{i_1 i_2}^{\mu_1 \mu_2} \delta_{\mu_1}^{v_{i_1}} \delta_{\mu_2}^{v_{i_2}} - \sum_{i_1 < i_2 < i_3} \sum_{\mu_1, \mu_2, \mu_3} J_{i_1 i_2 i_3}^{\mu_1 \mu_2 \mu_3} \delta_{\mu_1}^{v_{i_1}} \delta_{\mu_2}^{v_{i_2}} \delta_{\mu_3}^{v_{i_3}}.
\end{equation}
Without loss of generality, we assume that the above Hamiltonian is written in the lattice gas gauge as derived in Appendix~\ref{derivation_lattice_gauge_appendix}. We obtain effective pairwise couplings and fields in the zero-sum gauge by applying the following gauge transformations:
\begin{align}
    \hat{J}_{ i_1 i_2}^{\mu_1 \mu_2} 
    &\coloneqq J_{ i_1 i_2 }^{\mu_1 \mu_2} - \frac{1}{q} \left( \sum_{\mu'_1} J_{i_1 i_2}^{\mu'_1 \mu_2} + \sum_{\mu'_2} J_{ i_1 i_2}^{\mu_1 \mu'_2} \right) + \frac{1}{q^2} \sum_{\mu'_1, \mu'_2} J_{ i_1 i_2}^{\mu'_1 \mu'_2} \nonumber \\  
    & \qquad + \frac{1}{q} \sum_{\mu'_3} J_{ i_1 i_2 i_3}^{ \mu_1 \mu_2 \mu'_3} - \frac{1}{q^2} \left( \sum_{\mu'_1, \mu'_3} J_{ i_1 i_2 i_3}^{ \mu'_1 \mu_2 \mu'_3}  + \sum_{\mu'_2, \mu'_3} J_{ i_1 i_2 i_3}^{ \mu_1 \mu'_2 \mu'_3} \right) + \frac{1}{q^3} \!\! \sum_{\mu'_1, \mu'_2, \mu'_3} \!\! J_{ i_1 i_2 i_3}^{ \mu'_1 \mu'_2 \mu'_3} 
    \label{couplings_gauge_tranformation_n3} \\
    \hat{H}_{i_1}^{\mu_1} 
    &\coloneqq H_{i_1}^{\mu_1} - \frac{1}{q} \sum_{\mu'_1} H_{i_1}^{ \mu'_1} + \sum_{i_2} \left( \frac{1}{q} \sum_{\mu'_2} J_{i_1 i_2}^{\mu_1 \mu'_2} - \frac{1}{q^2} \sum_{\mu'_1, \mu'_2} J_{i_1 i_2}^{\mu'_1 \mu'_2} \right) \nonumber \\ 
    & \qquad + \frac{1}{q^2} \sum_{\mu'_2, \mu'_3} J_{i_1 i_2 i_3}^{ \mu_1 \mu'_2 \mu'_3} - \frac{1}{q^3} \!\! \sum_{\mu'_1, \mu'_2, \mu'_3} \!\! J_{i_1 i_2 i_3}^{\mu'_1 \mu'_2 \mu'_3},
    \label{fields_gauge_transformation_n3}
\end{align}
where, we always kept $i_1 \!\neq\! i_2, i_3$ and $i_2 \!\neq\! i_3$. Using the shift operation, introduced in \eqref{shift_opertor}, we find the following recurrence relation in the general expression for couplings in \eqref{n-body_formula}:
\begin{align}
    J_{i_1 \dots i_n}^{\mu_1 \dots \mu_n} 
    &= \left( e^{W_{i_n}^{\mu_n} \partial_{c}} - 1 \right) \sum_{K \subseteq [N-1] } \!\!\!\!\!  (-1)^{n - 1 - |K|}  \ln \left( 1 + e^{c +  \sum_{ k \in K } W_{i_k}^{\mu_k} }\right) =  \left( e^{W_{i_n}^{\mu_n} \partial_{c}} - 1 \right) J_{i_1 \dots i_{n-1}}^{\mu_1 \dots \mu_{n-1}}.
    \label{recurrence_relation_lg}
\end{align}
Using \eqref{recurrence_relation_lg} to operate in \eqref{couplings_gauge_tranformation_n3} gives
\begin{align}
    \hat{J}_{i_1 i_2}^{\mu_1 \mu_2} 
    &=  J_{i_1 i_2}^{\mu_1 \mu_2} - \frac{1}{q} \left( \sum_{\mu'_1} J_{i_1 i_2}^{\mu'_1 \mu_2} + \sum_{\mu'_2} J_{i_1 i_2}^{\mu_1 \mu'_2} \right) + \frac{1}{q^2} \sum_{\mu'_1, \mu'_2} J_{i_1 i_2}^{\mu'_1 \mu'_2} \nonumber \\
    & + \frac{1}{q} \sum_{\mu'_3} \left( e^{W_{i_3}^{\mu'_3} \partial_c} - 1 \right) \left( J_{i_1 i_2}^{\mu_1 \mu_2} - \frac{1}{q} \left( \sum_{\mu'_1} J_{i_1 i_2}^{\mu'_1 \mu_2} + \sum_{\mu'_2} J_{i_1 i_2}^{\mu_1 \mu'_2} \right) + \frac{1}{q^2} \sum_{\mu'_1, \mu'_2} J_{i_1 i_2}^{\mu'_1 \mu'_2} \right) \nonumber \\
    &= \frac{1}{q} \sum_{\mu'_3} e^{W_{i_3}^{\mu'_3} \partial_c}  \left( J_{i_1 i_2}^{\mu_1 \mu_2} - \frac{1}{q} \left( \sum_{\mu'_1} J_{i_1 i_2}^{\mu'_1 \mu_2} + \sum_{\mu'_2} J_{i_1 i_2}^{\mu_1 \mu'_2} \right) + \frac{1}{q^2} \sum_{\mu'_1, \mu'_2} J_{i_1 i_2}^{\mu'_1 \mu'_2} \right)  \nonumber \\
    & = \frac{1}{q^3} \!\! \sum_{\mu'_1, \mu'_2, \mu'_3} \!\! e^{W_{i_3}^{\mu'_3} \partial_c} \left( J_{i_1 i_2}^{\mu_1 \mu_2} - J_{i_1 i_2}^{\mu'_1 \mu_2} - J_{i_1 i_2}^{\mu_1 \mu'_2} + J_{i_1 i_2}^{\mu'_1 \mu'_2} \right).
\end{align}
Then, we obtain the expression for the pairwise couplings in the zero-sum gauge by replacing \eqref{2-body_couplings_formula} in the above
\begin{equation}
    \hat{J}_{i_1 i_2}^{\mu_1 \mu_2} =  \frac{1}{q^3} \!\! \sum_{\mu'_1, \mu'_2, \mu'_3} \!\! \ln \frac{\left( 1 + e^{c + W_{i_1}^{\mu_1} + W_{i_2}^{\mu_2} + W_{i_3}^{\mu'_3} }  \right)\left(1 + e^{c + W_{i_1}^{\mu'_1} + W_{i_2}^{\mu'_2} + W_{i_3}^{\mu'_3}} \right)}{\left(1 + e^{c + W_{i_1}^{\mu_1} + W_{i_2}^{\mu'_2} + W_{i_3}^{\mu'_3} } \right) \left( 1 + e^{c + W_{i_1}^{\mu'_1}  W_{i_2}^{\mu_2} + W_{i_3}^{\mu'_3}}. \right)}.
\end{equation}
Similarly, for the fields, we can write
\begin{align}
    \hat{H}_{i_1}^{\mu_1} 
    &= b_{i_1}^{\mu_1} - \frac{1}{q} \sum_{\mu'_1} b_{i_1}^{ \mu'_1} + J_{i_1}^{\mu_1} - \frac{1}{q} \sum_{\mu'_1} J_{i_1}^{ \mu'_1} \nonumber \\ 
    & + \frac{1}{q} \sum_{ \mu'_3} \left( e^{W_{i_3}^{\mu'_3} \partial_c} - 1 \right) \left( J_{i_1}^{\mu_1} - \frac{1}{q} \sum_{\mu'_1} J_{i_1}^{\mu'_1} \right) + \frac{1}{q} \sum_{ \mu'_2} \left( e^{W_{i_2}^{\mu'_2} \partial_c} - 1 \right) \left( J_{i_1}^{\mu_1} -
    \frac{1}{q} \sum_{\mu'_1} J_{i_1}^{\mu'_1} \right)  \nonumber \\
    & + \frac{1}{q^2} \sum_{\mu'_2, \mu'_3} \left( e^{W_{i_2}^{\mu'_2}\partial_c } - 1 \right) \left( e^{W_{i_3}^{\mu'_3}\partial_c } - 1 \right) \left( J_{i_1}^{ \mu_1} - \frac{1}{q} \sum_{\mu'_1} J_{i_1}^{\mu'_1 } \right) \nonumber \\
    &= b_{i_1}^{\mu_1} - \frac{1}{q} \sum_{\mu'_1} b_{i_1}^{ \mu'_1} + \frac{1}{q^3} \!\! \sum_{\mu'_1, \mu_2', \mu'_3} \!\! e^{\left(W_{i_2}^{\mu'_1} + W_{i_3}^{\mu'_2} \right) \partial_c } \left( J_{i_1}^{\mu_1} - J_{i_1}^{\mu'_1}  \right).
\end{align}
Replacing the lattice gas coupling definition given in \eqref{fields_formula_lg} in the above gives
\begin{equation}
    \hat{H}_{i_1}^{\mu_1} =  \hat{b}_{i_1}^{\mu_1} + \frac{1}{q^3} \!\!\sum_{\mu'_1, \mu'_2, \mu'_3} \!\! \ln \left( \frac{1 + e^{c_a + W_{i_1}^{\mu_1} + W_{i_2}^{\mu'_2} + W_{i_3}^{\mu'_3} }}{1+ e^{c_a + W_{i_1}^{\mu'_1} + W_{i_2}^{\mu'_2} + W_{i_3}^{\mu'_3} } } \right), 
\end{equation}
where we denote by $\hat{b}_{i_1}^{\mu_1}$ the visible biases in the zero-sum gauge. Following this kind of construction, we can write the expression for effective pairwise and fields for an RBM with an arbitrary number of hidden and visible nodes:
\begin{gather}
    \boxed{\hat{J}_{i_1 i_2}^{\mu_1 \mu_2} =  \frac{1}{q^\Nv} \!\! \sum_{\mu'_1\dots\mu'_\Nv} \!\! \sum_a \ln \frac{\left( 1 + e^{c + W_{i_1 a}^{\mu_1} + W_{i_2 a}^{\mu_2} + \sum_{k=3}^{\Nv} W_{i_k a}^{\mu'_k} }  \right) \left(1 + e^{c + W_{i_1 a}^{\mu'_1} + W_{i_2 a}^{\mu'_2} + \sum_{k=3}^{\Nv} W_{i_k a}^{\mu'_k} } \right)}{\left(1 + e^{c + W_{i_1 a}^{\mu_1} + W_{i_2 a}^{\mu'_2} + \sum_{k=3}^{\Nv} W_{i_k a}^{\mu'_k} } \right) \left( 1 + e^{c + W_{i_1 a}^{\mu'_1} +  W_{i_2 a}^{\mu_2} + \sum_{k=3}^{\Nv} W_{i_k a}^{\mu'_k} } \right)},} 
    \label{2-body_couplings_zs}
    \\
    \boxed{\hat{H}_{i_1}^{\mu_1} =  \hat{b}_{i_1}^{\mu_1} + \frac{1}{q^\Nv} \!\!\sum_{\mu'_1\dots\mu'_\Nv}\!\! \sum_a \ln \left( \frac{1 + e^{c_a + W_{i_1 a}^{\mu_1} + W_{i_2 a}^{\mu'_2} + \sum_{k=3}^\Nv W_{i_k a}^{\mu'_k} }}{1+ e^{c_a + W_{i_1 a}^{\mu'_1} + W_{i_2 a}^{\mu'_2} + \sum_{k=3}^\Nv W_{i_k a}^{\mu'_k} } } \right).}
\end{gather}
Moreover, we can further generalize the above by considering an arbitrary \(n\)-th order coupling, leading to Eq.~\eqref{n-body_formula_zs} in the main text, which we repeat here for completeness:
\begin{gather}
    \boxed{\hat J_{i_1 \dots i_n}^{\mu_1 \dots \mu_n} \!\! 
    = \!\!\! \sum_{K \subseteq [n] } \!\! (-1)^{n - |K|} \Bigg[
    \frac{1}{q^{\Nv}} \!\!\! \sum_{\mu'_1\dots\mu'_{\Nv}} \!\!\!  \sum_a  
     \ln \! \left( 1\! + \!e^{c_a +  \sum_{ k \in K } W_{i_k a}^{\mu_k} + \sum_{ l \in [\Nv] \setminus K} W_{i_l a}^{\mu'_l} } \right) \Bigg].
     \nonumber}
\end{gather}

\section{Alternative Gauge fixing of the Multi-Body Potts Model} 
\label{gauge_appendix}

In Refs.~\cite{feinauer2022interpretable, feinauer2022mean}, Feinauer et al. demonstrated that the Hamiltonians of energy-based models can be expressed as Potts-like Hamiltonians with multi-body interactions in the zero-sum gauge. In \cite{feinauer2022interpretable}, they also derived a sampling estimator to compute effective pairwise couplings and fields in such a gauge. Building upon this foundation, we extend those findings to establish a general mapping between probabilistic models for categorical variables and generalized Potts-like models, as defined in Eq.~\eqref{general_Potts}, with couplings given by Eq.~\eqref{n-body_formula_general}. To validate this approach, we now prove that this Potts-like formulation preserves the probability mass distribution, which is equivalent to proving the following theorem:
\begin{theorem}
\label{theo_1}
\textbf{Gauge Invariance}. Let \(\pi\) be a probability mass function defined over the sample space \(\boldsymbol{\Omega} \!\coloneqq\! \{1, 2, \dots, q \}^N\), such that \(\pi(\boldsymbol{v}) \!>\! 0\) for all \(\boldsymbol{v} \!\in\! \Omega\). Then, the relation  
\(
\pi(\boldsymbol{v}) \!\propto\! e^{-\mathcal{H}(\boldsymbol{v})}
\)
holds, given the Hamiltonian definition:
\begin{align}
    \mathcal{H}(\boldsymbol{v}) 
    &= -\sum_{n=1}^N \! \sum_{i_1=1}^{N-n+1} \sum_{i_2 = i_1 +1}^{N-n+2}  \!\!\! \dots \!\!\! \sum_{i_n= i_{n-1}+1}^{N} \sum_{\mu_1 = 1}^q \!\! \dots \!\! \sum_{\mu_n = 1}^q \! J_{i_1 \dots i_n}^{\mu_1 \dots \mu_n} \delta_{\mu_1}^{v_{i_1}} \! \cdots \delta_{\mu_n}^{v_{i_n}} \nonumber \\
    & = -\sum_{n=1}^N \! \sum_{i_1=1}^{N-n+1} \sum_{i_2 = i_1 +1}^{N-n+2}  \!\!\! \dots \!\!\! \sum_{i_n= i_{n-1}+1}^{N}  \sum_{K \subseteq [n]} (-\!1)^{n - |K|} \mathbb{E}_{\boldsymbol{u} \sim G }  \ln \pi \big( \boldsymbol{u} \big| u_{i_{k}} \! \!=\! v_{i_k} \!:\! k \!\in\! K \big).
    \label{general_Potts_2}
\end{align}
for any probability measure $G$ defined over $\boldsymbol{\Omega}$.
\end{theorem}
Note that in the above, we assumed $J_i^\mu \equiv H_i^\mu$; we will maintain this notation throughout the rest of this section.
\begin{proof}
\!Here, we will use an \textit{induction} argument in $N$. Let us choose $N\!=\!2$ as our \textit{base case} for didactical reasons. When $N\!=2$, it is then straightforward to obtain that
\begin{align}
    \mathcal{H}(\boldsymbol{v}) 
    &= -J_{1}^{v_1} - J_{2}^{v_2} - J_{1,2}^{v_1 v_2} \nonumber \\
    &= - \left[ \mathbb{E}_{\boldsymbol{\mu} \sim G}  \ln \pi \big(\boldsymbol{u} \big| u_1 \!=\! v_1 \big) - \mathbb{E}_{\boldsymbol{u} \sim G} \ln \pi (\boldsymbol{u}) \right] - \left[ \mathbb{E}_{\boldsymbol{u} \sim G} \ln \pi \big(\boldsymbol{u} \big| u_2 \!=\! v_2 \big) - \mathbb{E}_{\boldsymbol{u} \sim G} \ln \pi (\boldsymbol{u}) \right] \nonumber \\
    & \qquad - \left[ \mathbb{E}_{\boldsymbol{u} \sim G} \ln \pi \big( \boldsymbol{u} \big| u_1 \!=\! v_1, u_2 \!=\! v_2  \big) - \mathbb{E}_{\boldsymbol{u} \sim G} \ln \pi \big( \boldsymbol{u} \big| u_1 \!=\! v_1  \big) - \mathbb{E}_{\boldsymbol{u} \sim G} \ln \pi \big( \boldsymbol{u} \big| u_2 \!=\! v_2 \big) + \mathbb{E}_{\boldsymbol{u} \sim G} \ln \pi (\boldsymbol{u}) \right] \nonumber \\
    &= -\ln \pi (\boldsymbol{v}) + \mathbb{E}_{\boldsymbol{u} \sim G} \ln \pi (\boldsymbol{u}).
    \label{dem_1}
\end{align}
In the above we recognized that $\mathbb{E}_{\boldsymbol{u} \sim G}  \ln \pi \big(\boldsymbol{u} \big| u_1 \!=\! v_1, u_2 \!=\! v_2  \big) \!=\! \ln \pi (\boldsymbol{v})$, with $\boldsymbol{v} \!=\! (v_1, v_2)$. Since $\mathbb{E}_{\boldsymbol{u} \sim G} \ln \pi (\boldsymbol{u}) $ is independent of $\boldsymbol{u}$ (i.e., it is a constant), from the exponentiation of both sides of \eqref{dem_1} it follows that $\pi(\boldsymbol{v}) \!\propto\! e^{-\msfsl{H}(\boldsymbol{v})}$ holds for $N=2$. 

Next, assuming the following identity holds 
\begin{align}
    \mathcal{H}(\boldsymbol{v}^\ast) 
    &=\! -\sum_{n=1}^{N-1} \sum_{i_1=1}^{N-n} \sum_{i_2 = i_1 +1}^{N-n+1}  \!\!\! \dots \!\!\! \sum_{i_n= i_{n-1}+1}^{N-1}  \sum_{K \subseteq [n]} (-\!1)^{n - |K|} \mathbb{E}_{\boldsymbol{u}^\ast \sim G^\ast } \ln  \pi^\ast \big( \boldsymbol{u}^\ast \big| u_{i_{k}}^\ast \! \!=\! v_{i_k}^\ast \!:\! k \!\in\! K \big) \nonumber \\
    &=\! - \ln \pi^\ast (\boldsymbol{v}^\ast) + \mathbb{E}_{\boldsymbol{u}^\ast \sim G^\ast} \ln \pi^\ast (\boldsymbol{u}^\ast),
    \label{IH}
\end{align}
for any probability mass $\pi^\ast$ defined over a dimension $N-1$ sample space $\boldsymbol{\Omega}^\ast$, we will prove that it also holds when the dimension of the sample space is $N$. First, Let us separate couplings that include the site $i_1 \! = \! 1$ from those that do not in \eqref{general_Potts_2}
\begin{align}
    \mathcal{H}(\boldsymbol{v}) 
    &= -\sum_{n=1}^{N} \sum_{i_1 = 1}^1 \sum_{i_2=2}^{N-n+2} \!\!\! \dots \!\!\! \sum_{i_n=i_{n-1}+1}^{N}  \sum_{K \subseteq [n] } (-\!1)^{n - |K|} \mathbb{E}_{\boldsymbol{u} \sim G} \ln \pi \big( \boldsymbol{u} \big| u_{i_{k}} \! \!=\! v_{i_k} \!:\! k \!\in\! K \big)  \nonumber \\
    & \qquad - \sum_{n=1}^{N-1} \sum_{i_1 = 2}^{N-n+1} \sum_{i_2=i_1 + 1}^{N-n+2} \!\! \dots \!\! \sum_{i_n=i_{n-1}+1}^{N} \sum_{K \subseteq [n] } (-\!1)^{n - |K|} \mathbb{E}_{\boldsymbol{u} \sim G} \ln \pi \big( \boldsymbol{u} \big| u_{i_{k}} \! \!=\! v_{i_k} \!:\! k \!\in\! K \big).
    \label{dem_3}
\end{align}
Whereas, from the coupling definition given in Eq.~\eqref{n-body_formula_general}, we found the following relation
\begin{align}
    J_{i_1 \dots i_n}^{\mu_1 \dots \mu_n} \delta_{\mu_1}^{v_{i_1}} \cdots \delta_{\mu_n}^{v_{i_n}} 
    &= \sum_{K \subseteq [n] } (-\!1)^{n - |K|} \mathbb{E}_{\boldsymbol{u} \sim G} \ln \pi \big( \boldsymbol{u} \big| u_{i_{k}} \! \!=\! v_{i_k} \!:\! k \!\in\! K \big) \nonumber \\
    &= \!\!\!\! \sum_{K \subseteq [n] \setminus \{1\}} \!\!\!\!\! (-\!1)^{n - |K|} \Big[ \mathbb{E}_{\boldsymbol{u} \sim G}  \ln \pi \big( \boldsymbol{u} \big|u_{i_{k}} \!\!=\! v_{i_k} \!:\! k \!\in\! K \big) - \mathbb{E}_{\boldsymbol{u} \sim G}  \ln \pi \big( \boldsymbol{u} \big| u_{i_{k}} \!\!=\! v_{i_k} \!:\! k \!\in\! K \!\cup\! \{ 1 \} \big) \Big].
    \label{dem_4}
\end{align}
Replacing \eqref{dem_4} in the first line r.h.s. of \eqref{dem_3} line gives
\begin{gather}
     - \sum_{n=1}^{N} \sum_{i_1 = 1}^1 \sum_{i_2=2}^{N-n+2} \!\! \dots \!\!\!\!\sum_{i_n=i_{n-1}+1}^{N} \sum_{K \subseteq [n] \setminus \{1\}} \!\!\!\!\! (-\!1)^{n - |K|} \Big[ \mathbb{E}_{\boldsymbol{u} \sim G} \ln \pi \big( \boldsymbol{u} \big| u_{i_{k}} \!\!=\! v_{i_k} \!:\! k \!\in\! K  \big) - \mathbb{E}_{\boldsymbol{u} \sim G} \ln \pi \big( \boldsymbol{u} \big| u_{i_{k}} \!\!=\! v_{i_k} \!:\! k \!\in\! K \!\cup\! \{ 1 \} \big) \Big] \nonumber \\
    =- \sum_{n=0}^{N-1} \sum_{i_1 = 2}^{N-n+1} \!\!\!\! \dots \!\!\!\!\!\! \sum_{i_n=i_{n-1}+1}^{N} \sum_{K \subseteq [n] } \!\! (-\!1)^{n +1 - |K|} \Big[ \mathbb{E}_{\boldsymbol{u} \sim G} \ln \pi \big( \boldsymbol{u} \big| u_{i_{k}} \!\!\!=\! v_{i_k} \!:\! k \!\in\! K \big) \!-\! \mathbb{E}_{\boldsymbol{u}\setminus\{u_1\} \sim G \setminus \{u_1\} } \ln \pi \big( v_1, \boldsymbol{u}\!\setminus\! \{u_1\} \big| u_{i_{k}} \!\!=\! v_{i_k} \!:\! k \!\in\! K \big)  \Big],
    \label{dem_5}
\end{gather}
with $\boldsymbol{u}\!\setminus\!{\{u_1\} } \!\coloneqq\! (u_2, \dots, u_N)$, hence $ \pi ( v_1, \boldsymbol{u}\!\setminus\!{\{u_1\} }) \!=\! \pi (v_1, u_2, \dots, u_N)$. Additionally, in \eqref{dem_5} we denoted by $G\setminus \!\{u_1\}\!$ the corresponding marginal probability measure of $G$ for $\boldsymbol{u} \!\setminus\! \{u_1\} \in \boldsymbol{\Omega}\!\setminus\!\{u_1 \} $. Substituting \eqref{dem_5} in the first line of \eqref{dem_3} gives
\begin{align}
    \mathcal{H}(\boldsymbol{v}) 
    &= - \sum_{n=1}^{N-1} \sum_{i_1 = 2}^{N-n+1} \!\!\!\! \dots \!\!\!\!\!\! \sum_{i_n=i_{n-1}+1}^{N} \sum_{K \subseteq [n] } \!\! (-\!1)^{n - |K|} \mathbb{E}_{\boldsymbol{u} \setminus \{u_1\} \sim G\setminus \{u_1\}} \ln \pi \big( v_1, \boldsymbol{u} \!\setminus\! \{ u_1 \} \big| u_{i_{k}} \!\!=\! v_{i_k} \!:\! k \!\in\! K \big) \nonumber \\
    & \qquad + \mathbb{E}_{\boldsymbol{u} \sim G} \left[ \ln \pi (\boldsymbol{u}) \right] - \mathbb{E}_{\boldsymbol{u}\setminus\{u_1\} \sim G\setminus\{u_1\}} \left[ \ln \pi (v_1, \boldsymbol{u} \!\setminus\! \{u_1\}) \right].
    \label{dem_6}
\end{align}
Using the induction hypothesis \eqref{IH} in the first line of r.h.s. of \eqref{dem_6} readily gives
\begin{align*}
    \mathcal{H}(\boldsymbol{v}) 
    &= - \ln \pi ( v_1, \boldsymbol{v} \!\setminus\! \{ v_i \})  + \mathbb{E}_{\boldsymbol{u} \sim G}  \ln \pi (\boldsymbol{u}).
\end{align*}
\end{proof}

We note that the gauge freedom above manifests as the arbitrariness in the choice of $G$ when computing the couplings. Further, we can relate Eq.~\eqref{n-body_formula_general} with the more familiar gauge fixing conditions whenever the probability measure \( G \) is factorizable through the following theorem:
\begin{theorem}
    \textbf{Gauge Fixing.} Considering the effective couplings \( J_{i_1 \dots i_n}^{\mu_1 \dots \mu_n} \) defined in Eq.~\eqref{n-body_formula_general}, let \( g \) be the probability mass function of \( G \) such that \( g(\boldsymbol{v}) = \prod_{i=1}^{N} g_{i}(v_i) \). Then, it follows that  
\begin{equation}
     \qquad \sum_{\mu'} g_{i_n}(\mu') J_{i_1 \dots i_n}^{\mu_1 \dots \mu'} = 0,
    \label{general_condition}
\end{equation}
for all \( i_1, \dots, i_n \) and \( \mu_1, \dots, \mu_n \).
\label{theo_2}
\end{theorem}
\begin{proof}
    By simply replacing Eq.~\eqref{n-body_formula_general} in \eqref{general_condition} we can obtain
    \begin{align*}
        \sum_{\mu'} g_{i_n}(\mu') J_{i_1 \dots i_n}^{\mu_1 \dots \mu'} 
        & = \sum_{\mu'} g_{i_n} (\mu') \sum_{K \subseteq [n]} \!\! (-1)^{n - |K|} \mathbb{E}_{\boldsymbol{u} \sim G } \ln \pi \big( \boldsymbol{u} \big| u_{i_{k}} = \mu_{k} \!:\! k \!\in\! K \big) \ (\mathrm{with} \ \mu_n = \mu' ) \\
        &= \sum_{\mu'} g_{i_n}(\mu') \!\! \sum_{K \subseteq [n-1]} \!\!\!\! (-1)^{n-|K|} \Big [  \mathbb{E}_{\boldsymbol{u} \sim G } \ln \pi \big( u_1, \dots, u_{i_n}, \dots, u_N \big| u_{i_{k}} = \mu_{k} \!:\! k \!\in\! K \big) \\
        & \qquad - \mathbb{E}_{\boldsymbol{u} \sim G } \ln \pi \big( u_1, \dots, \mu',  \dots, u_N \big| u_{i_{k}} = \mu_{k} \!:\! k \!\in\! K \big)  \Big] \\
        &= \sum_{K \subseteq [n-1]} \!\!\!\! (-1)^{n-|K|} \Big [  \sum_{u_1} g_1 (u_1) \dots \sum_{u_{i_n}} g_{i_n} (u_{i_n}) \dots \sum_{u_N} g_{N} (u_N) \ln \pi \big( u_1, \dots, u_{i_n}, \dots, u_N  \big| u_{i_{k}} = \mu_{k} \!:\! k \!\in\! K \big) \\
        & \qquad - \sum_{\mu'} g_{i_n}(\mu') \sum_{u_1} g_1 (u_1) \dots \sum_{u_N} g_N (u_N) \ln \pi \big( u_1, \dots, \mu',  \dots, u_N \big| u_{i_{k}} = \mu_{k} \!:\! k \!\in\! K \big)  \Big] = 0.
    \end{align*}
\end{proof}
According to the above theorem, using a probability mass function in the coupling definition in Eq.~\eqref{n-body_formula_general} such that $g_i(u_i) \!=\! \delta_{q}^{u_i}$ or $g_i(u_i) \!=\! q^{-1}$ leads automatically to fix the lattice gas or zero-sum gauges, respectively. 

\section{More Details on the Large-$N$ Approximations}\label{CLT_appendix}
\subsection{Numerical Approximation of the State Average}
We begin with Eq.~\eqref{n-body_formula_zs}, repeated here for clarity:
\begin{gather}
    {\hat J_{i_1 \dots i_n}^{\mu_1 \dots \mu_n} \!\! 
    = \!\!\! \sum_{K \subseteq [n] } \!\! (-1)^{n - |K|} \Bigg[
    \frac{1}{q^{\Nv}} \!\!\! \sum_{\mu'_1\dots\mu'_{\Nv}} \!\!\!  \sum_a  
     \ln \! \left( 1\! + \!e^{c_a +  \sum_{ k \in K } \hat{W}_{i_k a}^{\mu_k} + \sum_{ l \in [\Nv] \setminus K} \hat{W}_{i_l a}^{\mu'_l} } \right) \Bigg].
     \nonumber}
\end{gather}
Our goal is to approximate the term in square brackets using the CLT, which requires $n \ll \Nv$. The first step reduces the full average over $q^{\Nv}$ configurations to a sum over $q^n$ terms:
\begin{align}
    &\frac{1}{q^{\Nv}} \!\!\sum_{\mu'_1\dots\mu'_{\Nv}} \!\! \ln \left(  1 + e^{c_a +  \sum_{k=1}^{\Nv} \hat{W}_{i_k a}^{\mu'_k} }\right) = \frac{1}{q^{n}} \!\!\! \sum_{\mu'_1\dots\mu'_n} \!\!\! \mathbb{E}_{x \sim X_a^{\setminus \{i_1, \dots, i_n \}} } \! \left[ \ln \left(  1 + e^{c_a +  \sum_{k=1}^{n} \hat{W}_{i_k a}^{\mu'_k} + x} \right) \right],\!\!
    \label{aproximation_couplings_ap}
\end{align}
where $X_a^{ \setminus \{ i_1, \dots, i_n \}} \! \coloneqq \! \sum_{l = n+1}^{\Nv} \! \hat{W}_{i_l a}^\ast$ is a random variable, with each $\hat{W}_{i_l a}^\ast$ drawn uniformly from $\{ \hat{W}_{i_l a}^\mu : \mu \in [q] \}$. In the zero-sum gauge, since $\sum_{\mu=1}^q \hat{W}_{i a}^\mu = 0$ $\forall i$, then $$\mathbb{E}[x] \!=\! 0 \ \ \mathrm{and}  \ \ \mathrm{Var} [x] \!=\! q^{-1} \sum_{l=n+1}^{\Nv} \sum_{\mu=1}^q  \left( \hat{W}_{i_l a}^{\mu} \right)^2. $$ Therefore, we approximate the expectation using a Gaussian integral, first normalizing $X_a^{\setminus \{i_1, \dots, i_n \}}$ to unit variance:
\begin{equation}
     \frac{ X_a^{\setminus\{i_1, \dots, i_n\}}}{  \sqrt{ q^{-1} \sum_{l=n+1}^\Nv \sum_{\mu=1}^q  \left( \hat{W}_{i_l a}^{\mu} \right)^2}}  \stackrel{\text{approx}}{\sim} \mathcal{N}(0,1).
     \label{CLT_assumption}
\end{equation}
Then, we can approximate the r.h.s. of Eq.~\eqref{aproximation_couplings_ap} as
\begin{align}
     & \frac{1}{q^{n}} \!\!\! \sum_{\mu'_1\dots\mu'_n} \!\!\! \mathbb{E}_{z \sim \mathcal{N}(0,1) }
     \! \left[ \ln \left(  1 + e^{c_a +  \sum_{k=1}^{n} \hat{W}_{i_k a}^{\mu'_k} + \sqrt{q^{-1} \sum_{l=n+1}^{\Nv} \sum_{\mu=1}^q \left( \hat{W}_{i_l a}^{\mu} \right)^2} z  } \right) \right]\nonumber\\
     &= \frac{1}{q^{n}} \!\!\! \sum_{\mu'_1\dots\mu'_n}\int_{-\infty}^{\infty} Dz  \ \ln \left( 1 + e^{ c_a + \sum_{k=1}^n \! \hat{W}_{i_k a}^{\mu'_k} + \sqrt{q^{-1} \sum_{l=n+1}^{\Nv} \sum_{\mu=1}^q \left( \hat{W}_{i_l a}^{\mu} \right)^2} z } \right) \!,\label{eq:CLT}
\end{align}
where $Dz = e^{-z^2/2} / \sqrt{2\pi}$. This integral can be evaluated numerically; in our tests, discretizing it with approximately 20 points yields accurate results. 

In our algorithm, the efficient computation of the $n$-th-order couplings is made possible through parallelization. To achieve this, we compute in parallel an $n$-rank tensor $\boldsymbol{T}_{i_1 \dots i_n}^{(n)}$, whose elements are defined as
\begin{equation}
    T_{i_1 \dots i_n }^{\mu'_1 \dots \mu'_n} = \sum_a \int_{-\infty}^{\infty} Dz \ \ln \left( 1 + e^{ c_a + \sum_{k=1}^n \! \hat{W}_{i_k a}^{\mu'_k} + \sqrt{q^{-1} \sum_{l=n+1}^{\Nv} \sum_{\mu=1}^q \left( \hat{W}_{i_l a}^{\mu} \right)^2} z } \right),
    \label{tensor_definition}
\end{equation}
where along the $k$-th dimension, the index $\mu'_{i_k}$ takes on all $q$ possible values. Note that the scale factor \eqref{tensor_definition} is the same for all elements of such a tensor. Thus, the rest of Eq.~\eqref{n-body_formula_zs} is approximated using the following relationship
\begin{equation*}
    J_{i_1 \dots i_n}^{\mu_1 \dots \mu_n} \approx \sum_{L \subseteq [n]}  \left[ \left(-{q} \right)^{-|L|} {\textstyle \sum_{\substack{\{\mu_{l_k} \colon l_k \in L} \}} T_{i_1 \dots i_n}^{\mu_1 \dots \mu_n }}  \right].
\end{equation*}
This procedure yields the $n$-rank tensor that contains all $n$-th order couplings among the visible sites $i_1, \dots, i_n$. The performance of this algorithm is summarized in Table~\ref{tab:computational_times}, and implementation details are available in the code at ~\href{https://github.com/DsysDML/couplings_inference}{DsysDML/couplings\_inference}~\cite{github:couplings_inference}.

\subsection{On the Validity of the Central Limit Theorem and Refinements}
\begin{figure}
    \centering
    \includegraphics[width=0.9\linewidth]{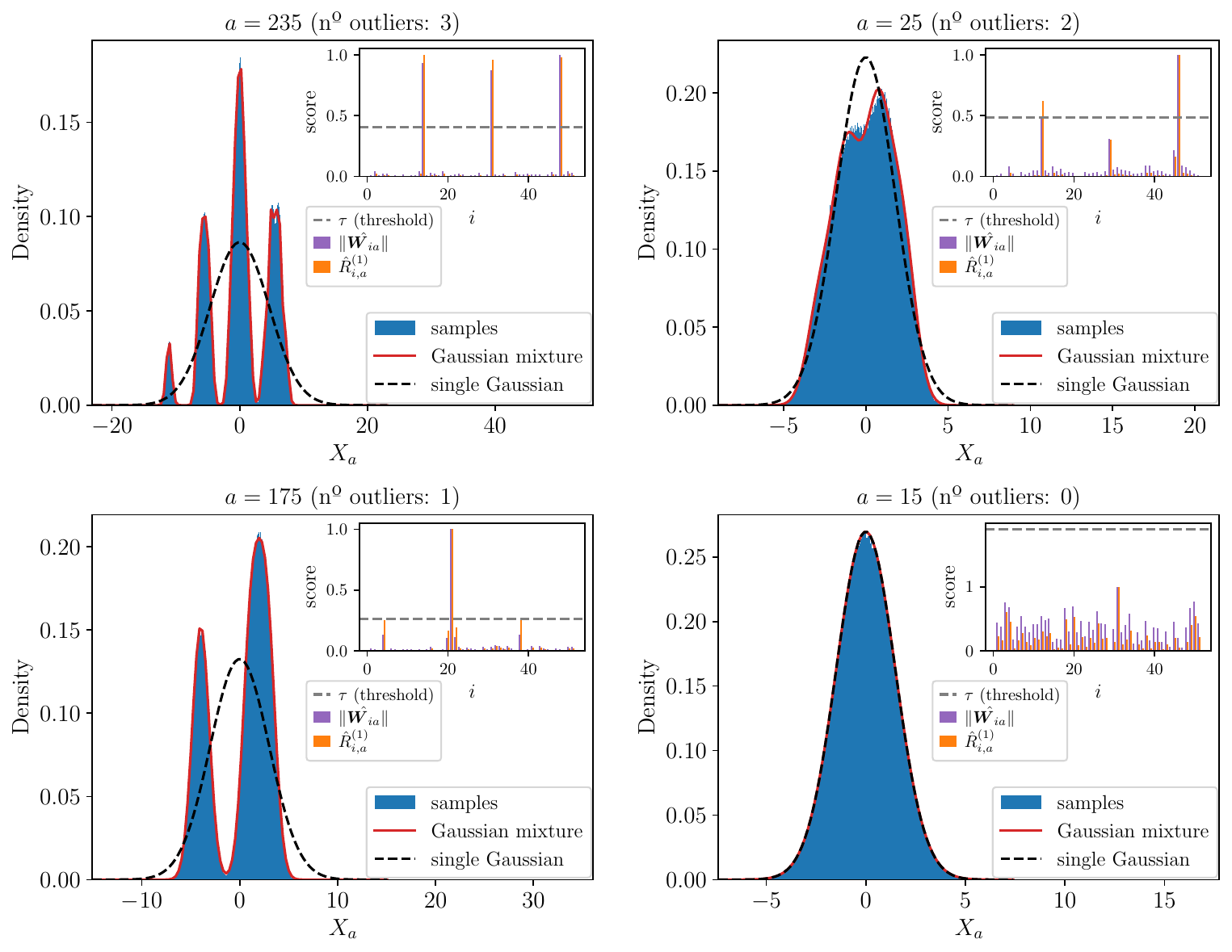}
    \caption{ Sample distribution of the sum of the weights randomly drawn from the {\bf (a)} 235-th, {\bf (b)} 22-th, {\bf (c)} 175-th, and {\bf (d)} 15-th columns of the weight matrix of the RBM trained with the modified Blume-Capel model samples (at $t=5 \cdot 10^6$). The dashed black line is $X_a$ under the single Gaussian approximation given in \eqref{eq:CLT}, while the solid red line is the approximation using a Gaussian mixture \eqref{CLT_assumption_2} with $\tau=2.5$. The inset shows a barplot with the $\| \boldsymbol{\hat{W}}_{ia} \|$ and the $\hat{R}_{i,a}^{(1)}$ computed for one hidden $a$ and visible $i$ node connection. These quantities are rescaled to 0.0 and 1.0. }
    \label{fig:gaussian_validation}
\end{figure}

\begin{figure}
    \centering
    \includegraphics[width=0.9\linewidth]{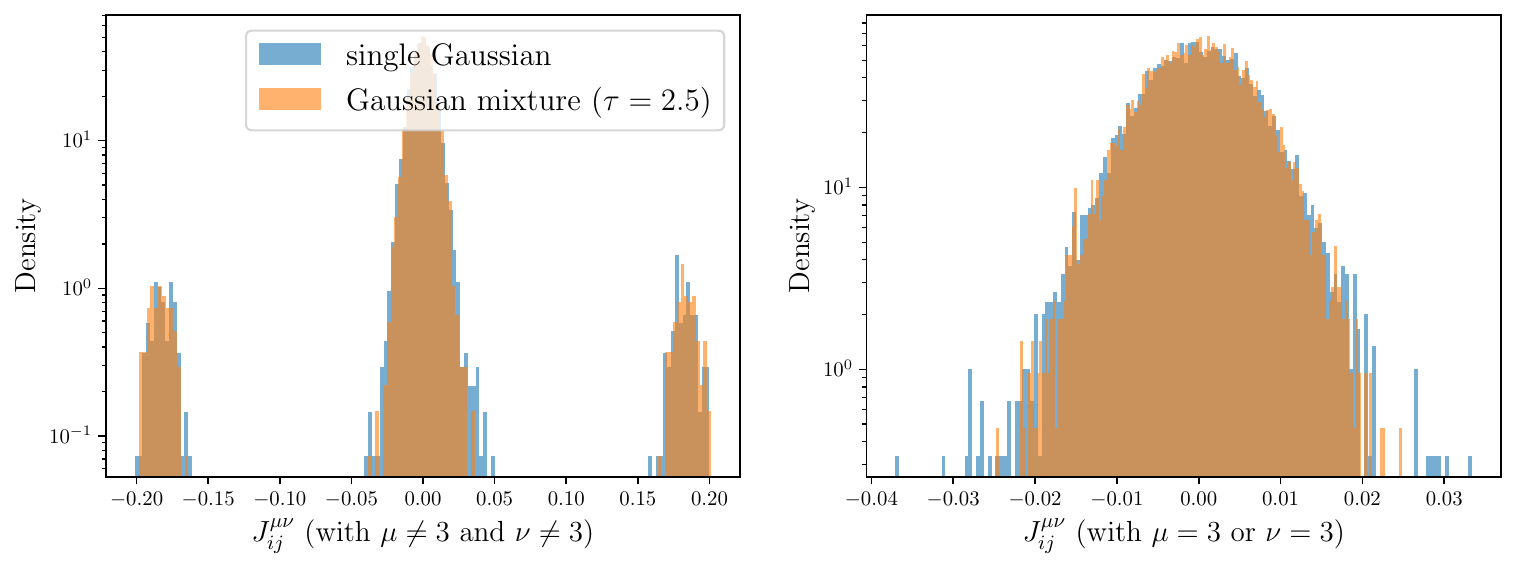}
    \caption{ 2-Body couplings extracted with a simple Gaussian and a Gaussian mixture approximation from the RBM trained with Blume-Capel model samples (at $t=5 \cdot 10^6$). }
    \label{fig:gaussian_validation_2}
\end{figure}

The approximation scheme described above relies on the assumption that the CLT applies to the random variable $X_a^{\setminus \{i_1, \dots, i_n \}} \! $. The validity of the CLT in this setting, discussed for instance in Appendix B of Ref.~\cite{coolen2005theory}, is based on Lindeberg’s theorem and cannot be guaranteed for arbitrary RBMs, as the structure of $\bm{W}$ depends on both the training method and the dataset. In typical RBM trainings, we found that deviations from Gaussianity can arise from either sparsity, which poses no difficulty, as the corresponding sums can be computed exhaustively, or from the presence of disproportionally large weight values, i.e., outliers, associated with certain hidden nodes. In these cases, the distribution of $X_a$ is no longer well approximated by a single Gaussian, but rather by a more general Gaussian mixture. This effect is illustrated in Fig.~\ref{fig:gaussian_validation} for the Blume-Capel test case.
 
To provide a construction of such a Gaussian mixture, let us first introduce a precise definition for the outliers in the model's weight matrix. Considering a hidden node $a$, one can compute $\| \boldsymbol{\hat{W}}_{ia} \| = \sqrt{ \sum_{\mu} \left( \hat{W}_{ia} ^\mu \right)^2 } $ for every $i \in [\Nv]$. Then, by defining a threshold $\tau$, we considered the set of outliers $I_a(\tau)$ after applying a $\log$ transformation, i.e.,  $ \forall i \in I_a(\tau)$, we have
\[
\frac{ \| \boldsymbol{\hat{W}}_{ia} \| - {\Nv}^{-1}\sum_{j =1}^{\Nv} \log  \| \boldsymbol{\hat{W}}_{ia} \| }{\sqrt{ {\Nv}^{-1} \sum_{k=1}^{\Nv} \bigg( {\log \| \boldsymbol{\hat{W}}_{ia} \| - {\Nv}^{-1}\sum_{j=1}^{\Nv} \log  \| \boldsymbol{\hat{W}}_{ia} \| } \bigg)^2 }} > \tau.
\]
Thus, a more general hypothesis for the distribution of $X_a$ than the simple Gaussian is that 
\begin{equation}
    { X_a }  \stackrel{\text{approx}}{\sim} \frac{1}{q^{|I_a(\tau)|}} \sum_{\mu_1\dots\mu_{|I_a(\tau)|}} \mathcal{N} \bigg( \textstyle \sum_{i_k \in I_a(\tau)} \hat{W}_{i_k a}^{\mu_k}, \ \sqrt{{q}^{-1} \sum_{i \in [\Nv] \setminus I_a(\tau)} \sum_{\mu'=1}^q\left( \hat{W}_{i a}^{\mu'} \right)^2} \ \bigg),
    \label{CLT_assumption_2}
\end{equation}
for a finite $\tau$ that keeps $\Nv \! \gg \! |I_a(\tau)| $. It is worth adding that the single Gaussian and the original distribution of $X_a$ are recovered in the limits as $\tau \to \infty$ and  $\tau \to 0$, respectively. Fig. \ref{fig:gaussian_validation} depicts that a reliable approximation for $X_a$ at $\tau = 2.5$ is obtained in the RBM training with modified Blume-Capel model samples. The average of a function $f(x)$ where $x \sim X_a$ using the $\eqref{CLT_assumption_2}$ approximation can be written as:
\begin{align*}
    \mathbb{E}_{x \sim X_a} [f(x)]
    = \frac{1}{q^{|I_a(\tau)|}} \sum_{\mu_1 \dots \mu_{|I_a(\tau)|}} \int_{-\infty}^{\infty} Dz \ f \bigg(  \textstyle \sum_{i_k \in I_a(\tau)} \hat{W}_{i_k a}^{\mu_k} + \sqrt{{q}^{-1} \sum_{i \in [\Nv] \setminus I_a(\tau)} \sum_{\mu'=1}^q\left( \hat{W}_{i a}^{\mu'} \right)^2} \ z \bigg).
\end{align*}
Then, one can consider that averaging a function over $X_a$ under the approximation in \eqref{CLT_assumption_2} is equivalent to explicitly summing over states $\mu_k$ of the sites $i_k \!\in\! I_a(\tau)$ and then averaging over $X_a^{\setminus\{i \in I_a(\tau) \} }$ using the single Gaussian approximation as it was introduced in the previous section. We used this fact to implement a refinement to our previous approach for pairwise couplings $J_{ij}^{\mu \nu}$, in which, for each hidden node $a$, we first identified $I_a(\tau)$ and then calculated its contribution to the couplings by evaluating the sum explicitly not only on states $\mu_k$ of interacting sites $i_k \in \{i_1, \dots, i_n \}$, but also on every $i_k \!\in\! I_a(\tau)$. 

Fig.~\eqref{fig:gaussian_validation_2} compares the histogram of reconstructed couplings using the simple and the Gaussian mixture approximation. While the improvement of the latter is modest, an increase in the precision of the estimates is clear. It is also reflected in a reduction of the root-mean-square reconstruction error from $8.8 \cdot 10^{-3}$ (under the single Gaussian approximation) to $8.3 \cdot 10^{-3}$ (under the mixed Gaussian approximation). The same procedure was applied to the RBM trained on the PF00072 dataset, yielding no changes in contact prediction performance; that is, the ROC and PPV curves remained identical. We attribute this negligible effect to the fact that hidden nodes with substantial deviations from the Gaussian assumption tend to encode non-zero higher-order interactions only when their associated outlier sites are involved. This behavior is reflected in the strong correlation between the Frobenius norm of the connection between a hidden node \(a\) and a visible site \(i\), \(\| \boldsymbol{\hat{W}}_{ia} \|\), and the quantity
\[
\hat{R}_{ia}^{(1)} \coloneq \sqrt{ \frac{1}{q} \sum_{\mu} \mathrm{Var}_{\boldsymbol{u} \sim U}\left[ \frac{1}{q} \sum_{\mu'} \ln \frac{\msfsl{H}_a \left( \boldsymbol{v} \,\big|\, v_i = \mu \right)}{ \msfsl{H}_a \left( \boldsymbol{v} \,\big|\, v_i =\mu' \right) } \right] },
\quad \text{with} \quad
\msfsl{H}_a \coloneq - \ln \left( 1 + e^{c_a + \sum_{i, \mu} W_{ia}^\mu \delta_{\mu}^{v_i} }  \right),
\]
as shown in the insets of Fig.~\ref{fig:gaussian_validation}. The quantity \(\hat{R}_{ia}^{(1)}\) measures how much the effective model encoded by a single hidden node \(a\) deviates from a purely non-interacting (fields-only) model: larger values of \(\hat{R}_{ia}^{(1)}\) indicate stronger higher-order couplings. Consequently, our initial Gaussian approximation, which already involves an exact summation over all combinations of interacting variables, implicitly incorporates, at least partially, a Gaussian mixture approximation for these non-negligible couplings. Owing to its computational efficiency and the possibility of parallelizing computations across hidden nodes, this approximation remains preferable for most practical purposes. For this reason, all results presented in this work rely on it, as it has proven sufficiently robust in practice. That said, cases requiring higher precision may benefit from a more refined treatment, which we leave for future work. The code to identify outliers and refine the effective model parameters is also available in ~\href{https://github.com/DsysDML/couplings_inference}{DsysDML/couplings\_inference}~\cite{github:couplings_inference}.

\section{Modified Blume-Capel Model} 
\label{appendix_blume}
\subsection{Definition of the Model}
The Blume-Capel (BC) Model \cite{blume1966theory, capel1966possibility}, given by
\begin{equation}
     \mathcal{H} (\boldsymbol{s}) = -J^{(2)} \sum_{\langle i, j \rangle} s_i s_j - D \sum_i s_i^2 - h \sum_i s_i,
 \end{equation}
 where $s_i \in \{-1, 0, 1 \}$, can be seen as a spin-1 version of the Ising model. In Eq.~\eqref{blume-capel_hamiltonian}, we introduced a modified version of the above model in which we set $D\!=\!h\!=\!0$ and included three-wise couplings $J^{(3)}$. Considering the following Potts-like Hamiltonian
 \begin{equation}
     \mathcal{H}_{\mathcal{D}} (\boldsymbol{v}) = - \!\!\! \sum_{ \langle i, j, k \rangle } \sum_{\mu, \nu, \eta}  J_{i j k}^{\mu \nu \eta} \delta_{\mu}^{v_{i}} \delta_{\nu}^{v_{j}} \delta_{\eta}^{v_{k}} - \!\! \sum_{ \langle i, j \rangle } \sum_{\mu, \nu}  J_{i j}^{\mu \nu} \delta_{\mu}^{v_{i}} \delta_{\nu}^{v_j},
 \end{equation}
 we obtain the same model defined by Eq.~\eqref{blume-capel_hamiltonian} using the following correspondence between $\boldsymbol{s}$ and $\boldsymbol{v}$
 \begin{align*}
     s_i \!=\! -1 &\leftrightarrow v_i \!=\! 1 \\
     s_i \!=\! 1 &\leftrightarrow v_i \!=\! 2 \\
     s_i \!=\! 0 &\leftrightarrow v_i \!=\! 3,
 \end{align*}
 and defining the pairwise and three-wise parameters of the Potts model for $i, j$, and $k$ neighbor sites as
 \begin{gather*}
     J_{i j}^{\mu \nu} =
     \begin{cases}
     J  & \mathrm{if} \ \mu \!=\! \nu \ \mathrm{and} \ \{\mu, \nu \} \! \not\ni \! 3 \\
     -J & \mathrm{if} \ \mu \! \neq \! \nu \ \mathrm{and} \ \{\mu, \nu \} \! \not\ni \! 3 \\
     0  & \mathrm{if} \ \{ \mu, \nu \} \!\ni\! 3,
     \end{cases}
\ \mathrm{and}, \\
     J_{i j k}^{\mu \nu \eta} =
     \begin{cases}
     J  & \mathrm{if} \ \{ \mu, \nu, \eta\} \!\not\ni \! 3 \ \mathrm{and} \ \mu \!+\! \nu \!+\! \eta\ \mathrm{even}, \\
     -J & \mathrm{if} \ \{ \mu, \nu, \eta\} \!\not\ni \! 3 \ \mathrm{and} \ \mu \!+\! \nu \!+\! \eta\ \mathrm{odd}, \\
     0  & \mathrm{if} \ \{ \mu, \nu, \eta\} \!\ni\! 3,
     \end{cases}
 \end{gather*}
where $\mu, \nu, \eta\!\in\! \{1,2,3\}$.
 Finally, we emphasize that in Fig. 1--b the main text, we showed the mean and standard deviation of the aggregated inferred couplings defined as
\begin{gather}
    \overline{J_{i j}^{(2)}} \coloneqq \frac{1}{4} \left\{ J_{i j}^{22} - J_{ij}^{21} - J_{ij}^{12} + J_{ij}^{11} \right\} + \sum_{\{ \mu, \nu\}\cap \{3\} \neq \emptyset} J_{i j}^{\mu \nu}, \label{eq:BC2}\\
    \overline{J_{i j k}^{(3)}} \coloneqq \frac{1}{4} \left\{ J_{i j k}^{222} - J_{ijk}^{221} - J_{ijk}^{212} - J_{ijk}^{122} + J_{ijk}^{112} + J_{ijk}^{121} + J_{ijk}^{211} - J_{ijk}^{111} \right\} + \sum_{\{ \mu, \nu, \eta\} \cap \{3\} \neq \emptyset} J_{i j k}^{\mu \nu \eta}.\label{eq:BC3}
\end{gather}

\subsection{Extra Figures of the Blume-Capel Inverse Experiment}\label{app:Blumeextra}

In the main text, we analyzed an inverse experiment in which the two- and three-body interaction terms in the ground truth model had equal strength, i.e., $J^{(2)} = J^{(3)} = 1$. In that setting, we observed a clear hierarchy in the learning dynamics: the model consistently learned second-order interactions before higher-order ones. A natural question is whether this behavior is merely a consequence of the specific choice of interaction strengths. To address this, we now consider three additional experiments in which $J^{(2)} = 1$ is fixed, while $J^{(3)}$ takes values 1, 2, and 3. Furthermore, we adopt a fast training regime with $\gamma = 0.1$.

As before, we analyze the structure of the effective model as a function of training time. In Fig.~\ref{fig:BlumeJ3}, we present the counterparts to Fig.~1--\figpanel{b} in the main text, corresponding to the three training runs: dots indicate $J^{(3)} = 3$, dashed lines $J^{(3)} = 2$, and solid lines $J^{(3)} = 1$. We observe that the inferred two-body couplings are similar across all runs, while the main differences emerge only in the three-body terms. In each case, the inferred three-body couplings converge to the expected line $J^{(3)} \beta$, confirming that the earlier learning of lower-order terms is robust to changes in the relative interaction strengths. Perhaps even more striking is that the hierarchy in the learning dynamics remains unchanged: the three-body interactions are consistently learned at the same stage of training, regardless of their absolute strength.

\begin{figure}[ht]
\centering
\includegraphics[width=0.5 \linewidth]{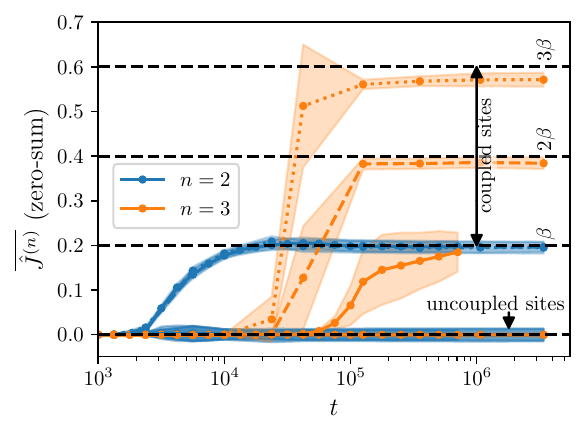}
\caption{Same analysis as in Fig.~1--\figpanel{b} in the main text, now applied to data generated from three modified Blume-Capel models at $\beta = 0.2$, each with fixed two-body interaction strength $J^{(2)} = 1$ and varying three-body strength $J^{(3)} = 1, 2, 3$. The models are trained using a fast schedule ($\gamma = 0.1$). The plots show the mean inferred couplings as a function of training time (no. of parameter updates), averaged separately over coupled and uncoupled sites. Solid lines correspond to $J^{(3)} = 1$, dashed lines to $J^{(3)} = 2$, and dots to $J^{(3)} = 3$. Shaded areas indicate $\pm$ the standard deviation across all corresponding interaction terms. Two-body couplings (in blue) collapse across the three runs, while three-body couplings evolve differently and stabilize at their expected values $J^{(3)} \beta$. The learning hierarchy remains intact: second-order interactions are consistently inferred before third-order ones, regardless of their relative magnitudes.
}
\label{fig:BlumeJ3}
\end{figure}

\section{Direct Coupling Analysis}

\begin{figure}[ht]
\centering
\includegraphics[width=0.9 \linewidth]{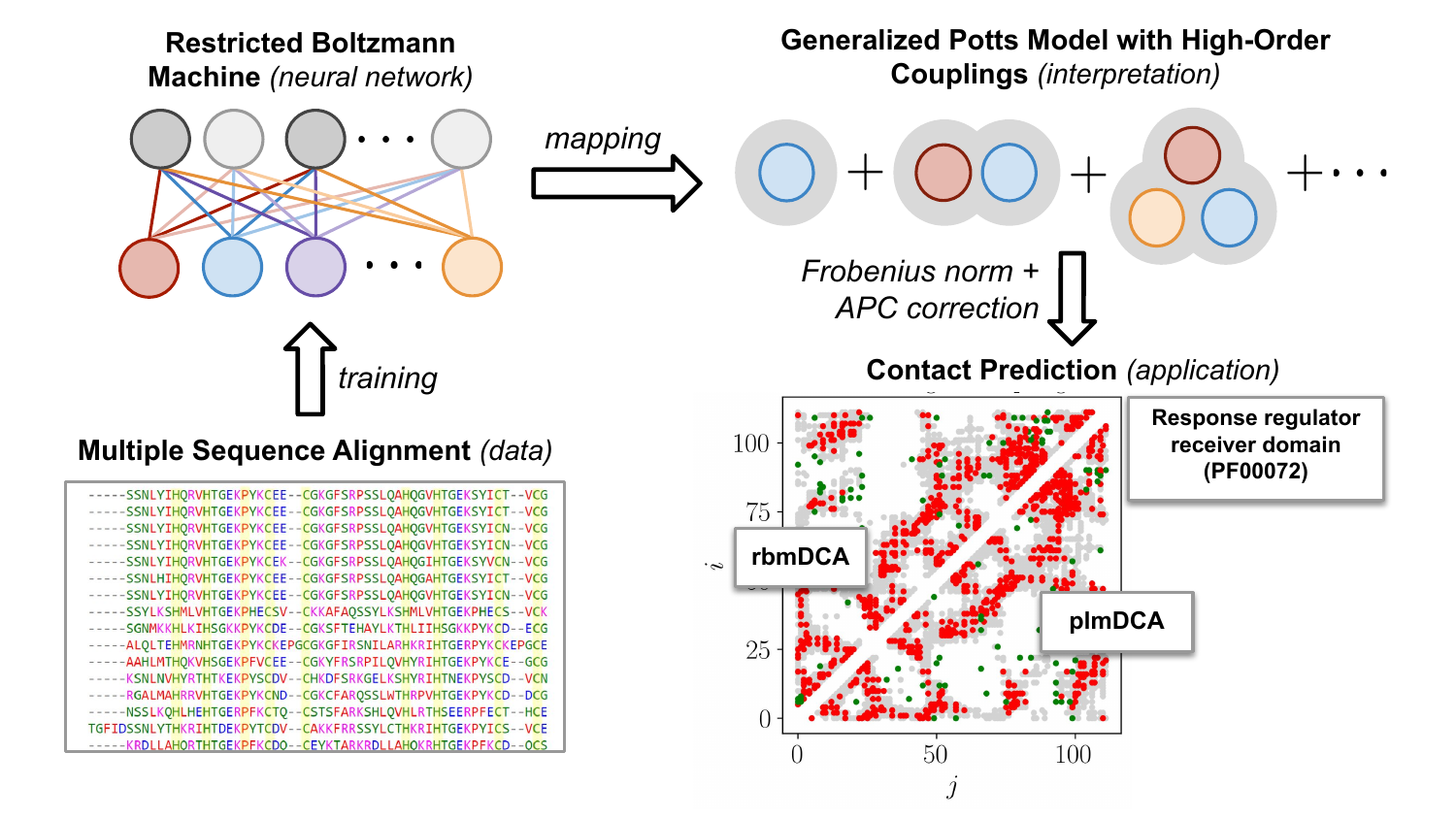}
\put(-460, 115){(\textbf{a})}
\put(-460, 240){(\textbf{b})}
\put(-255, 240){(\textbf{c})}
\put(-255, 140){(\textbf{d})}
\caption{\textbf{Epistatic coupling extraction from Neural Networks}. The above sketches the protein contact prediction task from Multiple Sequence Alignment (MSA) as an example of how mapping neural networks onto the Potts Model improves interpretability. The above \textit{pipeline} depicts, first, the training of the neural network (e.g., an RBM) (\textbf{b}) with data (e.g., MSA) (\textbf{a}). Then, the trained model (\textbf{b}) is mapped onto a Potts-like model (\textbf{c}) using the mapping provided by this work. Finally, parameters of (\textbf{c}) can be used to predict epistatic contacts in the tertiary structure of the protein. (\textbf{d}) shows the contact prediction for the strongest 500 couplings obtained for the Response Regulator Receiver Domain family (Pfam entry: PF00072), where light-gray dots are the contacts of the protein, red dots are true positives, and green dots are false positives. We showed the prediction obtained with our RBM-based inference (rbmDCA) in the upper-left part of the matrix, while the prediction obtained with the well-established pseudo-likelihood inference (plmDCA) is shown in the lower-right part.}
\label{fig:rbmDCA}
\end{figure}

\label{appendix_DCA}
A homologous protein family consists of a group of proteins that share similar biological functions, three-dimensional structures, and a common evolutionary origin. Given a collection of \(M\) homologous protein sequences, these can be arranged into a matrix with \(M\) rows and \(N\) columns, known as a \textit{Multiple Sequence Alignment} (MSA). Constructing an MSA from large ensembles of biological sequences is a complex task, and a detailed discussion falls beyond the scope of this manuscript (for a comprehensive treatment, see the classic reference \cite{durbin1998biological}). 

Each position in an MSA typically assumes one of 21 possible states, corresponding to the 20 standard amino acids and a \textit{gap} that accounts for insertion or deletion mutations occurring throughout evolutionary history. While sequence variation within each column naturally arises from evolution, some positions exhibit significantly less variability due to \textit{stabilizing} selection, which preserves critical functional or structural sites, such as a protein’s binding or active site.

In addition to conserved sites, certain pairs of distant columns exhibit correlated mutations, where a change in one position systematically accompanies a shift in the other. These \textit{epistatic} interactions suggest structural or functional constraints, often indicating physical contacts in the protein’s tertiary structure. Such co-evolving sites arise because interacting amino acids can be replaced by another pair that maintains similar physico-chemical interactions, thereby preserving the protein’s overall structure and function.

Constructing a maximum entropy model to capture the statistical properties of conserved positions and pairwise epistatic contacts in MSA data naturally leads to the  Potts model \cite{cocco2018inverse}:
\begin{equation}
    \mathcal{H}(\boldsymbol{v}) \!=\! - \!\!\!\!\!\!\sum_{ 1 \le i<j \le N} \sum_{\mu, \nu} J_{i j}^{\mu \nu} \delta_{\mu}^{v_i} \delta_{\nu}^{v_j} - \sum_{i, \mu} H_i^{\mu} \delta_{\mu}^{v_i},
    \label{PottsDCA}
\end{equation}
where a strong field $H_i^\mu$ in any state $\mu$ indicates a conserved site in position $i$ and significantly large epistatic couplings values $J_{ij}^{\mu \nu}$ suggest contact between sites $i$ and $j$. The inference of the parameters of \eqref{PottsDCA} from MSA data is normally referred as the \textit{Direct Coupling Analysis} (DCA). Existing DCA methods employ approaches such as pseudo-likelihood maximization \cite{ekeberg2013improved, ekeberg2014fast}, Boltzmann machine learning \cite{muntoni2021adabmdca, rosset2025adabmdca}, and auto-regressive models \cite{trinquier2021efficient}, among others. This work introduces a new method to this framework: RBM-DCA.

As is standard in DCA analysis~\cite{cocco2018inverse}, after inferring the epistatic couplings in Eq.~\eqref{PottsDCA}, the coupling strength between two sites \(i\) and \(j\) is computed as the Fröbenius norm of the corresponding coupling matrix, given by:
\begin{equation}
    F_{ij}^{(2)} = \sqrt{\sum_{\mu, \nu} \left( J_{ij}^{\mu \nu} \right)^2 }.
\end{equation}
Then, we can achieve an ever better contact prediction performance by implementing the \textit{average-product correction} (APC) \cite{dunn2008mutual}:
\begin{equation}
    F_{ij}^\mathrm{APC} = F_{ij}^{(2)} - \frac{\sum_k F_{ik}^{(2)} \sum_{k} F_{kj}^{(2)}}{\sum_{k,l}F_{kl}^{(2)}},
\end{equation}
which intends to minimize any background noise \cite{burger2010disentangling, feinauer2016statistical}.

In the figure \ref{fig:rbmDCA}, we showed the general pipeline followed to conduct a contact prediction from MSA. The MSA data we used correspond to the companion data provided in Ref. \cite{trinquier2021efficient} for the PF00072 protein family, including structural data of a tertiary structure in this family.

The training code used has been incorporated into \href{https://github.com/DsysDML/rbms}{DsysDML/rbms}~\cite{github:rbms}~\cite{github:rbms}, a Python-based library for RBM training. Besides, the coupling extraction and contact routines were implemented in the public repository ~\href{https://github.com/DsysDML/couplings_inference}{DsysDML/couplings\_inference}~\cite{github:couplings_inference}. One of our most significant practical achievements was implementing a parallelization algorithm that makes the computation of high-order couplings achievable; we read typical computation times obtained for the model trained with PF00072 MSA data in table~\ref{tab:computational_times}.
\begin{table}[h]
\caption{Typical computation times obtained for a system with $\Nv =112, \ \Nh = 1000$ and $q=21$. }
\centering
\begin{tabular}{ccc}
\hline
 object & number of parallelized operations& time (ms)  \\
\hline
all fields matrix & $q \!\times\! \Nv $ & $14.20 \ \pm$ 0.210   \\
2-body coupling matrix & $q \!\times\! q$ & $3.14 \ \pm$ 0.0627 \\
3-body coupling 3-rank tensor & $q \!\times\! q \!\times\! q$ & $55.1 \ \pm \ 1.73 $ \\
\hline
\end{tabular}
\label{tab:computational_times}
\end{table}

The above computations were done on an NVIDIA Geforce RTX 3090 graphic card with 24 GB memory. All external fields can be computed in parallel, while 2-body and 3-body couplings were calculated for each pair or triad sequentially. The total time required to compute all the 2-body and 3-body couplings was approximately 20s and 3h 20min, respectively.
\\

\begin{figure}[ht]
    \centering
    \includegraphics[width=1\linewidth]{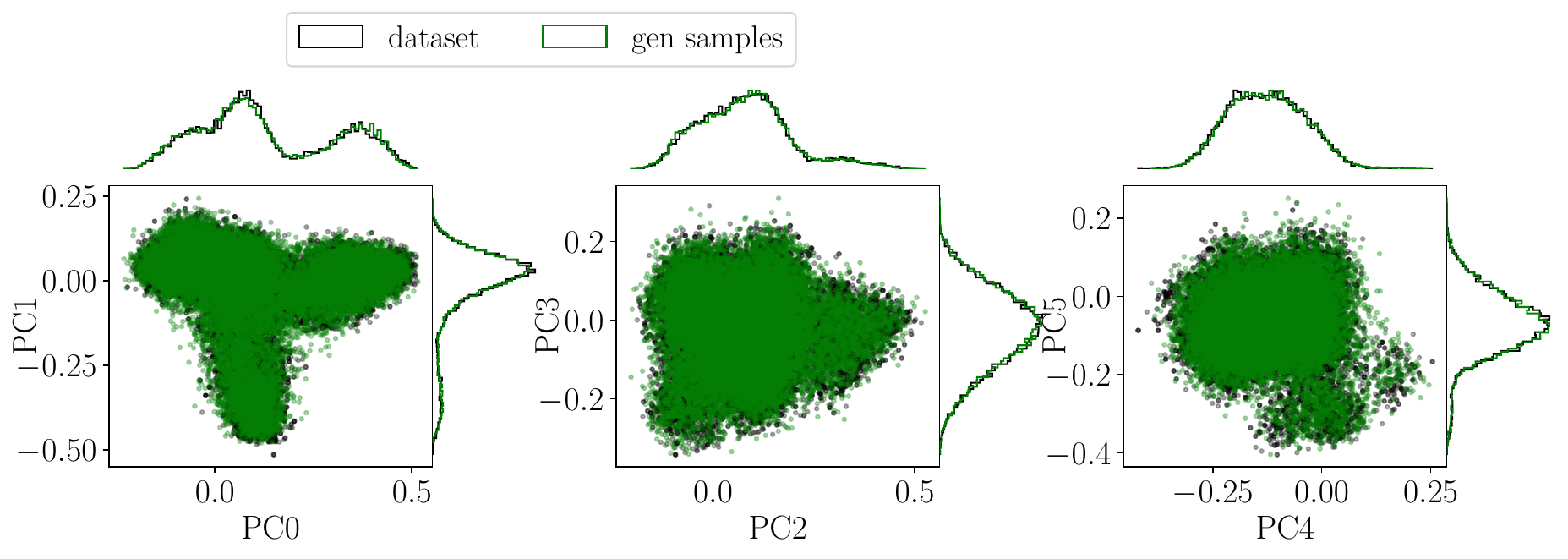}
    \put(-500, 140){\figpanel{a}}
    
    \includegraphics[width=1\linewidth]{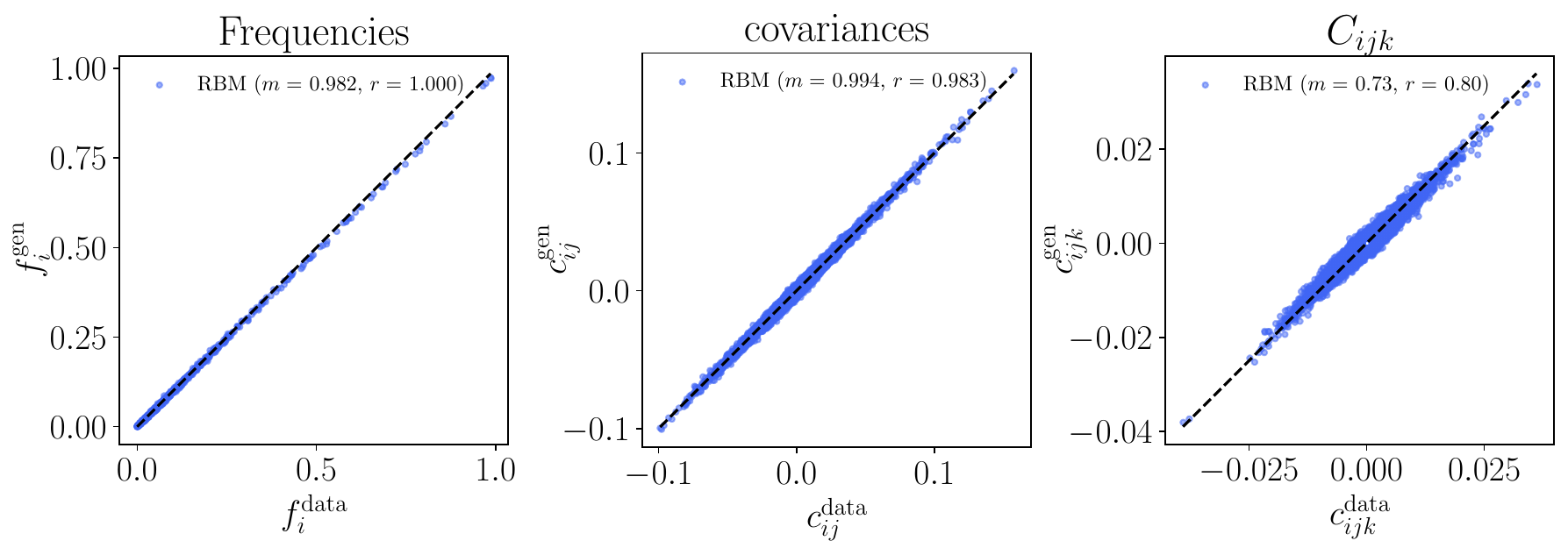}
    \put(-510, 170){\figpanel{b1}}
    \put(-330, 170){\figpanel{b2}}
    \put(-170, 170){\figpanel{b3}}
    
    \caption{{\bf Generative performance of RBMs.} In~\figpanel{a}, we compare the generated data (green) and the real data (black), each consisting of 20,000 samples, projected onto the first six principal components of the dataset. Marginal histograms along the sides display the distribution of data along each principal direction. In~\figpanel{b}, we assess the agreement between  the generated and the real data using scatter plots that compare single-site frequencies \figpanel{b1}, pairwise covariances \figpanel{b2}, and three-body correlations \figpanel{b3}. Each point represents a site, a pair, or a triplet of sites, respectively. The diagonal line marks perfect agreement between the empirical and model-generated statistics.
    }
    \label{fig:sampling}
\end{figure}

\begin{figure}[ht]
    \centering
    \includegraphics[height=5.1cm]{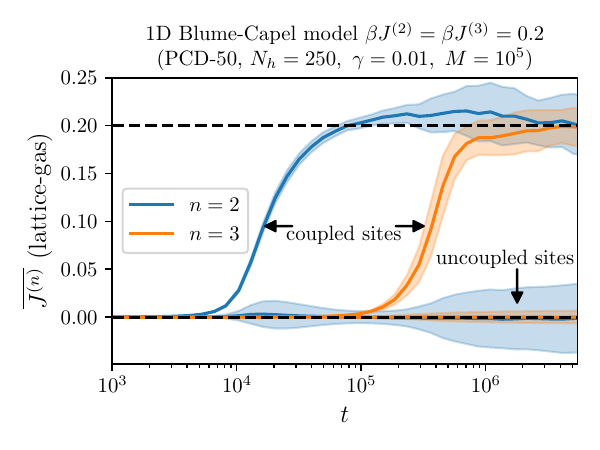}
    \includegraphics[height=5.1cm]{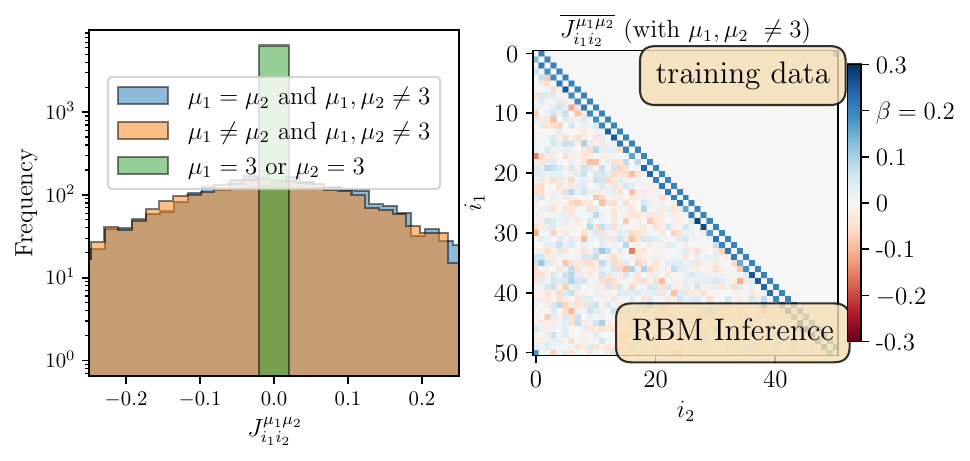}
    \caption{\textbf{Inference with the lattice-gas model in the Blume-Capel model.} We reproduce the same figures shown in Fig. 1--\figpanel{b} to \figpanel{d} of the main text, but using the lattice-gas gauge to infer the effective model parameters. While this approach yields reasonable average estimates once the average over the colors is computed, the fluctuations are significantly larger than in the zero-sum gauge, and the method fails to capture the color-dependent couplings accurately.
}
    \label{fig:latticegas}
\end{figure}

\begin{figure}[ht]
    \centering
    \includegraphics[height=4.5cm]{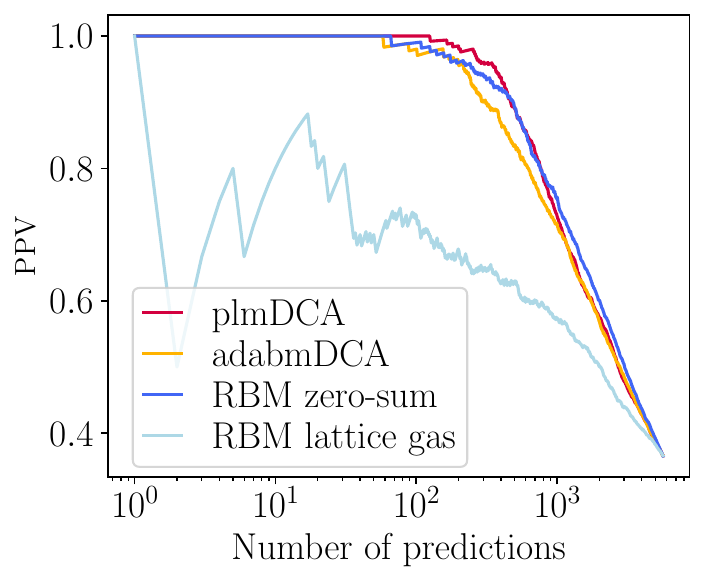}
    \includegraphics[height=4.5cm]{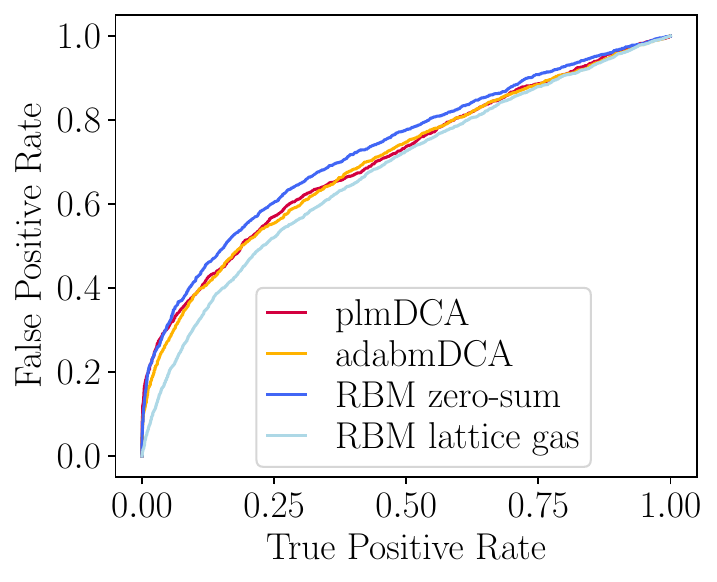}
    \includegraphics[height=5.1cm]{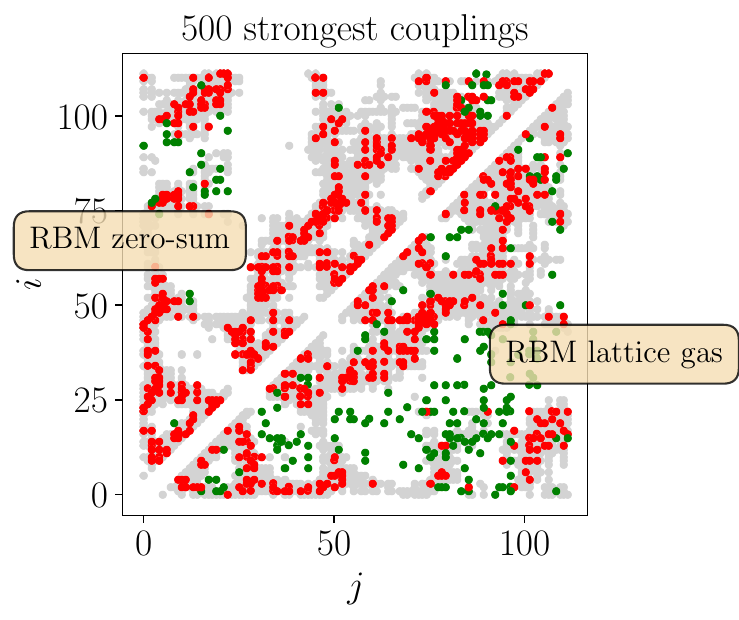}
    \put(-490, 135){(\textbf{a})}
    \put(-330, 135){(\textbf{b})}
    \put(-160, 135){(\textbf{c})}
    \caption{\textbf{Inference with the lattice-gas model in the protein family.} In \figpanel{a} and \figpanel{b}, we show the PPV and ROC curves, respectively, for contact predictions obtained using the RBM in the lattice-gas gauge (light blue), and compare them with the results from Fig.~2--\figpanel{b} in the main text for the other inference methods. The predictive performance is clearly inferior in the lattice-gas gauge. In \figpanel{c}, we directly compare the contact maps inferred with the RBM in the zero-sum (upper triangle) and lattice-gas (lower triangle) gauges. Grey dots indicate physical contacts from the three-dimensional structure, red marks true positives, and green marks false positives.
}
    \label{fig:latticegasprotein}
\end{figure}

\section{Generation Performance of the RBM}\label{generation_appendix}
As shown in the main section, the inference of pairwise couplings using RBMs provides a more informative description of structural contacts than that obtained with standard Inverse Potts methods, or Boltzmann Machines (BMs), and is comparable in quality to pseudo-likelihood maximization (plm) methods. While plm approaches are highly effective at identifying the strongest contacts, they are known to perform poorly when used as generative energy-based models, as they fail to reproduce the statistical properties of the original dataset. In contrast, BMs excel at generating protein family data. Here, we demonstrate that RBMs offer a compelling compromise: they are capable of inferring accurate interaction networks while simultaneously generating realistic synthetic data that faithfully captures the dataset's statistics. 

To this end, in Fig.~\ref{fig:sampling}--\figpanel{a}, we compare 20,000 training datapoints with an equal number of samples generated by the model, projected onto the first six principal components of the dataset. In Fig.~\ref{fig:sampling}--\figpanel{b}, we present scatter plots comparing key statistical observables computed from the real and generated data: site frequencies (Fig.~\ref{fig:sampling}--\figpanel{b1}), pairwise covariances $c_{ij}$ (Fig.~\ref{fig:sampling}--\figpanel{b2}), and three-body correlations $c_{ijk}$ (Fig.~\ref{fig:sampling}--\figpanel{b3}). Each panel reports the Pearson correlation coefficient and the slope of the best-fit linear regression between the model and empirical values. To avoid excessive memory usage, for the three-body correlations we restrict the analysis to triplets where $c_{ijk} > 0.005$ in either the real or the generated dataset.

\section{Inference with the Lattice-Gas Gauge}
\label{latticegas_appendix}

As mentioned in the main text, we observe that the zero-sum gauge yields more accurate inference of the model parameters from the RBM than the lattice-gas gauge, despite the practical advantage of the latter, which does not require large-$N$ approximations for its computation. Nevertheless, the effective parameter formulas derived for the lattice-gas gauge can also be applied. We compare the resulting estimates with the ground truth in the Blume-Capel model and with structural contacts in the protein case.

In Fig.~\ref{fig:latticegas}, we show the analogue of Figs.~1--\figpanel{b} to \figpanel{d} of the main text, this time using the lattice-gas gauge to infer the model parameters. Although the ground-truth parameters of the Blume-Capel model are identical in both gauges, the inference is visibly more accurate in the zero-sum gauge, as evidenced by the reduced fluctuations observed in Fig.~1--\figpanel{b} in the main text compared to Fig.~\ref{fig:latticegas}. Notably, the correct 1-D connectivity pattern is recovered only after averaging the contributions of each pair or triplet over all colors (as in Eqs.~\eqref{eq:BC2} and ~\eqref{eq:BC3}). The lattice-gas gauge fails to infer the color-resolved couplings accurately.

In Fig.\ref{fig:latticegasprotein}, we repeat the analysis of Fig.~2--\figpanel{b} from the main text, assessing the ability of the inferred coupling matrix to predict structural contacts in the protein. This time, we also evaluate the performance of the effective model obtained in the lattice gas gauge. As before, the predictive performance deteriorates markedly, as seen in the reduced accuracy across both statistical tests. The PPV curve further highlights a complete reorganization of the couplings: the strongest inferred interactions no longer align with physical contacts in the structure. The same deterioration of the contact prediction is clearly observed in the contact map as illustrated in Fig.\ref{fig:latticegasprotein}-\figpanel{c}.

\end{widetext} 

\end{document}